\documentclass[a4paper,10pt]{article}
\pdfoutput=1 

\usepackage{jheppub} 

\usepackage[utf8]{inputenc}
\usepackage[T1]{fontenc}
\usepackage[english]{babel}
\usepackage{fancyhdr}
\usepackage{booktabs}
\usepackage{hyperref}
\usepackage{amssymb,amsfonts,amsmath}
\usepackage{graphics,graphicx,pstricks,color,colortbl}
\usepackage{slashed}
\usepackage{cleveref}
\usepackage{subcaption}
\usepackage{multirow}
\usepackage{booktabs}

\newcommand{\ab}{{\rm\ ab}}
\newcommand{\fb}{{\rm\ fb}}

\newcommand{\GeV}{{\rm\ GeV}}
\newcommand{\TeV}{{\rm\ TeV}}

\title{\boldmath Vector-like quarks decaying into singly and doubly charged bosons at LHC}

\author[a]{Gennaro Corcella}

\author[b]{, Antonio Costantini}

\author[c]{, Margherita Ghezzi}

\author[d,e]{, Luca Panizzi}

\author[f]{, Giovanni Marco Pruna}

\author[d,g]{and Jakub \v{S}alko}

\affiliation[a]{INFN, Laboratori Nazionali di Frascati, Via E. Fermi 54, 00044, Frascati, Italy}
\affiliation[b]{Centre for Cosmology, Particle Physics and Phenomenology {\rm (CP3)},\\
Universit\'e catholique de Louvain, Chemin du Cyclotron, B-1348 Louvain-la-Neuve, Belgium}
\affiliation[c]{Institut f\"{u}r Theoretische Physik, Eberhard Karls Universit\"{a}t T\"{u}bingen, Auf der Morgenstelle 14, 72076 T\"{u}bingen, Germany}
\affiliation[d]{Department of Physics and Astronomy, Uppsala University, Box 516, SE-751 20 Uppsala, Sweden}
\affiliation[e]{School of Physics and Astronomy, University of Southampton, Highfield, Southampton SO17 1BJ, UK}
\affiliation[f]{Department of Theoretical Physics, Maynooth University,
  Maynooth, Co.~Kildare, Ireland}
\affiliation[g]{Albert Einstein Center for Fundamental Physics, Institut f\"{u}r Theoretische Physik, Universit\"{a}t Bern, Sidlerstrasse 5, CH-3012 Bern, Switzerland.}

\emailAdd{gennaro.corcella@lnf.infn.it}
\emailAdd{antonio.costantini@uclouvain.be}
\emailAdd{margherita.ghezzi@itp.uni-tuebingen.de}
\emailAdd{luca.panizzi@physics.uu.se}
\emailAdd{giovanni.marco.pruna@lnf.infn.it}
\emailAdd{salko@itp.unibe.ch}

\abstract{
  We investigate the production of vector-like quarks with charge $5/3$ at the LHC and their subsequent decays into new singly or doubly charged bosons
  plus a heavy quark (top or bottom). In particular, we explore final states
  with same-sign di-leptons (electron or muon pairs), with the leptons
  coming from the decay of the new bosons and, in the case of production
  of singly charged bosons, from top quarks as well.
  These processes are predicted by classes of models based on extensions
  of the gauge group of the Standard Model, such as the 331 Model, where the electroweak symmetry is described by $SU(3)_L\times U(1)_X$, $X$ being a new abelian charge.
  For this purpose, a CMS search for vector-like partners with charge 5/3 decaying into $Wt$ is recast to obtain model-independent bounds and projected reaches at future luminosity stages of the LHC.
    The results are then interpreted
    as mass bounds for the new particles predicted in the 331 Model and
    as a constrain on the scale of its spontaneous symmetry breaking.
    The complete set of model-independent results are provided as recast
    efficiencies, to allow for 
  reinterpretation in different scenarios.
}

\preprint{CP3-21-47}

\begin{document}


\maketitle
\flushbottom

\section{Introduction}\label{sec:intro}
In recent years, the Large Hadron Collider (LHC) has confirmed the description of the dynamics of elementary particles given by the Standard Model (SM). The discovery of the Higgs boson~\cite{Aad:2012tfa,Chatrchyan:2012ufa} and the subsequent measurement of its properties~\cite{Sirunyan:2018koj} represent the most considerable achievement of the LHC Physics programme so far. Nonetheless, it is well-known that theoretical and experimental arguments call for an extension of the SM. The issue of naturalness, the Dark Matter and baryogenesis problems, and the unknown mechanism behind the origin of neutrino masses are some of these reasons. 

New vector-like quarks (VLQs) are often present in many of the extensions of the SM. VLQs are heavy quarks whose left- and right-handed chiral components transform under the same representation of the SM gauge group. They are predicted for example in composite Higgs model, especially in the framework of partial compositeness~\cite{Kaplan:1983fs,Kaplan:1991dc,Gripaios:2009pe,Panico:2016ull}, or can be related to Kaluza--Klein excitations of the known SM quarks \cite{Antoniadis:1990ew,Csaki:2003sh,Cacciapaglia:2009pa}. They can also appear in models with extended gauge symmetry \cite{ArkaniHamed:2002qx,Abbas:2017vle}. The phenomenology of VLQs interacting only with SM particles has been widely studied in literature \cite{Lavoura:1992np,delAguila:2000rc,AguilarSaavedra:2005pv,Cynolter:2008ea,AguilarSaavedra:2009es,Mrazek:2009yu,Cacciapaglia:2012dd,Berger:2012ec,Okada:2012gy,DeSimone:2012fs,Falkowski:2013jya,Buchkremer:2013bha,AguilarSaavedra:2013qpa,Matsedonskyi:2014mna,Matsedonskyi:2015dns,Panella:2017spx,Barducci:2017xtw} and both ATLAS and CMS have recently devoted a considerable effort in the analyses establishing bounds on this type of new particles, both in pair production \cite{Aaboud:2017zfn,Aaboud:2017qpr,Aaboud:2018xuw,Aaboud:2018saj,Aaboud:2018xpj,Aaboud:2018wxv,Aaboud:2018pii,Sirunyan:2017pks,Sirunyan:2018qau,Sirunyan:2018omb,Sirunyan:2019sza,Sirunyan:2018yun,Sirunyan:2020qvb} and single production \cite{Aaboud:2018saj,Aaboud:2018ifs,Sirunyan:2017ynj,Sirunyan:2018fjh,Sirunyan:2018ncp,Sirunyan:2019xeh}.

In specific new Physics (NP) realisations, VLQs can have different charge assignments under the SM electroweak (EW) gauge group $SU(2)_L\times U(1)_Y$. Hence, there is the possibility of having VLQs with exotic electric charges, \emph{e.g.} a new state with electric charge $5/3$ ($X_{5/3}$) interacting with the third generation SM quarks. In minimal extensions of the SM, the $X_{5/3}$ can only be part of VLQ doublets or triplets and decay only to the $W$ boson and an up-type quark or (if the mass splitting allows it) another VLQ with charge $2/3$. The phenomenology of the $X_{5/3}$ in minimal SM extensions has also been studied in literature~\cite{Mrazek:2009yu,Cacciapaglia:2012dd,DeSimone:2012fs,Matsedonskyi:2014mna}.

The dominant mechanism for $X_{5/3}$ production at the hadron colliders is via quantum-chro\-mo\-dy\-na\-mics (QCD) processes, which yield particle-antiparticle pairs. Since the pair production involves exclusively the SM QCD coupling, the tree-level cross section is independent of the $X_{5/3}$ properties, other than its mass. The $X_{5/3}$ VLQ can also be singly produced via EW processes: 
such a production mode has a smaller phase-space suppression even at large VLQ masses, but the cross section is more model-dependent,
as it depends on the EW coupling of the $X_{5/3}$ with the $W$ boson, which can be reduced due to the mixing
angles regulating the quark field rotations.
Experimental searches have been performed in both channels~\cite{Aaboud:2018uek,Aaboud:2018xpj,Sirunyan:2018yun}.

Recently, exotic decays of VLQs have been the subject of intense study~\cite{Serra:2015xfa,Aguilar-Saavedra:2017giu,Chala:2017xgc,Bizot:2018tds,Han:2018hcu,Xie:2019gya,Benbrik:2019zdp,Cacciapaglia:2019zmj,Aguilar-Saavedra:2019ghg}, driven on the one hand by the 
non-observation of SM decays of such particles, which have been searched
for at the LHC, on the other hand
by the fact that decays into new scalars or vectors are often predicted to be dominant in many realisations of physics beyond the Standard Model. In this context, exotic decays of the $X_{5/3}$ have also been considered, mostly justified by composite models, which favour final states with the presence of heavy SM fermions~\cite{Bizot:2018tds,Xie:2019gya}. 

Extensions of the gauge group of the SM, however, can allow decays of such new states into lighter families.
As a case study, the $331$ Model~\cite{Singer:1980sw,Valle:1983dk,Pisano:1991ee,Frampton:1992wt,Foot:1994ym,Hoang:1995vq},
where the EW symmetry group of the SM, $SU(2)_L \times U(1)_Y$, is replaced with $SU(3)_L \times U(1)_X$,
contains a heavy quark with electric charge $5/3$ as well as new singly and doubly charged vector bosons in its spectrum,
together with other heavy/exotic new particles. The new vector bosons can decay with equal branching ratios into any SM lepton family
(see References~\cite{Corcella:2017dns,Corcella:2018eib,Corcella:2021upj} for recent phenomenological analyses of
the 331 Model in the version of Reference~\cite{Frampton:1992wt}), thus allowing to fully exploit the
same-sign di-lepton channels, whenever they are originated from the decay of heavy $X_{5/3}$ quarks \footnote{In the context of 331 Models, charged vectors
are often called bileptons since they have lepton number $L=\pm 2$.}. 
One can envisage that such processes are characterised by very low SM
backgrounds, in such a way that
strong bounds on new particles' masses can be set.

The aim of the paper is to present a model-independent analysis of exotic decays of the $X_{5/3}$ into singly and doubly  charged bosons, either scalars or vectors, and to reinterpret the mass bounds in terms of constraints on the parameters of the 331 Model. 
The model-independent analysis is performed through the recast of a search for heavy fermionic partners of the top quark (with charge $5/3$). Specifically, we have recast the CMS search in Reference~\cite{Sirunyan:2018yun}, corresponding to $35$ fb$^{-1}$ of integrated luminosity, targeting the pair production of $X_{5/3}$ and its decay into $W^+\, t$. 
The model-independent results are presented as optimal bounds for the extreme hypotheses of 100\% branching ratios of the new particles into specific channels, and as a database of recast efficiencies which can then be used to reconstruct any scenario where the $X_{5/3}$ decays into singly or doubly charged bosons, leading to a final state with light leptons. 
These results are then reinterpreted in terms of bounds on the masses of the new particles in the context of the 331 Model, where the $X_{5/3}$ has mass-dependent BRs and the new bosons can decay into any family of SM leptons. The bounds are then ultimately used to determine the scale of the spontaneous symmetry breaking of the 331 gauge symmetry. 
Both model-independent bounds and limits on the 331 Model are provided for the luminosity of the CMS search and as exclusion reaches for higher luminosity stages of the LHC, specifically at $300$ fb$^{-1}$ and $3$ ab$^{-1}$ (HL-LHC). The possibility to discriminate between different scenarios if excesses are observed is also briefly discussed. Although this analysis is performed in the narrow-width approximation (NWA), the adopted strategy is straightforwardly adaptable to scenarios where the new particles have a finite width.

The paper is organized as follows. In Section~\ref{sec:LHCpheno} the simplified model of the $X_{5/3}$ and the model-independent analysis of its production and decay into charged and doubly charged bosons are presented. The reinterpretations of the results in the framework of the 331 Model is given in Section~\ref{sec:331}. Finally, we provide conclusions and perspective for further analyses in Section~\ref{sec:conclusion}.

\section{Model-independent analysis}\label{sec:LHCpheno}
The phenomenology of the $X_{5/3}$ and of the singly and doubly charged exotic bosons (scalars and vectors) can be studied in a model-independent way by considering a simplified Lagrangian, in which the SM is augmented with the new particles and the new interactions, neglecting any further new physics state. A reinterpretation of the results of this analysis assumes therefore that the signal events in the final states are almost entirely originated by the VLQ and the exotic bosons with negligible contributions from other sources.

The simplified Lagrangian can be written as
\begin{equation}
\mathcal L = \mathcal L_{\rm{SM}} + \mathcal L_{\rm{kin}} + \mathcal L_{X} +  \mathcal L_{S} + \mathcal L_{V}\, ,
\label{eq:ex_VLQ_SymbolicLagrangian}
\end{equation}
where $\mathcal L_{\rm SM}$ is the SM Lagrangian, $\mathcal L_{\rm kin}$ contains the kinetic and mass terms for the $X_{5/3}$ and for the exotic bosons. All the masses of new particles, namely $M_X$ for the VLQ, $M_{S^{(+,++)}}$ for the new scalars and $M_{V^{(+,++)}}$ for the new vectors\footnote{In the following text, charges will not be written explicitly unless required to avoid ambiguities.} are considered as free independent parameters in the simplified model. The interaction terms of the Lagrangian are:
\begin{eqnarray}
\mathcal L_{X} &=& \frac{g_w}{\sqrt{2}}\kappa^{XW}_{L,R}\, \bar X_{5/3}\, \slashed W\, P_{L,R}\, t + h.c. \nonumber\\ 
&+&  \kappa^{XSpp}_{L,R}\, \bar X_{5/3}\, S^{++}\, P_{L,R}\, b + \kappa^{XSp}_{L,R}\, \bar X_{5/3}\, S^{+}\, P_{L,R}\, t + h.c. \nonumber\\
&+& \kappa^{XVpp}_{L,R}\, \bar X_{5/3}\, \slashed V^{++}\, P_{L,R}\, b + \kappa^{XVp}_{L,R}\, \bar X_{5/3}\, \slashed V^{+}\, P_{L,R}\, t + h.c.\, ,
\label{eqn:ex_VLQ_XtoBoson_lagrangian}
\end{eqnarray}
\begin{equation}
\mathcal L_{S} = \sum_{l,l' = e,\mu} (\lambda^{S}_{L,R})^{ll'}\, \bar l^c\, S^{++}\, P_{L,R}\, l'  + \sum_{l,l' = e,\mu} (\lambda^{S})^{l\nu_{l'}}\, \bar l\, S^{-}\, \nu_{l'}  + h.c.\, ,
\label{eqn:ex_VLQ_ScalaLagrangian}
\end{equation}
\begin{equation}
\mathcal L_{V} = \sum_{l,l' = e,\mu} (g^{V}_{L,R})^{ll'}\, \bar l^c\, \slashed V^{++}\, P_{L,R}\, l' + \sum_{l,l' = e,\mu} (g^{V})^{l\nu_{l'}}\, \bar l\, \slashed V^{-}\, \nu_{l'} + h.c.\, .
\label{eqn:ex_VLQ_VectorLagrangian}
\end{equation}
Here, $g_w$ is the weak coupling constant and the superscript $c$ denotes charge conjugation. All the coupling strengths $\kappa_{L/R}^{X}$ associated with the interactions of $X_{5/3}$ to singly  and doubly charged bosons are real parameters while the coupling strengths $\{\lambda,g\}^{S/V}$ ($\{\lambda,g\}_{L/R}^{S/V}$) of singly charged (doubly charged) bosons to leptons are real matrices in the flavour space. In this analysis the decays of the exotic bosons are limited to the first two SM lepton families. This choice is meant to describe classes of theoretical models where new bosons can decay with sizable branching ratios to light leptons, such as the 331 Model described in the following, and is furthermore dictated by the similarity in the analysis of final states with electrons or muons, for which the strategies (reconstructions, selections, cuts, \emph{et cetera}) are distinct with respect to channels
involving taus (especially when they are reconstructed through their hadronic decays). 

The model allows for multiple straightforward generalizations. Firstly, the VLQ could decay into quarks of the first or second generations, for which experimental searches focusing mostly on decays into third generation would likely be less sensitive.
Secondly, the new bosons can decay into quarks (of any generation, depending on their mass). Unless the top quark is involved in these decays, however, hadronic decays of the bosons would reduce or completely eliminate same-sign lepton signature in the final state. For these reasons such generalisations described above are not considered in our analysis, but are nevertheless worth exploration in future works.

\subsection{Simulation, recast and analysis strategies}
\label{subsec:simstrategy}

The simplified Lagrangian has been implemented in {\sc FeynRules}~\cite{FeynRules} extending the model described in Reference~\cite{Fuks:2016ftf} and exported in the Universal FeynRules Output (UFO) format~\cite{UFO} to be used within the Monte Carlo event simulator {\sc MadGraph5\_aMC@NLO} \cite{Alwall:2014hca,Frederix:2018nkq}. 

Event simulations have been performed for the generic process
\begin{equation}\label{simu}
  pp\to q_3 \bar q_3 l^\pm l^\mp + \{l^\pm l^\mp,\nu \bar\nu\},
\end{equation}  
where $q_3=\{t,b\}$, $l^\pm=\{e^\pm,\mu^\pm\}$ and $\nu=\{\nu_e,\nu_\mu\}$
and the flavour (charge) of $q_3$ depends on whether an intermediate
$X_{5/3}$ decays into either a singly or a doubly charged boson.
In the simulation of the process in Equation~\ref{simu},
we impose the propagation of $X_{5/3}$ as well as of 
singly and doubly charged bosons, but we do not require them to be
resonant\footnote{In this way, one can have, e.g., subprocesses
  where an off-shell
  $X_{5/3}$ quark is exchanged in the $t$-channel, and not only
  the pair production $X_{5/3}\bar X_{5/3}$.}. 
The presence of BSM particles in the explored topologies suppresses
the irreducible SM contributions in the signal samples.
In this way, {\sc MG5\_aMC} treats all the intermediate particles as off-shell states and correctly deals with the finite width effects and spin correlations, however small they might be. 
For the purposes of an exploratory study of the sensitivity in these channels, the width-over-mass ratio of all the new particles $X_{5/3}$, $S^{\pm,\pm\pm}$ and $V^{\pm,\pm\pm}$ has been fixed to a small value (0.1\%), corresponding to a narrow-width approximation (NWA), and therefore also interference between signal and irreducible SM background has been considered negligible. 
This approximation is strictly valid only when the mass difference between the decaying particle and the decay products is large enough~\cite{Berdine:2007uv}.

Hereafter, our investigation will be carried out at leading order (LO): the LO
matrix-elements computed by  {\sc MG5\_aMC} are convoluted with
the NNPDF 3.0 LO set of parton densities~\cite{Ball:2014uwa}
to give LO cross-sections.
Parton-level events are subsequently provided with 
parton showers and hadronisation by means of {
  \sc Pythia 8.2}~\cite{Pythia8}, while
the detector simulation will be accounted for using
{\sc Delphes 3} \cite{Delphes}. 

\begin{figure}[!htbp]
\centering
\includegraphics[width=.485\textwidth]{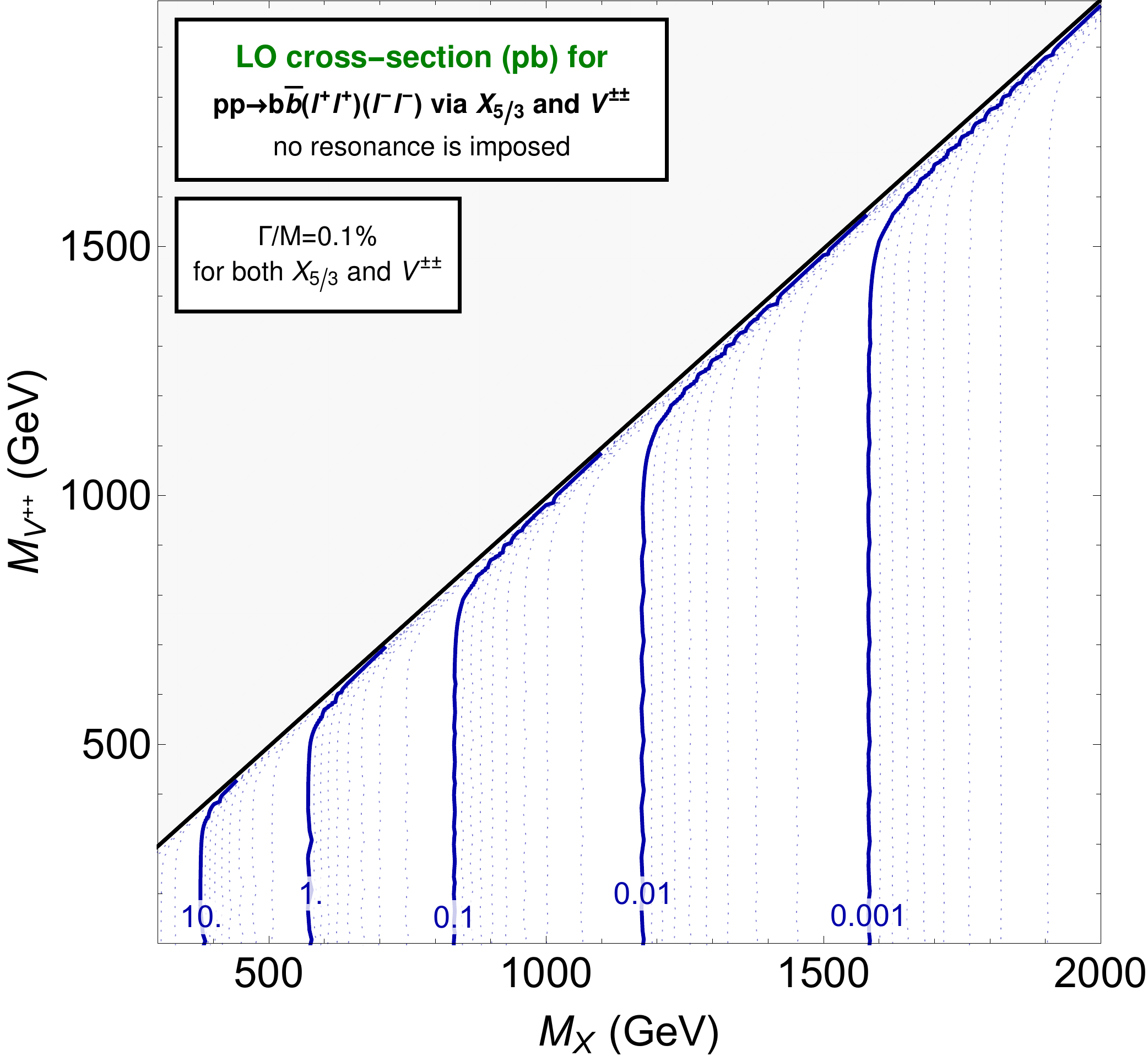}
\caption{\label{fig:xs} LO cross-sections (blue lines)
in the $\{M_X,M_{V^{\pm\pm}}\}$ plane for the process of Equation~\ref{bb4l}
  at for LHC for $\sqrt{s}=13\TeV$, assuming 100\% BRs. }
\end{figure}

In Figure~\ref{fig:xs} we show in the $(M_X,M_{V^{\pm\pm}})$ plane
the values of the cross-sections (blue solid and dotted lines) 
at the leading order (LO) for a center-of-mass energy of $13\TeV$ and the
$2\to 6$ process
\begin{equation}\label{bb4l}
  pp\to b\bar b\ (l^+l^+)\ (l^-l^-),
  \end{equation}
where, once again, $X_{5/3}$ and $V^{\pm\pm}$ are present as intermediate
particles, but are not forced to be resonant.
From Figure~\ref{fig:xs} one can learn that 
the cross-section is essentially independent of the mass of the charged boson and that the NWA breaks as the mass difference approaches the kinematic limit corresponding to the bottom mass. Larger deviations from this behaviour would arise if the width of the $X_{5/3}$ or of the bosons were larger
with respect to the corresponding masses.

The {\sc MadAnalysis5} \cite{MadAnalysis5} framework is then used for recasting experimental results from a CMS search at the centre-of-mass energy $\sqrt{s}=13$ TeV targeting the production of $X_{5/3}$ decaying to $Wt$~\cite{Sirunyan:2018yun}.\footnote{The code is available in the {\sc Dataverse} repository~\cite{DVN/DQZWYL_2021}. Validation documentation about the recasting is provided in Appendix~\ref{app:recastingvalidation}.} This search considers final states characterised by same-sign di-leptons or a single lepton in a dataset corresponding to an integrated luminosity of 35.9 fb$^{-1}$. As the final states considered in our scenarios always contain at least two light leptons (of opposite sign), the single-lepton analysis has not been recast. Furthermore, results from the same-sign di-lepton final state are based on a cut-and-count analysis, while results from the single-lepton final state exploit a shape analysis, which would make the validation of our implementation more challenging in exchange for a potentially low sensitivity due to the large reconstruction efficiency of light leptons.
Three signal regions (SRs) are defined in the same-sign di-lepton analysis, sharing the same set of cuts and differing only by the flavour of the signal lepton pair, namely $ee$, $\mu\mu$ and $e\mu$. Details on the cuts and the
characterisation of the signal regions will be provided in Appendix~\ref{app:recastingvalidation}.

This simulation and recasting procedure was used to scan over the $\{M_X,M_B\}$ mass plane with the kinematical constraint $M_X > M_B+m_q$, where $M_B$ generically corresponds to the mass of the (doubly ) charged boson into which the $X_{5/3}$ decays 

The efficiencies $\varepsilon_{\rm SR}(M_X,M_B)$, corresponding to the different SRs and function of the masses of $X$ and $B$,\footnote{In general $\varepsilon_{\rm SR}$ depends also on the total widths, but as mentioned above, in our analysis the widths are not free parameters, as they are computed depending on the masses to respect the relation $\Gamma/M=0.1\%$ for both $X_{5/3}$ and the exotic bosons.} are obtained from the recast, and used to compute the corresponding fiducial cross sections of the signal as $\hat \sigma_{\rm SR} = \varepsilon_{\rm SR}\, \sigma\,$, where $\sigma$ is the theoretical cross section of the signal process. 
The number of signal events at a corresponding integrated luminosity, $s_{\rm SR}(\mathcal L) = \mathcal L\, \hat \sigma_{\rm SR}$, is then used in combination with the number of expected background events $b_{\rm SR} \pm \Delta b_{\rm SR}$ in each signal region to compute the exclusion confidence level for the signal. We have adopted the `exact Asimov significance' for exclusion and discovery, proposed in Reference~\cite{Bhattiprolu:2020mwi}, and we have used the corresponding software code {\sc Zstats}~\footnote{\url{https://github.com/prudhvibhattiprolu/Zstats}} for our estimations. 


Besides obtaining the current recast bounds for the various scenarios, we provide projection for luminosities at future experiments.
We consider two representative luminosities: $\mathcal L = 300 \fb^{-1}$, corresponding to the nominal luminosity at the end of Run 3 of LHC and $\mathcal L = 3 \ab^{-1}$, corresponding to the nominal luminosity for the High-Luminosity phase of the LHC (HL-LHC). In principle, both LHC Run 3 and HL-LHC will run with a center-of-mass energy $\sqrt{s}=14\TeV$  but the projections using events generated at $13\TeV$ represent a good approximation at this level of the analysis.
Systematic uncertainties in the expected numbers of signal and background events are known to dominate over statistics at high luminosities, as the relative contribution of the former stays constant while for the latter it scales as $1/\sqrt{\mathcal L}$.
Therefore, for the projections we assume that the total uncertainties are dominated by systematics and their relative contribution is fixed to an optimistic value of 5\%.

\subsection{Results in the $\mathbf{ee}$ SR and for $\mathbf{S^{\pm(\pm)},V^{\pm(\pm)}}$ decays into electrons}

Using the simplified Lagrangian in Equation~\ref{eq:ex_VLQ_SymbolicLagrangian}, we show in this section model-independent results, obtained considering the scenario in which the (doubly ) charged scalar/vector bosons emerging from the $X_{5/3}$ decay into electrons, {\it i.e.}
\begin{equation}
  \{ S^\pm,V^\pm \} \to e^\pm \overset{\text{\tiny(--)}}{\nu}_e\ ;\ \{ S^{\pm\pm},V^{\pm\pm} \} \to e^\pm e^\pm .
\end{equation}
  The flavour of the neutrino is of course not strictly relevant, but we impose that the new particles decay with BRs of 100\% in one specific channel to allow for reinterpretation of the results in the general case: if decays such as $\{ S^\pm,V^\pm \} \to e^\pm \overset{\text{\tiny(--)}}{\nu}_\mu$ are also allowed, it is enough to sum the branching ratios for decays into $e^\pm$ and all allowed neutrino flavours.
\begin{figure}[t!]
\centering
\includegraphics[width=.465\textwidth]{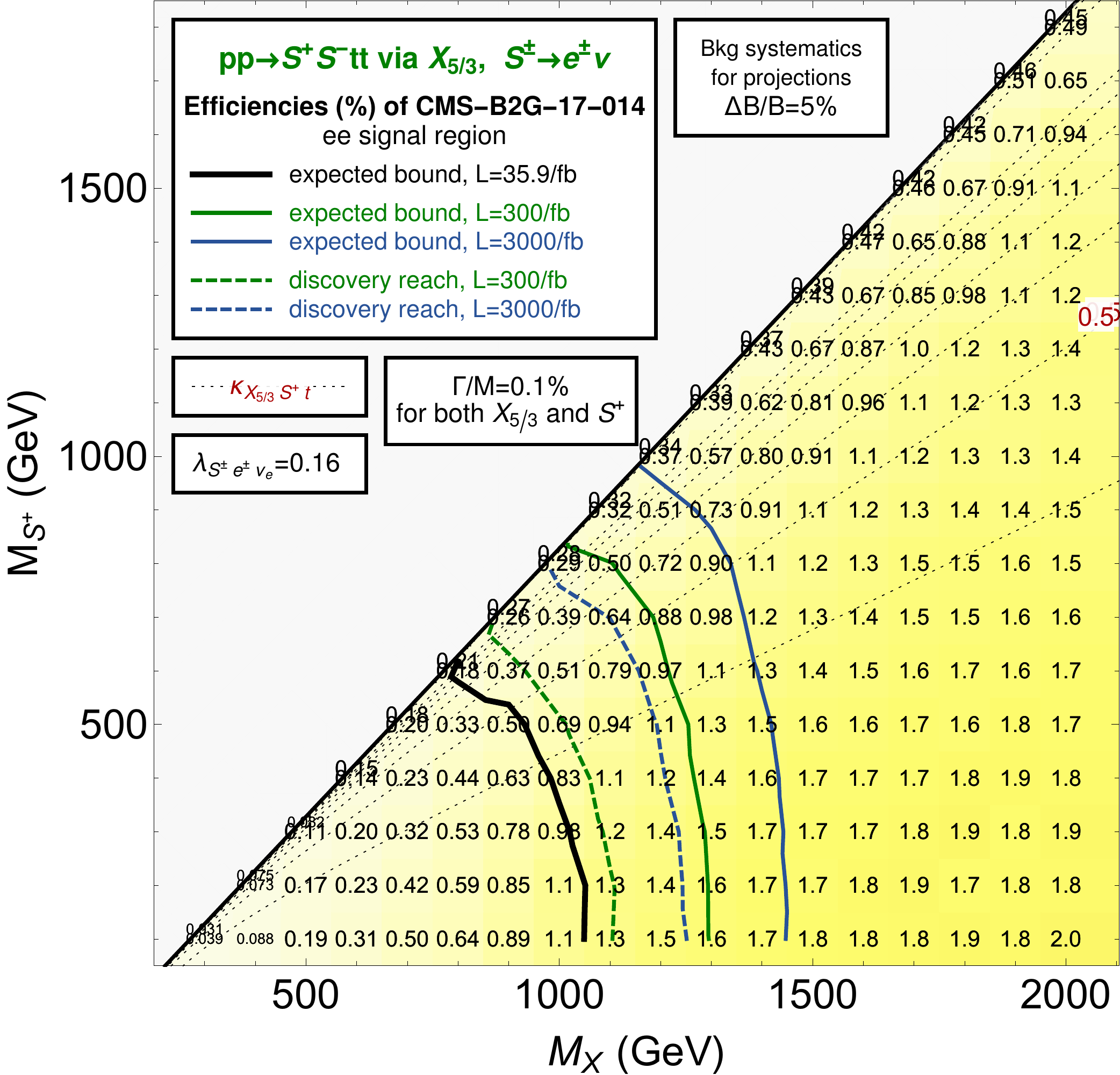}
\includegraphics[width=.465\textwidth]{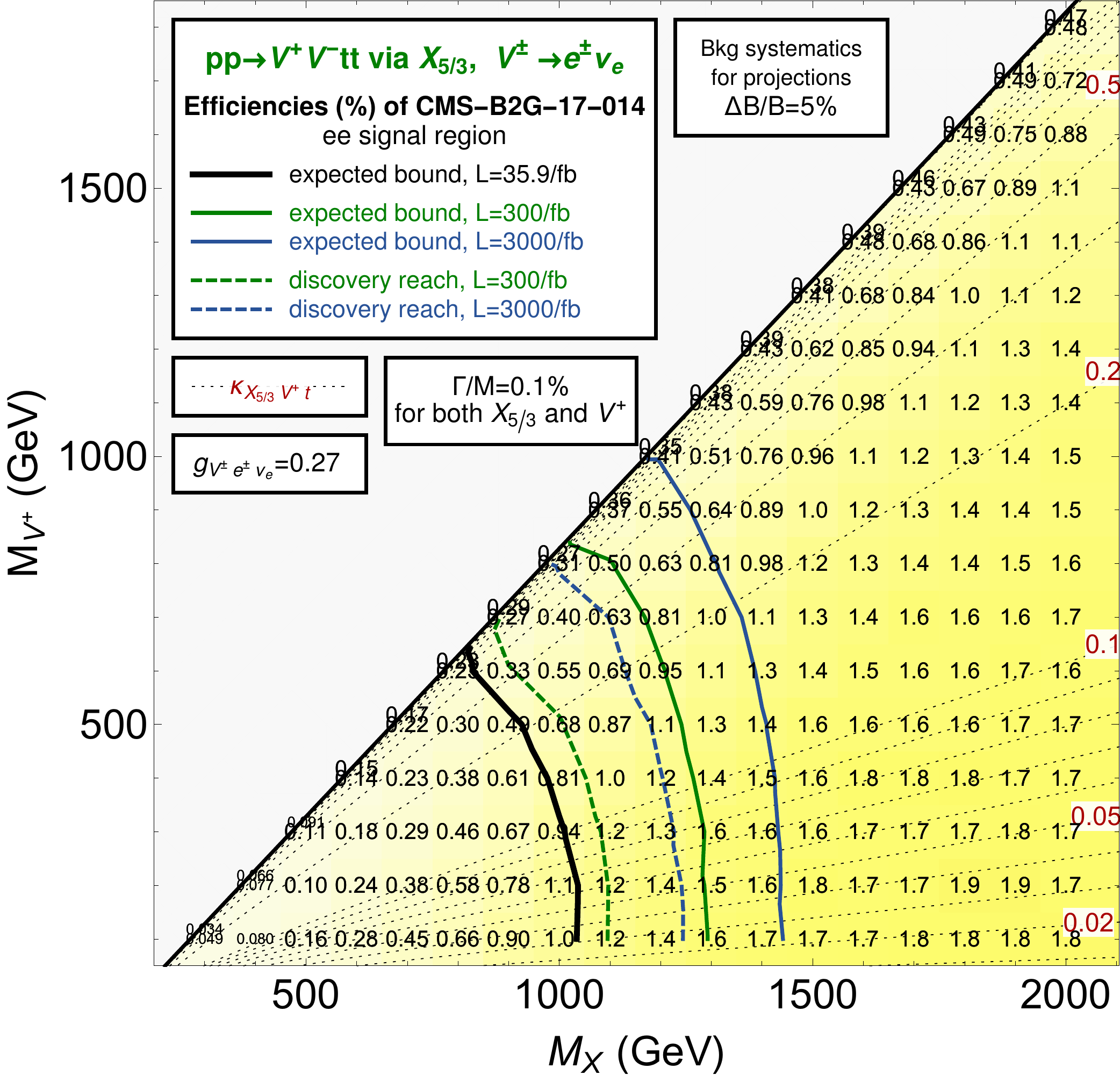}
\includegraphics[width=.465\textwidth]{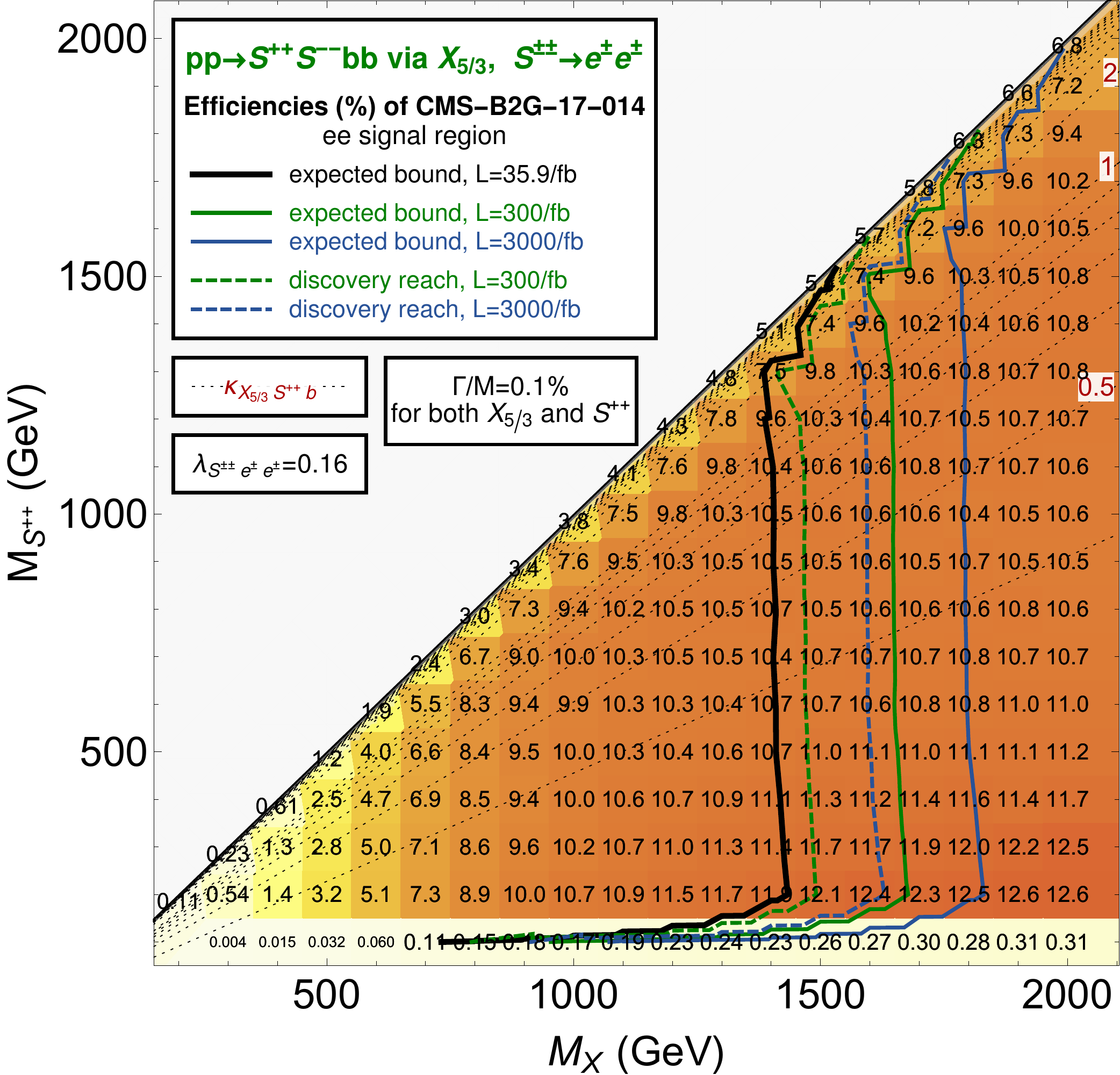}
\includegraphics[width=.465\textwidth]{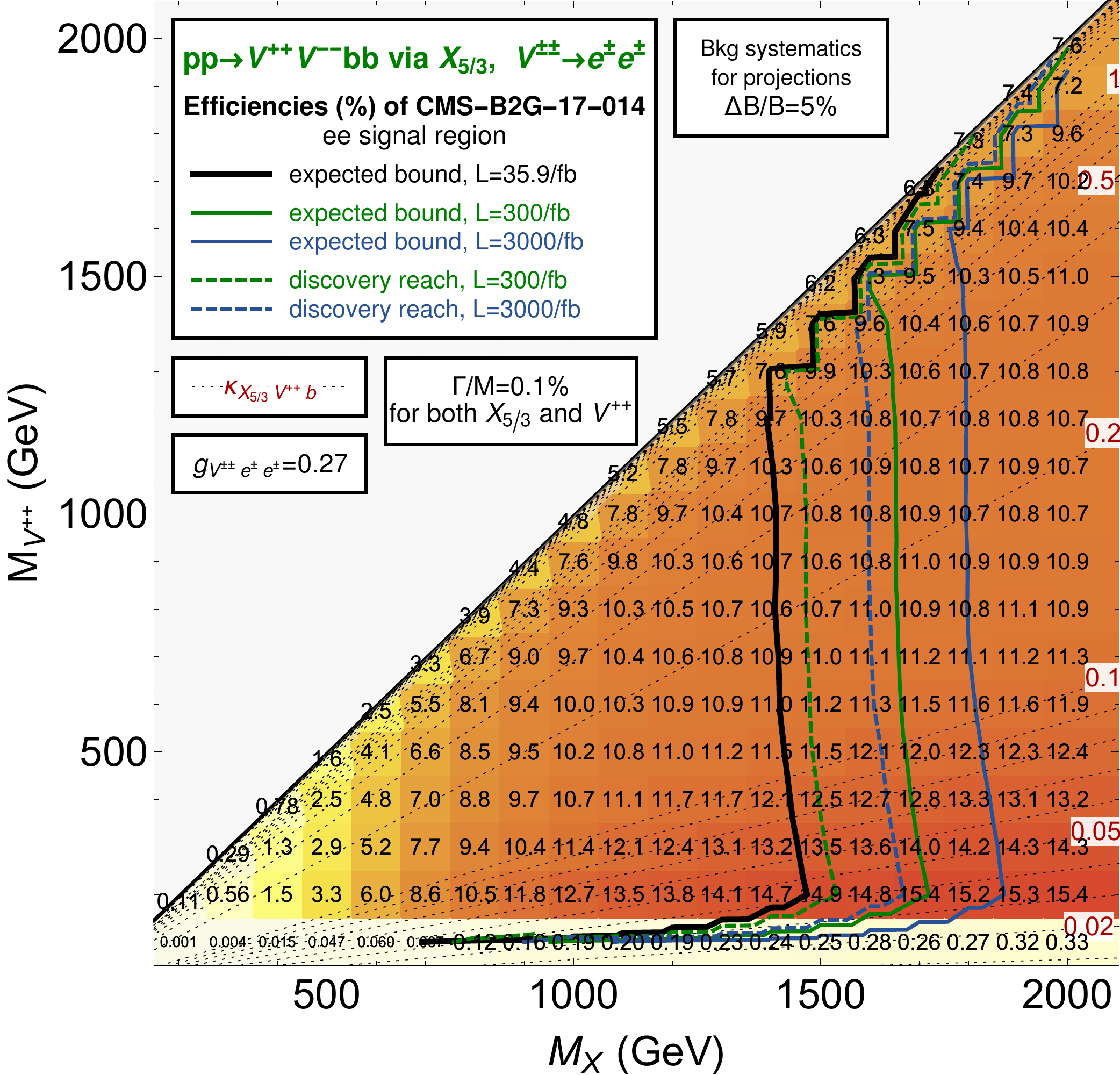}
\caption{\label{fig:boundseSRee} Recast efficiencies, bounds and projections in the $\{M_X,M_B\}$ plane ($B=S^{\pm(\pm)},V^{\pm(\pm)}$) for the $ee$ SR of the CMS search~\cite{Sirunyan:2018yun} and assuming that the (doubly ) charged bosons decay 100\% into charged electrons (and one neutrino flavour for singly charged boson). 
In the \textbf{top row} the results for singly charged bosons are shown, while the \textbf{bottom row} shows the results for doubly charged bosons. In the \textbf{left column} results for the scalars $S^{\pm(\pm)}$ are presented, while the \textbf{right column} shows the results for the vectors $V^{\pm(\pm)}$. 
Each plot shows the values of the recast efficiencies (in \%) for each scanned mass point, and the results are also visualised as a colour gradient. 
The 95\% CL expected exclusion bounds (corresponding to a significance $z=2$ in Equation~\ref{eq:Asimov}) are superimposed as solid lines of different colours for three different luminosities: $L=35.9\fb^{-1}$ (the luminosity of the CMS search, and the current bounds on this scenario), and the projected results for $L=300\fb^{-1}$ (Run 3) and $L=3\ab^{-1}$ (HL-LHC), assuming a systematic uncertainty for the background of 5\%. For the projected results also the discovery reaches ($z=5$) are shown as dashed lines. 
The couplings of $X_{5/3}$ needed to give $\Gamma_X/M_X=0.1\%$ are shown as dotted contours. For bosons, couplings are constant in the limit of  massless electron. The couplings of $X_{5/3}$ and doubly charged bosons are chosen to be purely left-handed, as in the NWA this choice does not impact the results. Analogous results for different decay channels and different signal regions are presented in Appendix~\ref{app:otherbounds}.}
\end{figure}

In Figure~\ref{fig:boundseSRee} we show the recast efficiencies in the $\{M_X,M_B\}$ plane, with $B=S^{\pm(\pm)},V^{\pm(\pm)}$, for different final states, corresponding to the exclusive decays of $X_{5/3}$ into either $S^\pm$, $V^\pm$, $S^{\pm\pm}$ or $V^{\pm\pm}$, and the subsequent exclusive boson decays into electrons or electron-neutrino pairs in the singly charged case. The $W$ bosons emerging from top
decays are always allowed to decay inclusively.
The efficiencies are shown both as a grid of numbers and as a colour code
and are evaluated in the $ee$ SR of the CMS search, which is the most sensitive to the selected decays.

The values of the couplings between $X_{5/3}$ and the bosons, necessary to obtain the fixed width-over-mass ratio $\Gamma/M=0.1\%$, are represented by thin dashed lines in the same plane, while the couplings of the bosons to electrons and muons are approximately constant in the whole plane due to the negligible masses of the decay products. With such a small value of the width-over-mass ratio the couplings remain well within the perturbative limit in the whole phenomenologically interesting region of the parameter space, and become non-perturbative only
very closely to the kinematic limit, where the mass difference between $X_{5/3}$ and the singly/doubly charged boson is equivalent to the top/bottom mass.

As already mentioned, the results describe a scenario corresponding to the NWA for both $X_{5/3}$ and the charged bosons in the region where their mass-gap is not too small. 
In this limit the process of pair production and decay of the VLQs and the charged bosons can be factorised and spin correlations between the intermediate- and final-state particles can be safely neglected. This has two consequences: 1) the choice of the chirality of the couplings of the $X_{5/3}$ and of the doubly charged boson does not impact the kinematics of the final states; 2) the results for scalars or vectors are expected to be completely analogous in all the parameter space. Both effects apply to the whole parameter space, with the possible exception of the region with small mass gap. As mentioned in Section~\ref{subsec:simstrategy}, we have not applied the factorization between production and decays in our simulations, to numerically estimate the region of validity of the NWA in the parameter space of our scenario and evaluate differences between the different coupling chiralities and different spins of the charged bosons. For the spin case, Figure~\ref{fig:boundseSRee} clearly shows that differences between scalar or vector bosons are negligibly small in the whole parameter space, including the small region of almost degenerate masses for the doubly charged case. For this reason, analogous results considering different final states, and non-diagonal decays of both $X_{5/3}$ and the bosons, are shown in Appendix~\ref{app:otherbounds} only for the scenarios where the $X_{5/3}$ decays to vector bosons, as the results for the scalars with same charge are expected (and have been verified) to be always qualitatively and quantitatively similar. Analogous checks regarding the chirality of the couplings have been performed, confirming numerically that this choice does not affect significantly the results. For this reason, in all the plots of this paper, the couplings of $X_{5/3}$ and of doubly charged bosons are chosen to be purely left-handed. 

For each scenario, the expected exclusion limit corresponding to the luminosity of the CMS search are provided as black contours. 
The projected $2\sigma$ expected exclusion reaches for the nominal luminosities at the end of Run 3 ($300\fb^{-1}$) and the high-luminosity phase of the LHC (HL-LHC) ($3\ab^{-1}$) are presented as green and blue contours, respectively. 
The corresponding projections of $5\sigma$ expected discovery reaches are represented as dashed lines with the same colours. For projected results, a fixed (optimistic) systematic uncertainty of 5\% on the background yields is assumed.

For the singly charged bosons the same-sign leptons in the final state are ensured by the leptonic decay of the $W$ boson emerging from the top quark. However, due to the $W$ BR into leptons (around 11\% for either electrons or muons), the number of events passing the selections of the CMS search
(see Appendix~\ref{app:recastingvalidation}) is limited. For events which pass the cuts, the results show that the efficiencies increase for increasing $M_X$ and remain approximately constant with respect to the mass of $S^+$ or $V^+$, except in the small mass-gap region, where they reduce, showing poorer sensitivity of the search when leptons in the final state are softer. 
Due to the fact that the cross-section is practically independent of the mass of the singly charged boson in the whole parameter space (see Figure~\ref{fig:xs}), the exclusion and discovery reaches are almost entirely affected by the different efficiencies for same $M_X$: while it is possible to exclude $X_{5/3}$ up to approximately $1.1\TeV$ when the singly charged boson has $M_B=100\GeV$, the bound becomes weaker as the mass gap decreases, allowing to exclude $M_X\lesssim800\GeV$ around the kinematical limit. 
With the assumption of 5\% systematics on the SM background, the exclusion limits on $M_X$ for the luminosities of Run 3 and HL-LHC improve by approximately $150\GeV$ and $400\GeV$, respectively, when $M_B=100\GeV$, and the improvement is larger for smaller mass differences.

For the doubly charged bosons the efficiencies follow an analogous pattern as in the singly charged case, but they are around one order of magnitude higher in the whole $\{M_X,M_B\}$ plane, except when $M_B$ is lighter than $100\GeV$. This is expected, as the decays of the doubly charged bosons always result in final states with same-sign leptons, unlike in the singly charged boson cases where the second same-sign charged lepton could only arise from the $W$ boson. 
The efficiencies in the region where the mass of the doubly charged boson is lighter than $100\GeV$ exhibit a sharp drop: in this region the CMS search imposes a veto both on opposite-sign leptons and on same-sign electrons with an invariant mass within $15\GeV$ of the $Z$ mass. The veto on the same-sign lepton pair is only imposed in the di-electron final state, and indeed the area with small efficiencies in the low $M_B$ region does not appear when the final state contains same-sign leptons other than $ee$ coming from the new charged bosons (see results in Appendix~\ref{app:otherbounds}).
The exclusion and discovery reaches for the doubly charged bosons are strongly affected by the interplay between the cross-section (depending only on $M_X$ for large mass gaps and increasing as the kinematical limit of almost degenerate $M_X$ and $M_B$ masses is approached) and the efficiencies. The excluded masses for the $X_{5/3}$ are in general higher than in the singly charged boson case, approaching $1.4\TeV$ when the mass gap is large and $M_B$ is heavier than $100\GeV$: the better limits in this region are entirely due to the higher efficiencies, as the cross-section is almost exclusively driven by the QCD pair production of $X_{5/3}$, which depends only on $M_X$.
When $M_B$ is light, however, the bounds become much weaker, as the search is poorly sensitive in this region due to the $Z$ veto for same-sign electrons. The bounds in the small mass-gap region become stronger due to the increase of the cross-section when the NWA breaks. Results in this region should nevertheless
be taken with caution, as the effects due to the interference of signal and background (not included in this analysis) might affect the determination of the limits. The projected exclusion and discovery reaches in the NWA region increase analogously as in the singly charged boson case under the assumption of the same systematic uncertainty of 5\% on the background yield.

The reinterpretation of results for a generic model relies on the combination of the entire set of efficiency tables presented in Appendix~\ref{app:otherbounds} and can be done for scenarios predicting $X_{5/3}$ interacting with singly and/or doubly charged bosons which decay into SM light leptons with generic BRs, as long as the widths of the new particles are relatively small with respect to their masses. In fact, assuming that the width does not exceed values for which the NWA is questionable, we can safely assume that the efficiencies computed with our simulations are valid with good approximation, such that the effective cross-section can be computed as
\begin{align}
\label{eq:sigmaeff}
 \sigma_{\rm eff}(SR) = \sum_{\scriptsize \begin{array}{c}B_i,B_j=B^{\pm\pm},B^\pm\\ l_a,l_b=e,\mu\end{array}} 
 &
 \sigma_{\rm QCD}(M_X)\ \epsilon_{SR}(M_X,M_B;B_i,l_a;B_j,l_b)\nonumber\\
 &\times BR(X\to q\ B_i) BR(\bar X\to \bar q\ B_j)BR(B_i\to l_a) BR(B_j\to l_b) \;,
\end{align}
where $B=\{S,V\}$ and $l_a,l_b$ are charged
leptons with flavours $a$ or $b$ produced in boson decays. This equation will be applied in the next section to reinterpret our model-independent results for the 331 Model, where $B=V$ only.

A final aspect to be considered in the model-independent analysis is the possibility to discriminate different scenarios in case of observation of an excess in future searches. A quantitative treatment of the discrimination between different signal hypotheses is beyond the scopes of this analysis, and is left for a future work, but a qualitative treatment is possible and is presented in the following, as a guideline for a more detailed phenomenological analysis.

Clearly, discriminating different spins of the charged boson (scalar or vector) is not feasible in the NWA regime: the dominant contribution arising from the QCD pair production of the $M_X$, and the possibility to factorise the production and decay of both $X_{5/3}$ and the charged boson with excellent approximation make the subdominant effects related to the different spins of the bosons out of reach with the sensitivity of the search we have recast. If the new particles have non-negligible widths, the contribution of topologies which are sub-dominant in the NWA, and the interference between different signal topologies and between signal and SM background can potentially enhance the dependence of the properties of final state objects on parameters other than the mass of the $X_{5/3}$ and the BRs of the new particles~\cite{Moretti:2016gkr,Moretti:2017qby,Carvalho:2018jkq,Crivellin:2018ahj}. Considering the presence of a coloured particle, large-width aspects should also be considered alongside QCD corrections for a precise determination of the kinematics of the final state~\cite{Cacciapaglia:2018qep,Deandrea:2021vje}.

The discrimination of different points in the $\{M_X,M_B\}$ plane corresponding to the same effective cross-section is however easier even in the NWA. In the case of interactions between the $X_{5/3}$ and doubly charged bosons,  reconstructing both masses of the new particles is trivially possible due to the absence of invisible objects. If, however, the $X_{5/3}$ interacts with singly charged bosons, the mass reconstruction becomes more challenging due to the neutrinos in the final state, but the reduction of the efficiencies for same $M_X$ and larger mass of the boson means that points with analogous effective cross-section are characterised by different values of both masses, and in particular by different values of $M_X$; since distributions of objects in the final state would be mostly sensitive to $M_X$, a discrimination of different points should be possible, and simply requires enough events to reconstruct distributions with reasonable accuracy.

\section{Interpretation for the 331 Model with $\beta_{Q}=\pm\sqrt3$}\label{sec:331}
The $331$ Model~\cite{Singer:1980sw,Valle:1983dk,Pisano:1991ee,Frampton:1992wt,Foot:1994ym,Hoang:1995vq} is an extension of the SM characterized by the gauge symmetry pattern $SU(3)_c\times SU(3)_L \times U(1)_X$. 
The promotion of the SM $SU(2)_L$ symmetry to an $SU(3)_L$ results in a redefinition of the hypercharge as $\mathbb{Y}=\beta_Q \mathbb{T}^8+X\mathbb{I}$, where $\mathbb{T}^8$ is one of the Gell--Mann matrices, $X$ is a new abelian charge assignment, $\mathbb{I}$ is the identity matrix and $\beta_Q$ is a free parameter of the model. When the latter is not specified, the setup is called `general $331$ Model'. 

\begin{figure}[t!]
  \centering 
 \includegraphics[width=.485\textwidth]{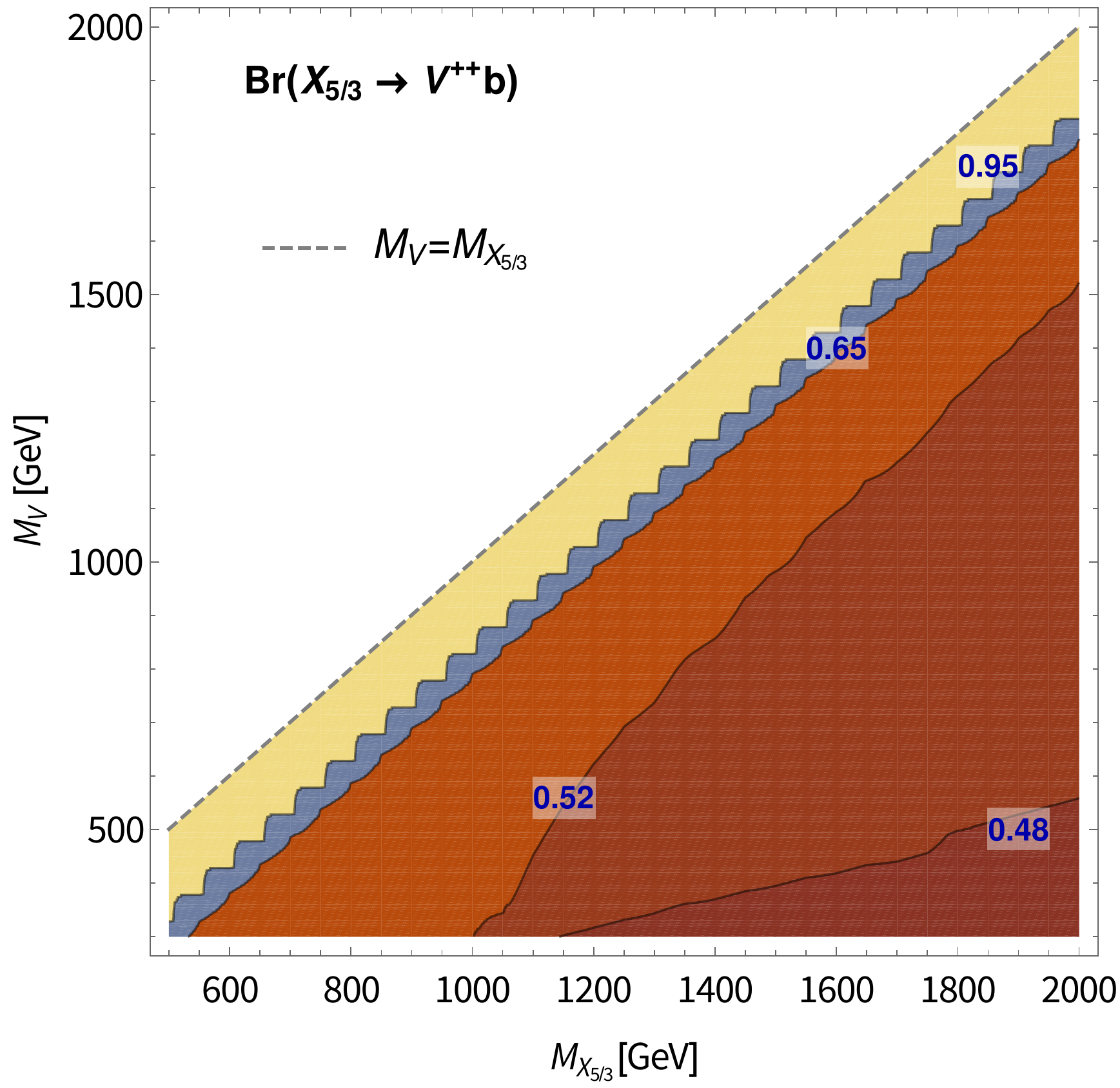}\hspace{.2cm}
 \includegraphics[width=.485\textwidth]{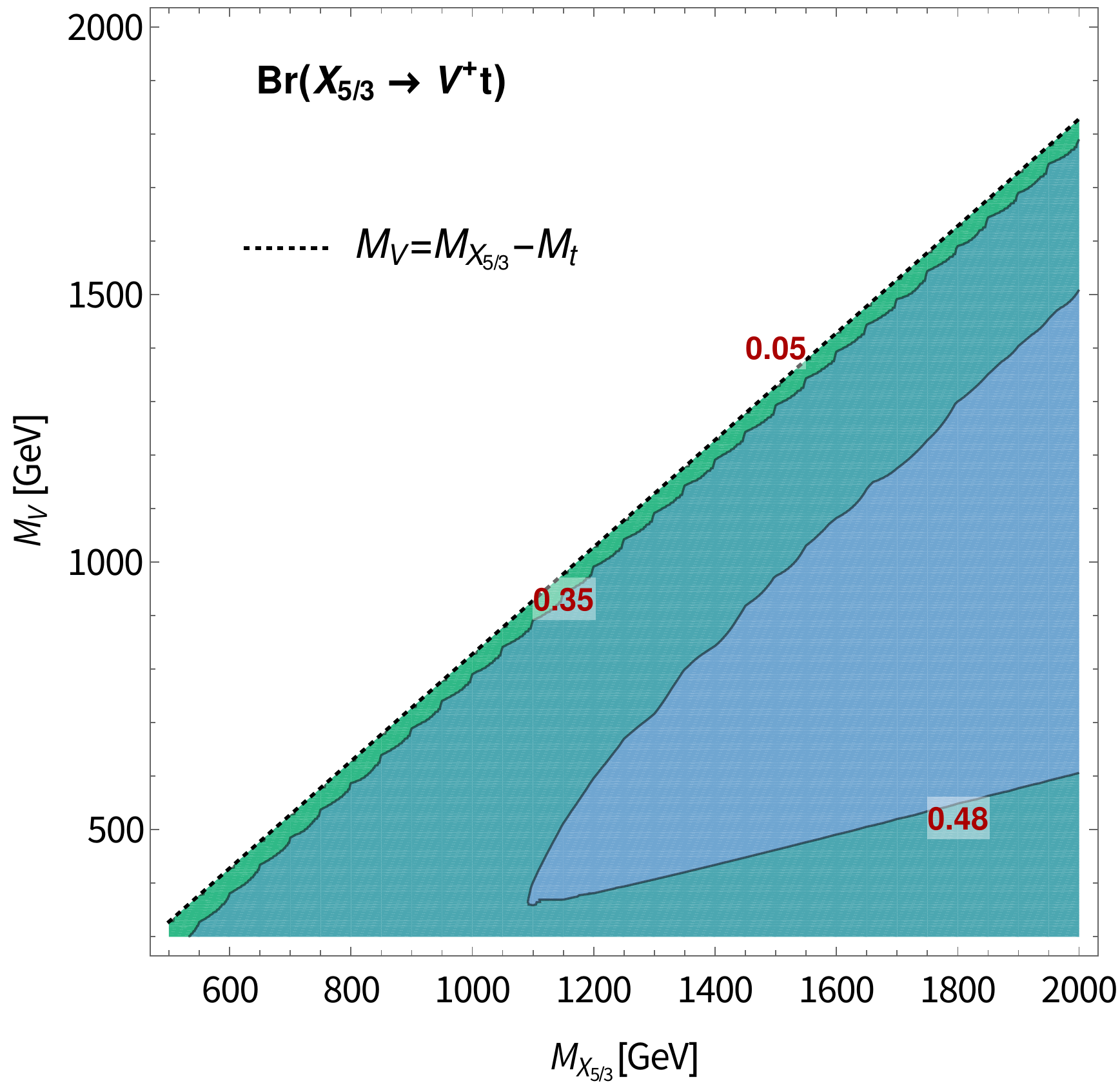}
  \caption{\label{fig:br331} Branching ratio of $X_{5/3}\to V\,q$ in the 331 Model with $\beta_Q=\pm\sqrt3$. The left panel shows the contours of $BR(X_{5/3}\to V^{++}b)=0.48,\,0.52,\,0.65,\,0.95$. The right panel shows the contours of $BR(X_{5/3}\to V^{+}t)=0.05,\,0.35,\,0.48$.}
\end{figure}

The $\beta_Q$ parameter is constrained by theoretical and phenomenological arguments. In particular, the
positiveness of the extra neutral gauge boson $Z'$ mass implies that
\begin{equation}
|\beta_Q|\le \sqrt{\frac{1}{\sin^{2}\theta_W}-1}
\end{equation}
where $\theta_W$ is the Weinberg angle \cite{Buras:2013dea}. Explicit requirements on the electric charges of the new particles, such as the extra quarks predicted by the $331$ Model, will select definite values for $\beta_Q$. In order to reproduce the case analysed in the previous sections, one has to select a version of the 331 Model that accommodates an exotic quark with $Q=5/3$ in the third family. This can be achieved by the choice $\beta_Q=\pm\sqrt3$. Beside the exotic quarks, the same version of the $331$ Model features doubly charged vector bosons in the spectrum.
The spontaneous electroweak symmetry breaking occurs
via three scalar triplets of $SU(3)_L$, usually
named $\chi$, $\rho$ and $\eta$,
and the 
masses of the exotic states originate from the breaking 
\begin{equation}\label{eq:ssb331}
SU(3)_L \times U(1)_X \to SU(2)_L \times U(1)_Y,
\end{equation}
which, following Reference \cite{Costantini:2020xrn}, 
is driven by $\chi$ and its vacuum expectation value (VEV) $v_\chi$.
The mass expressions of the exotic quark and singly /doubly charged vectors
read
\begin{equation}\label{eq:massXY}
M_{X_{5/3}} = \frac{1}{\sqrt2}Y_{X} v_\chi\,,\quad M_{V} = \frac{1}{2} g_L v_\chi\,,
\end{equation}
where the mass splitting between $V^+$ and $V^{++}$ (which is of the order of
$\sqrt{v_\rho^2+v_\eta^2}/v_\chi$, $v_\rho$ and $v_\eta$ being the VEVs of the
other two scalar triplets) is neglected.
The parameters $Y_X$ and $g_L$ appearing in Equation~\ref{eq:massXY} are the Yukawa coupling of the $X_{5/3}$ and the gauge coupling of the left-handed symmetry group, respectively. Moreover, the strength of the interaction vertex among the doubly charged vector, the exotic quark and the SM quark is
\begin{equation}\label{eq:k331}
\kappa_{X_{5/3}\,V}=\frac{g_L}{\sqrt2}.
\end{equation} 
In Figure~\ref{fig:br331}, the branching ratios of the exotic-quark
decay $BR(X_{5/3}\to V\,q)$ are plotted in the $(M_{X_{5/3}},M_{V})$
plane. In particular the left (right) plot is the $V^{++}b$ ($V^+ t$) channel. These have been computed numerically by employing a UFO implementation \cite{FeynRules}
of the $331$ Model in {\sc MadGraph5\_aMC@NLO}~\cite{Alwall:2014hca,Frederix:2018nkq}.

In the left plot of Figure~\ref{fig:br331} the contours of the $BR$ with
$$BR(X_{5/3}\to V^{++}b)=0.48,\,0.52,\,0.65,\,0.95$$
are plotted: the dashed-gray line corresponds to the mass degeneracy $M_V=M_{X_{5/3}}$. When $M_{V}\ll M_{X_{5/3}}$, the branching ratio approaches its lowest value $(\sim0.48)$. Conversely, this value increases as the mass splitting between the exotic quark and the vector decreases, reaching the limiting value $BR(X_{5/3}\to V^{++}b)\sim1$ at $M_{X_{5/3}}-M_{V}<M_t$.

The right plot of Figure~\ref{fig:br331} shows the contours with
$$BR(X_{5/3}\to V^{+}t)=0.05,\,0.35,\,0.48\,,$$ while the dashed-black line corresponds to the threshold $M_V=M_{X_{5/3}}-M_t$.
 However, the overall behaviour is basically complementary to the previous case. When $M_{X_{5/3}}-M_{V}<M_t$ there is no phase space allowed for the two-body decay $X_{5/3}\to V^{+}t$, therefore the branching ratio vanishes. On the other hand, it increases as the mass splitting between the exotic quark and the vector becomes larger. However, as the $BR(X_{5/3}\to V^{+}t)=0.48$ contour suggests, we are not in presence of a monotonically increasing function in the considered mass range.

The signal yields for the 331 Model can then be computed at a given luminosity after determining the effective cross-section. In the specific case of the 331 scenario, where the $X_{5/3}$ can only decay to vector bosons of either charge (and where $V^\pm$ and $V^{\pm\pm}$ have same masses) and the vector bosons decay into any SM lepton flavour with same $BR=1/3$, the effective cross-section for the CMS signal region $SR$ can be obtained from Equation~\ref{eq:sigmaeff} considering the BRs of the $X_{5/3}$ reported in Figure~\ref{fig:br331}. 

It needs to be specified at this point how $\tau$ leptons are treated in our analysis. The contribution to the total signal yield from the decays of vector bosons into $\tau$ leptons is not negligible: if the $\tau$'s decay leptonically, there can be direct contributions to same-sign di-lepton final states from chains like $V^{++}\to\tau^+\tau^+\to (e^+\nu_e\bar\nu_\tau)(e^+\nu_e\bar\nu_\tau)$. On the other hand, if the taus decay hadronically, there can still be contributions if the other vector boson decays to a same-sign lepton pair. In the model-independent analysis these channels were not treated explicitly. Hence, a conservative estimation of the effects of the $\tau$ decays has been considered: the contribution from leptonic $\tau$ decays only, with the same efficiencies of the light-lepton decays, reported in Appendix~\ref{app:otherbounds}. This estimation does not take into account two competing effects: 1) the energy loss of the charged leptons in the three-body $\tau$ decay (compared to the two-body decay of the vector boson into light-leptons);
2) the contributions of hadronic $\tau$ decays. The first effect is expected to play a role especially for small mass-gap between the $X_{5/3}$ and the vector boson, when the decay products are softer and might not pass the thresholds of kinematic cuts, and its contribution would reduce the signal yield in this region. The second effect would increase the number of signal events in the whole parameter space.
In a scenario where the vector bosons fully decay into $\tau$ leptons, a comparison with the results of Reference~\cite{Xie:2019gya} exhibits an overall lowering of the bounds by about $\sim 300\GeV$. We therefore expect that in the context of our framework, where branching ratios into taus are not 100\%, the bounds are likely underestimated by an amount $\mathcal O(100\GeV)$.
As a matter of fact, it has to be stressed that there are other factors which 
in principle should be taken into account for the sake of a 
precise determination of the bounds, for example the inclusion of 
NLO corrections to the cross-sections,
while we have assumed LO rates everywhere, or large-width effects.
Therefore, since we are mostly interested in estimating the range
of the bounds on the mass of vector-like quarks, 
rather than in their exact computation, 
the effects of tau decays will not be further investigated, as they are
beyond the scopes of this analysis.

\begin{figure}[th!]
  \centering
  \includegraphics[width=.325\textwidth]{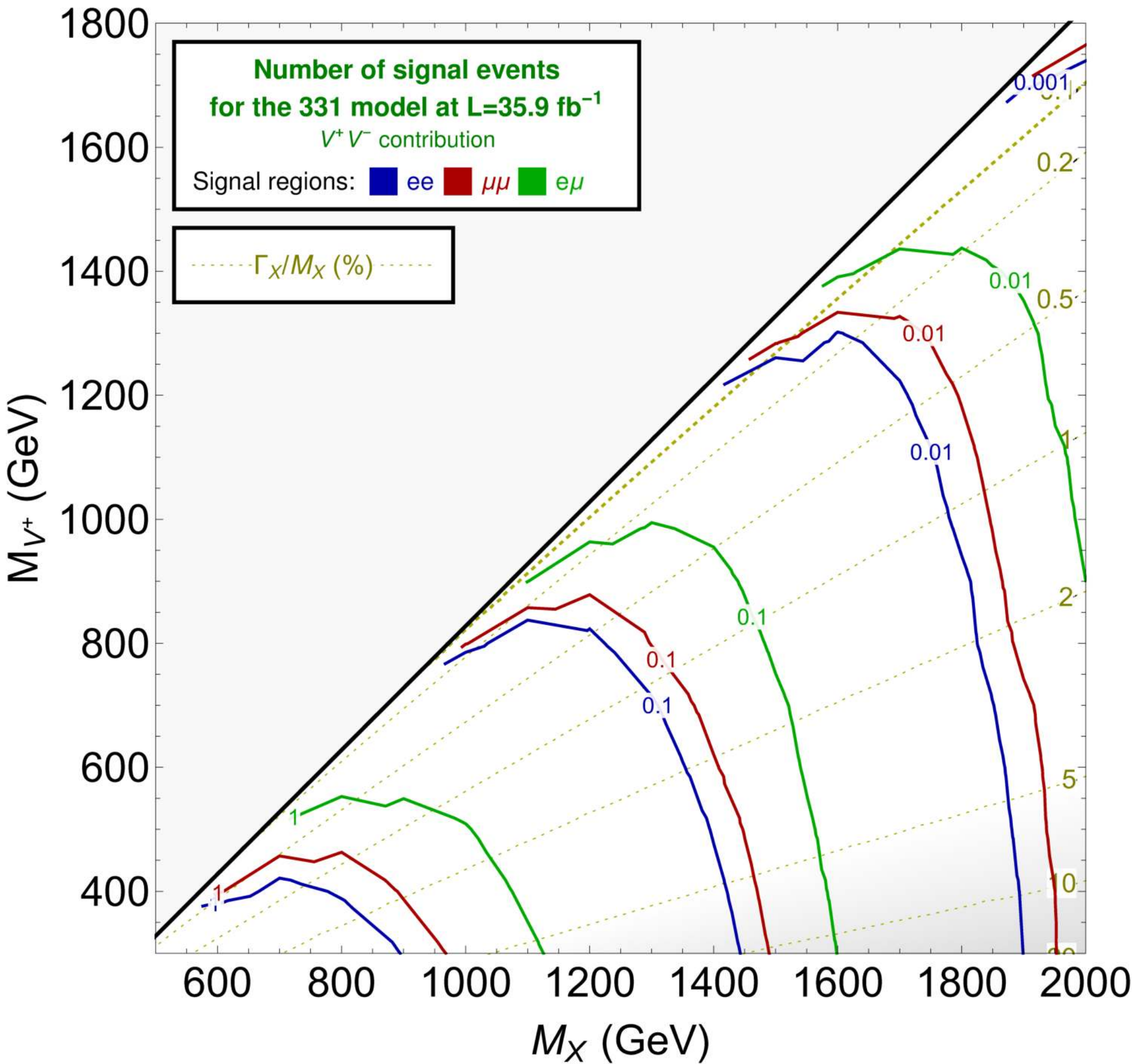}
  \includegraphics[width=.325\textwidth]{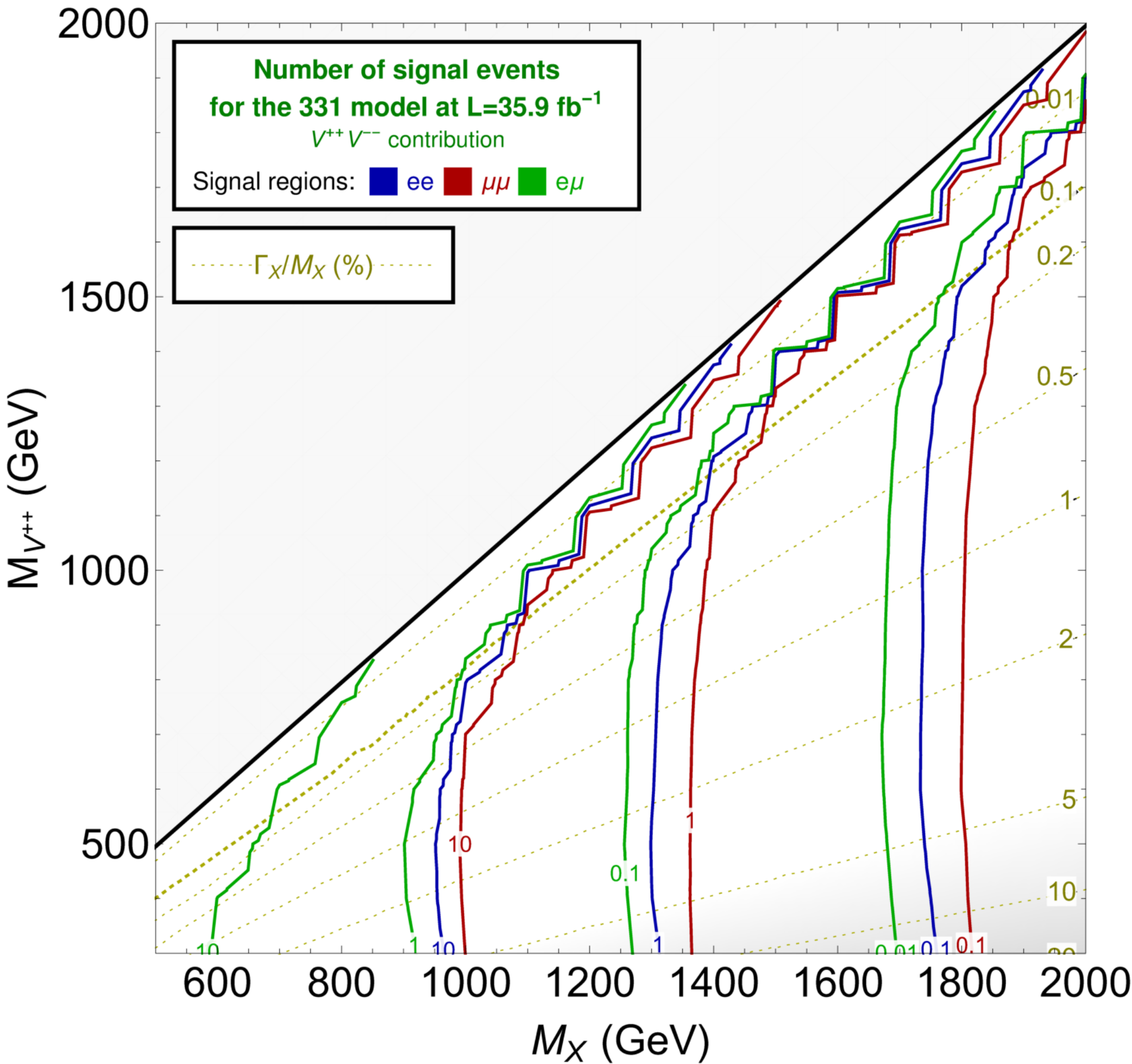}
  \includegraphics[width=.325\textwidth]{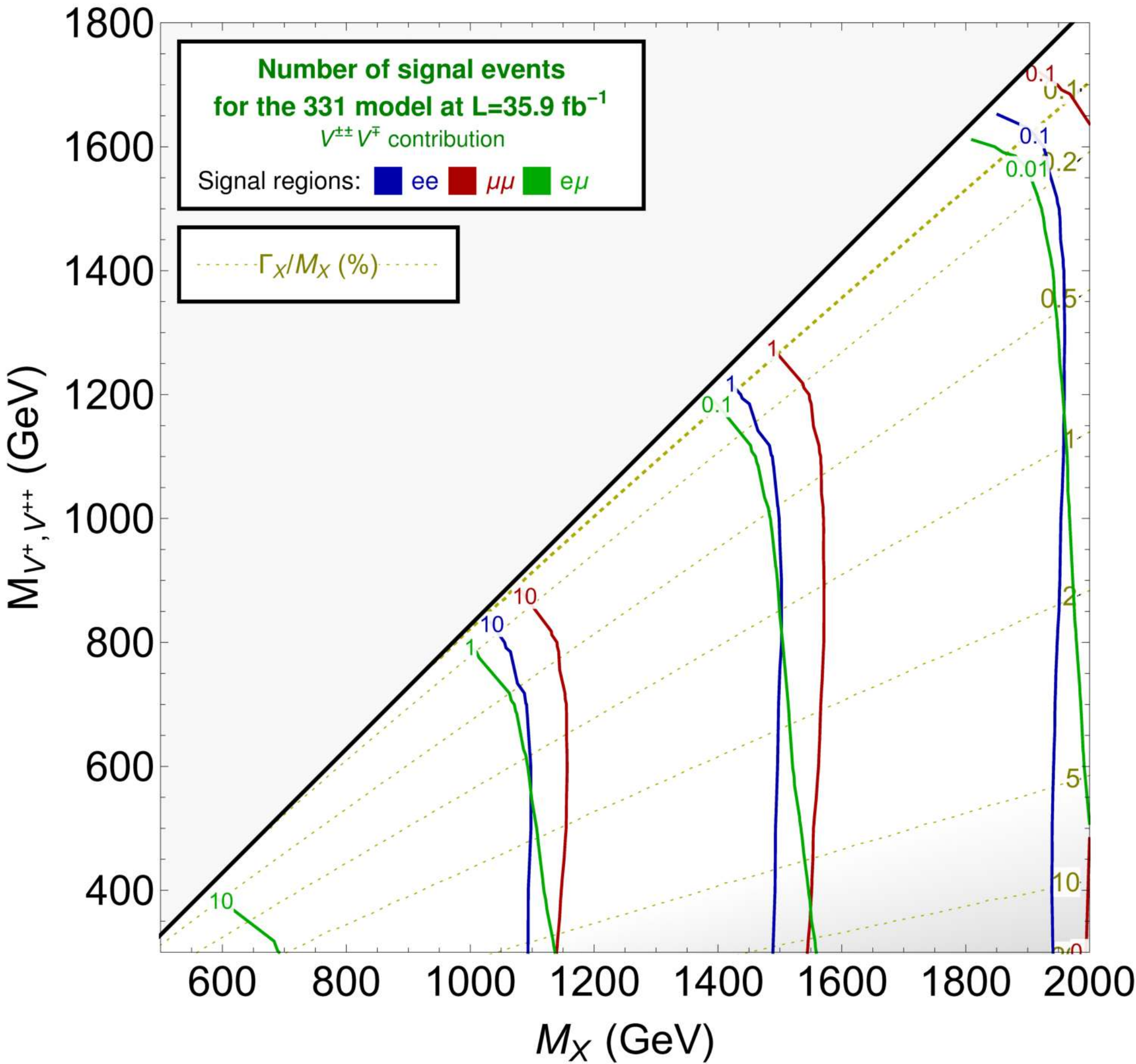}
  \caption{\label{fig:331yields} Contributions to the signal yield at a luminosity $\mathcal L=35.9\fb^{-1}$ for the three possible decay channels of the $X_{5/3}$: both to $V^\pm$ \textbf{(left panel)}, both to $V^{\pm\pm}$ \textbf{(central panel)}, and one to $V^{\pm\pm}$ and the other to $V^{\mp}$ \textbf{(right panel)}. The contours with constant $\Gamma_X/M_X$ are shown as yellow dotted lines. The line corresponding to the width-over-mass ratio used for the model-independent results is highlighted, and the region where the width becomes large is shaded, as the validity of the reinterpretation of the model-independent results is not accurate outside the NWA.}
\end{figure} 

After considering these effects, the yields corresponding to the luminosity of the CMS search~\cite{Sirunyan:2018yun} ($\mathcal L = 35.9\fb^{-1}$) are reported in Figure~\ref{fig:331yields} for each signal region, where it is possible to see that the lower efficiencies and branching ratios of the $V^+V^-$ decay make this contribution negligible with respect to $V^{++}V^{--}$ and $V^{\pm\pm}V^{\mp}$ for the $ee$ and $\mu\mu$ SRs, but not for the $e\mu$ SR, which is mostly fed by the $\tau$ decays of the vector bosons. 

Generally, the yields related to two of the analysed physics cases (pure electron and pure muon lepton states) that are represented by the blue and red lines display a quantitatively similar behaviour, hence the lines corresponding to same yields are very close to each other. On the other hand, the yields corresponding to the mixed lepton state are considerably smaller for the $V^{++}V^{--}$ and $V^{\pm\pm} V^\mp$ contributions, while they dominate for the $V^+V^-$ contribution. The reason behind this is again to be found in the lepton-flavour universality of the 331 Model (in its minimal realisation), implying a ``democratic'' decay of vector bosons into any lepton family, \emph{i.e.} $BR(V\to \ell\ell)\simeq 1/3$ provided the exotic quarks are heavier than the vector bosons and such a decay window is closed. The almost identical phenomenological outcomes in the pure electron and muon channels displayed in Figure~\ref{fig:331yields} is thus justified by the analogous efficiencies of these channels. Conversely, since the mixed lepton state is not generated at the matrix-element level ($V^{++}\not\to e^+ \mu^+$), its yields come from a combination of \emph{e.g.} $X_{5/3}\,\bar X_{5/3}\to V^{+}t \;V^-\bar t$, with $V^+\to e^+ \nu_e$ and $t\to W^+\, b \to \mu^+\nu_\mu\; b$ and from the contribution of the $\tau$ decays where different $\tau$'s decay to different light leptons. As the decay chain of the $W$ boson is suppressed by the $BR(W^+\to e^+\nu_e)\sim BR(W^+\to \mu^+\nu_\mu)\sim 11\%$,  the main contribution to the $e\mu$ channel is provided by the $\tau$ decays.

\begin{figure}[ht!]
  \centering
  \includegraphics[width=.6\textwidth]{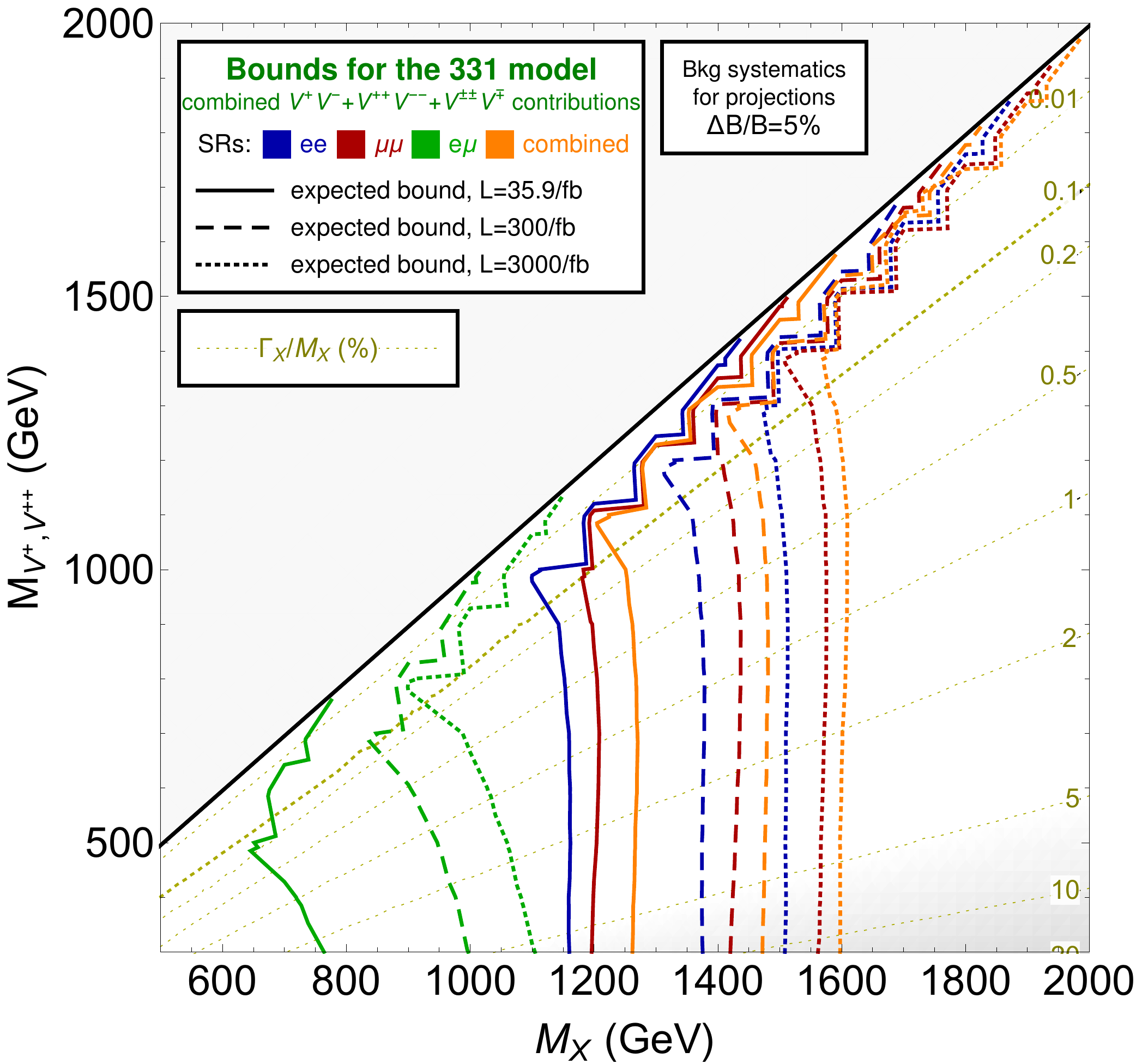}
  \caption{\label{fig:bounds331} Expected bounds from the recast of the CMS search~\cite{Sirunyan:2018yun} for the 331 Model with $\beta_Q=\pm\sqrt3$ plotted in the $\{M_{X_{5/3}},M_V\}$ plane. The contributions from $V^+V^-$, $V^{++}V^{--}$ and $V^{\pm\pm}V^{\mp}$ are summed at different luminosities for the $ee$ (blue line), $\mu\mu$ (red line) and $e\mu$ (green line) lepton states and the statistical combination of the signal region with the CLs procedure is shown in orange. The meaning of the width-over-mass curves, shading, and region of validity are the same as Figure~\ref{fig:331yields}.}
\end{figure}

The bounds for the 331 Model with $\beta_Q=\pm\sqrt3$, obtained with a stand-alone analysis on the results of the model-independent case presented in Section~\ref{sec:LHCpheno}, are shown in Figure~\ref{fig:bounds331} in the $\{M_{X_{5/3}},M_V\}$ plane. 
Blue, red and green colours are adopted for the $ee$, $\mu\mu$ and $e\mu$ signal regions, respectively, and solid/dashed/dotted line stands for the $L=35.9/300/3000$ fb$^{-1}$ luminosity stages. The orange lines correspond to the statistical combination of the three signal regions, done through the CLs method~\cite{Read:2000ru,Read:2002hq}. The systematic uncertainties of the background are those of the CMS search for $L=35.9\fb^{-1}$, while they are optimistically assumed to be 5\% for projections at future luminosities. The systematic uncertainty for the signal is always assumed to be 10\%, but results are however very weakly dependent on this choice.

Overall, the exclusion limits are dominated by the $V^{\pm\pm}$ contribution and, for each luminosity stage, they are roughly independent of $M_V$, if $M_V\ll M_{X_{5/3}}$, as expected from the model-independent analysis. 
The combined results, together with the mass expression of the vectors (cf. Eq~\ref{eq:massXY}) allow us to conclude that the spontaneous symmetry breaking pattern conceived by the 331 Model with $\beta_Q=\pm\sqrt3$ occurring at a scale of $v_\chi=2\,M_V/g_L\simeq 6$ TeV (or higher) will not be excluded by the LHC data, being $M_V\gtrsim 1.95$ TeV the strongest achievable bound from our analysis. Nevertheless, the NWA is less accurate when $M_V\sim M_{X_{5/3}}$. A more conservative bound on $v_\chi$ can be obtained considering the highest mass value for which the bounds in Figure~\ref{fig:bounds331} are still independent of $M_V$, \emph{i.e.} 
\begin{equation}
v_\chi= 2\, M_V/g_L\sim 4\; {\rm TeV}.
\end{equation} 

The reinterpretation of the model-independent results for the 331 Model is valid in almost the whole parameter space, considering the assumption of NWA for both $X_{5/3}$ and the charged bosons. The only region for which the width of the $X_{5/3}$ is not small has a negligible impact on the determination of the bounds. In Figure~\ref{fig:bounds331} this region is shaded in gray, and if we consider a $\Gamma_X/M_X$ ratio larger than 10\% as the threshold for breaking the NWA, the current bounds and future projections can be used with excellent approximation even for low masses of the charged vectors. In the opposite limit of small mass-gap, the bounds for the 331 Model start to be questionable only in the region of the parameter space where the $X_{5/3}$ and the charged bosons masses imply $\Gamma_X/M_X\lesssim0.1\%$ (see also Figure~\ref{fig:xs} for a quantitative estimate of the corresponding region).

\section{Conclusions}\label{sec:conclusion}
In this work we have analysed signatures at the LHC arising from a new exotic vector-like quark with charge 5/3 ($X_{5/3}$) interacting with new singly 
and doubly charged bosons, either scalars or vectors, which in turn decay to SM electrons or muons. The analysis is motivated by classes of theoretical scenarios in which new particles can interact with light SM states, and the results have been interpreted as bounds on the parameters of the the $331$ Model with $\beta_Q=\pm\sqrt 3$ where new singly  or doubly charged vector bosons can decay to any SM lepton with same branching ratios.

The phenomenology of the $X_{5/3}$ VLQ has been firstly studied from a model-independent perspective in the framework of a simplified model. The bounds on VLQ and new boson masses have then been obtained by recasting a CMS analysis targeting the SM decay of $X_{5/3}$ into $Wt$, leading to a final state with same-sign dileptons. The {\sc MadAnalysis 5} code used for the recast analysis has been
made available in the {\sc Dataverse} repository~\cite{DVN/DQZWYL_2021}. The model-independent results have been expressed as recast efficiencies in the mass-mass plane of the $X_{5/3}$ and new bosons, for all possible final states arising from different decay chains. This exploratory analysis has been performed at LO and in the narrow-width approximation for both $X_{5/3}$ and bosons. 

Optimal bounds, assuming 100\% branching ratios in individual channels, have been extracted for different phases of the LHC, namely the luminosity of the CMS analysis, $35.9\fb^{-1}$, the nominal one expected at the end of Run 3 and the high-luminosity phase (HL-LHC), i.e. $300\fb^{-1}$ and $3000\fb^{-1}$, respectively. In the case of exclusive interactions with the doubly charged bosons, the $X_{5/3}$ mass has a current lower limit around $1.5\TeV$ almost independent from the mass of the boson in the whole parameter space, with an exclusion reach for the HL-LHC phase reaching $\sim 1.8\TeV$. If the $X_{5/3}$ interacts exclusively with singly charged bosons, the bounds are more strongly affected by the decays of the $W$ boson arising from the top quark. When the singly charged boson mass is of order $100\GeV$, the current lower limit on the mass of the $X_{5/3}$ is around the TeV, and the bound reduces to $\sim 800\GeV$ if the mass-gap is small. The bounds for the two cases, $M_V=\mathcal O(100\GeV)$ and small mass-gap between $X_{5/3}$ and $V^+$, increase to $\sim1.4\TeV$ and $\sim1.2\TeV$ respectively in the HL-LHC phase. In both cases, the bounds do not significantly change depending on the decay channel of the boson, except for the same-sign di-electron final states where a $Z$-boson veto on the invariant mass of any electron pair considerably reduces the sensitivity of the search for boson masses around $100\GeV$ or less. Discovery reaches for future luminosities have also been provided and, in this context, a discussion has been presented about the possibility to discriminate
between scenarios, characterised by either different mass combinations or different spins of the bosons, in case of a future discovery.

The results of the model-independent analysis have been suited for different BSM interpretations, and for the first time the bounds for the light-lepton decays of the charged vector/scalar boson have been provided. In the case of the 331 Model the bounds are dominated by the doubly charged vector contribution, in complete analogy with the model-independent case. The projected bounds at the highest luminosity considered ($3$ ab$^{-1}$) call for the spontaneous symmetry breaking $SU(3)_L\times U(1)_X \to SU(2)_L\times U(1)_Y$ to occur at a scale higher than $4\TeV$ (conservative case).

The extension to NLO processes and the inclusion of the large-width effects are possible future direction to be addressed. Quantitative information coming from the theoretical and phenomenological analysis of the full scalar potential of the 331 Model will also be included to improve the determination of the constraints on the model.

\section*{Acknowledgements}
AC acknowledges support from FRS–FNRS under project T.04142.18.
LP's work is supported by the Knut and Alice Wallenberg foundation under the SHIFT project, grant KAW 2017.0100. LP acknowledges the use of the IRIDIS 4 HPC Facility at the University of Southampton. The work of JS has received funding from the Swiss National Science Foundation (SNF) through the Eccellenza Professorial Fellowship “Flavor Physics at the High Energy Frontier” project number 186866.

\newpage

\appendix

\section{Implementation and validation of the recasting of CMS-B2G-17-014}\label{app:recastingvalidation}
The CMS analysis~\cite{Sirunyan:2018yun} targets two different kinds of final states, same-sign di-leptons and single-lepton. As mentioned in Section~\ref{sec:LHCpheno}, only the former has been recast and implemented in the {\sc MadAnalysis 5}~\cite{MadAnalysis5,MadAnalysis5:Expert} framework, as it is the most sensitive to the final states of the processes treated in our analysis. \\

The analysis relies on the identification of so-called tight leptons and AK4 jets. Their reconstruction has been done through {\sc Delphes 3}~\cite{Delphes}, used within {\sc MadAnalysis 5}. The default CMS detector card for {\sc Delphes 3} has been modified to include the pileup subtraction provided in {\sc MadAnalysis 5}. Jets have been clustered following the anti-$k_T$ algorithm~\cite{Cacciari:2008gp} implemented in the {\sc FastJet} package~\cite{Cacciari:2011ma} within {\sc Delphes 3}. The distance parameter for jet clustering has been set to 0.4, corresponding to the definition of AK4 jets.

The lepton reconstruction at the Delphes level has been done using the {\tt Electron/MuonIsolation} modules, with the default parameter values to find
isolated leptons. Such leptons have been then checked against isolated photons
and AK4 jets for the uniquely identified objects. However, more stringent isolation criteria for leptons have been implemented at the analysis level,
and therefore the Delphes output already contains identified leptons even before using the isolation module. For the purposes of the calculation of the isolation criteria at the analysis level, the tracker and calorimeter information have been included in the output {\sc ROOT} file.

The isolation variable $I$ for leptons is calculated at the analysis level as the scalar sum of the transverse momenta ($p_T$) of all particle-flow candidates within the cone of radius $R$ around the lepton, divided by the lepton's $p_T$. The cone radius $R$ is a function of the lepton's $p_T$ defined as
\begin{equation}
 R=\frac{10\GeV}{\min\left[\max(p_T,50\GeV),200\GeV\right]}.
\end{equation}
Tight leptons are required to have the isolation variable $I < 0.1$;
also, all particle-flow candidates contributing to $I$ have to satisfy $p_T > 20\GeV$ in order to successfully validate the recast. This seemingly large value has been set to achieve a good agreement with the CMS results for the expected number of events and the $H_T^{\rm lep}$ distribution (described later). Changing this single value in the analysis makes our results significantly different from the CMS ones.

Further requirements have been set on the pseudorapidity $\eta$ and transverse momentum $p_T$ of leptons and jets. Tight electrons are required to have $|\eta| < 2.5$ and those with $1.44 < |\eta| < 1.57$ (barrel-endcap transition region) were excluded from the analysis. Tight muons have instead pseudorapidity requirement $|\eta| < 2.4$. 
In the same-sign di-muon channel, the signal muons are additionally required to not be both within $|\eta| > 1.2$, unless they are in opposite sides of the detector (different signs of $\eta$) or well separated in azimuthal angle $\varphi$ ($\Delta\varphi > 1.25~{\rm rad}$). 
The cuts $p_T > 30\GeV$ and $|\eta| < 2.4$ have also been set on the
reconstructed AK4 jets.

The CMS analysis also defines two analysis-specific variables: the number of constituents in the event, namely the number of AK4 jets summed with the number of tight leptons, excluding the signal pair, and $H_T^{\rm lep}$, the scalar $p_T$ sum of all constituents, including the same-sign lepton pair.

After the object reconstruction, the following cuts are applied:
\begin{itemize}
 \item[-] The signal lepton pair is the pair of same-sign tight leptons maximizing the scalar $p_T$ sum: the leading lepton is required to have $p_T > 40\GeV$, the subleading one $p_T > 30\GeV$.
 \item[-] The invariant mass of the signal lepton pair is required to be greater than $20\GeV$ to veto quarkonium states.
 \item[-] Events containing any opposite-sign lepton pair of the same flavour or same-sign electron pair with invariant mass within 15 GeV of the mass of the $Z$ boson are discarded.
 \item[-] Each event is required to contain at least two AK4 jets passing the definitions above.
 \item[-] The number of constituents is required to be $N_{\rm const} \geq 5$.
 \item[-] $H_T^{\rm lep}$ has to satisfy $H_T^{\rm lep} \geq 1200$  GeV.
\end{itemize}

The validity of our implementation of the CMS analysis has been checked by simulating the process of pair production of $X_{5/3}$ quarks and their subsequent on-shell decays into $W t$. Following \cite{Sirunyan:2018yun},
events have been simulated at LO using {\sc MG5\_aMC}~\cite{Alwall:2014hca} and the decays performed through {\sc MadSpin}~\cite{Artoisenet:2012st}. For the validation we used the {\sc NNPDF23\_NLO\_as\_0119} PDF set~\cite{Ball:2012cx}
and, as in  \cite{Sirunyan:2018yun}, the cross section is rescaled to the
NNLO+NNLL value computed by the Top++ code \cite{Czakon:2011xx}.

The UFO model used for the validation is described in~\cite{Fuks:2016ftf} and is publicly available in the {\sc Feynrules} repository. Events have been simulated for purely left-handed (LH) and for purely right-handed (RH) couplings between the $X_{5/3}$ and the $W$ boson; in each case, the value of the coupling has been set to 0.1 to ensure the NWA, as shown in Table~\ref{tab:NWAvalidation}.
Hadronisation and parton showering have been simulated through {\sc Pythia 8.2}~\cite{Sjostrand:2014zea}. 
\begin{table}[h]
\centering
\begin{tabular}{ccccccccc}
\toprule
 $M_X~(\rm GeV)$ & 800 & 900 & 1000 & 1100 & 1200 & 1300 & 1400 & 1500 \\
 $\Gamma_X/M_X~(\%)$ & 0.18 & 0.24 & 0.30 & 0.37 & 0.44 & 0.53 & 0.61 & 0.71 \\
\bottomrule
\end{tabular}
\caption{\label{tab:NWAvalidation} $X_{5/3}$ width-over-mass ratio (in \%) as function of $M_X$ for purely LH or RH couplings of the $X_{5/3}$ to $W^+ t$.
The numerical value of the coupling has been set to 0.1, but $\Gamma_X/M_X$
is independent of this choice. }
\end{table}

The benchmark scenario $M_X = 1\TeV$ has been considered for both coupling hypotheses. Using the cut-flow information, the total efficiencies $\varepsilon_{\rm SR}(M_X)$ have been obtained for all signal regions $SR \in \{ee , e\mu , \mu\mu\}$. The number of expected events $s_{SR}$ in the given signal region has been then calculated using the standard expression
\begin{equation}\label{yield}
s_{SR}(M_X) = \mathcal L \sigma_{pp\to X_{5/3} \bar X_{5/3}} (M_X) \varepsilon_{SR}(M_X),
\end{equation} where $\mathcal L = 35.9\fb^{-1}$ is the integrated luminosity of the CMS search. 
The calculated values of $s_{SR}(1~\rm{TeV})$
for the right-handed coupling in different SRs are quoted in Table~\ref{tab:Svalidation}, and compared to the values presented in Table 1
of~\cite{Sirunyan:2018yun}.
The error associated to the obtained expected number of events has been estimated to be 10\%, to account for systematic uncertainties of the simulated dataset. Our results are in excellent agreement with the CMS reported yields.
\begin{table}[h!]
\centering
\begin{tabular}{ccc}
\toprule
Channel & CMS result & recast result \\
\midrule 
$ee$     & $11.6 \pm 0.8$ & 11.63\\ 
$\mu\mu$ & $16.1 \pm 1.2$ & 16.24\\
$e\mu$   & $26.9 \pm 1.9$ & 26.87\\
\bottomrule 
\end{tabular}
\caption{\label{tab:Svalidation} The expected yields for a $X_{5/3}$ with mass $M_X = 1\TeV$ and purely right-handed couplings in three different SRs, along with the CMS results (Table 1 of~\cite{Sirunyan:2018yun}). The uncertainty of our results is estimated to be 10\%.}
\end{table}

\begin{figure}[!htbp]
\centering
\includegraphics[width=.325\textwidth]{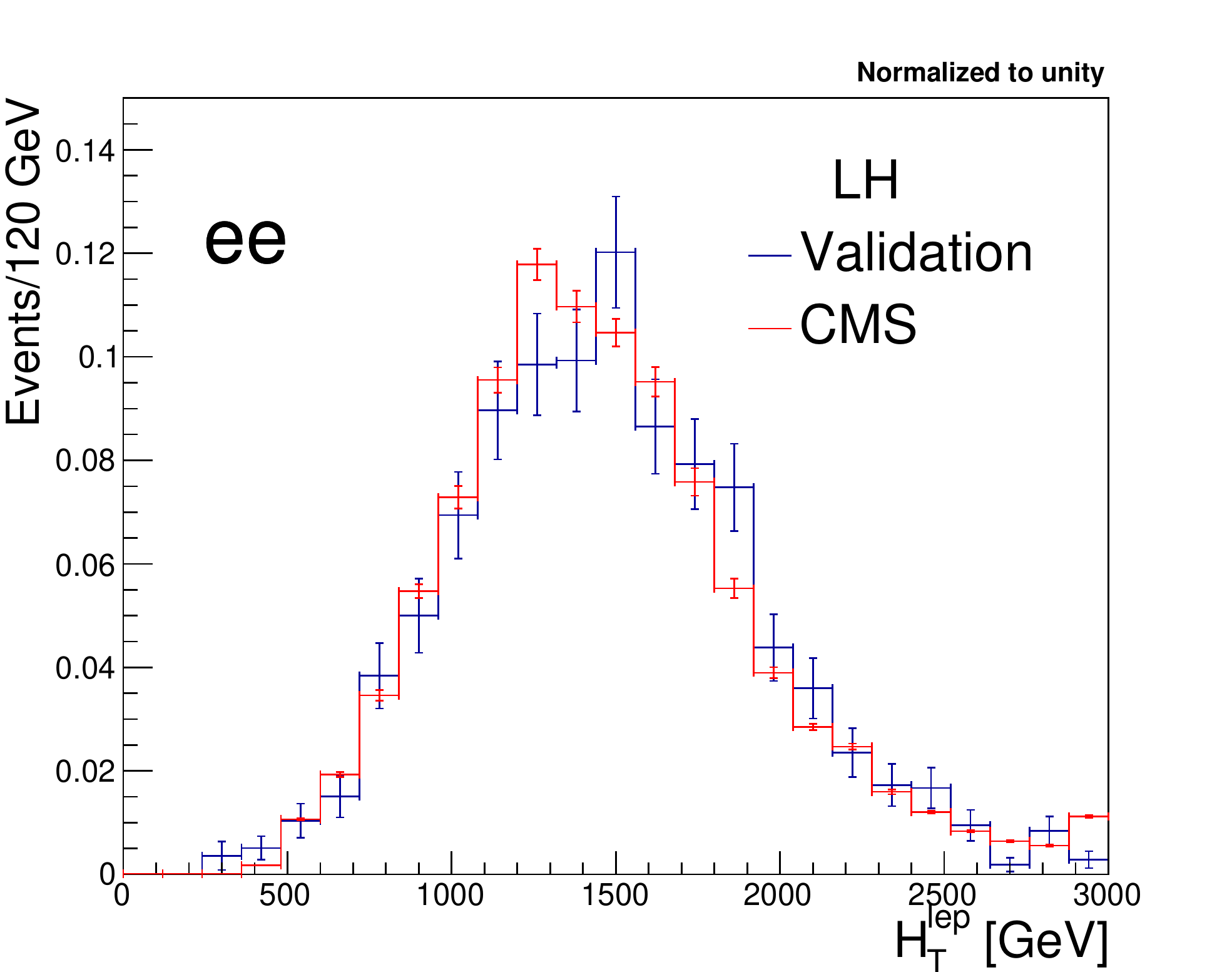}
\includegraphics[width=.325\textwidth]{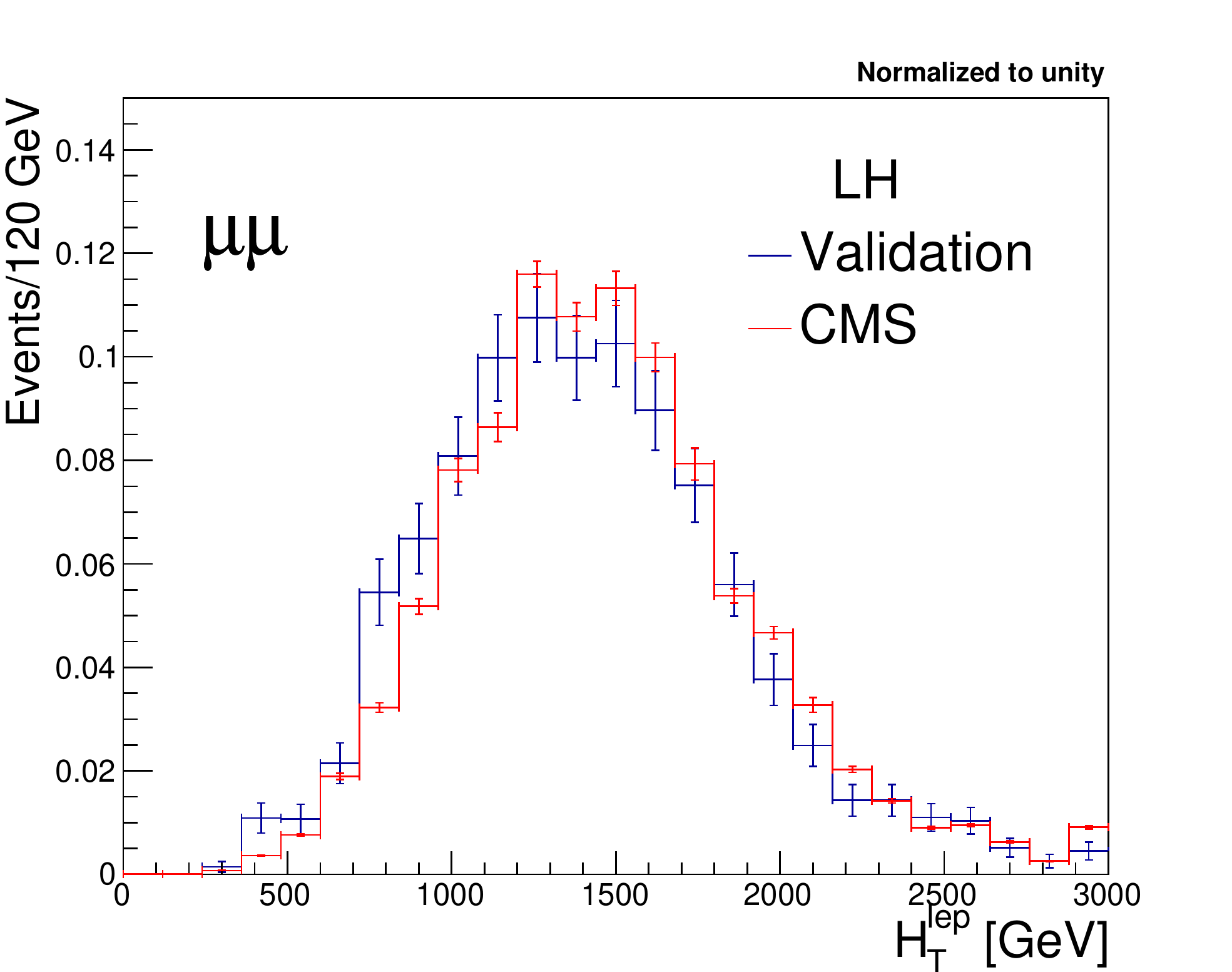}
\includegraphics[width=.325\textwidth]{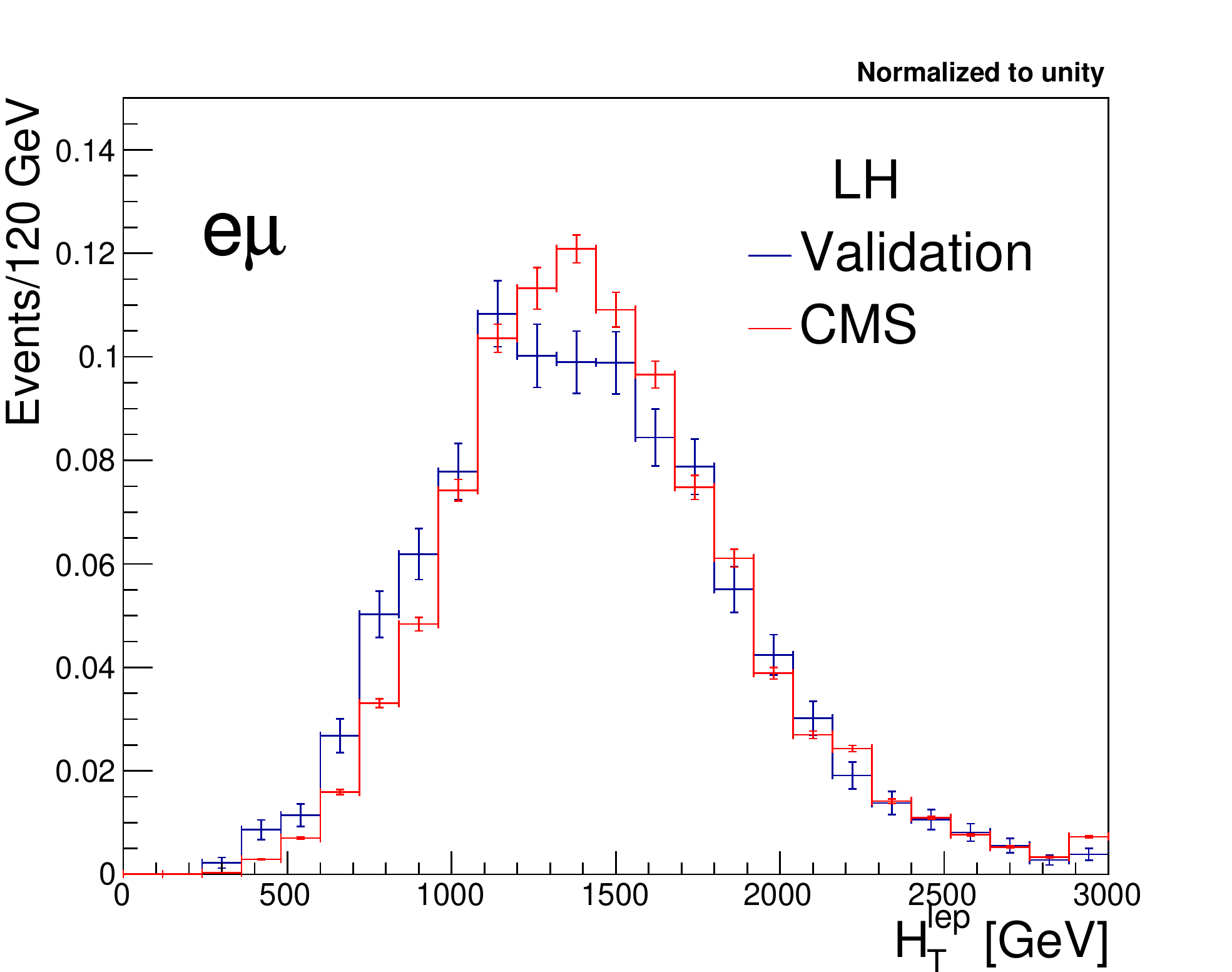}\\
\includegraphics[width=.325\textwidth]{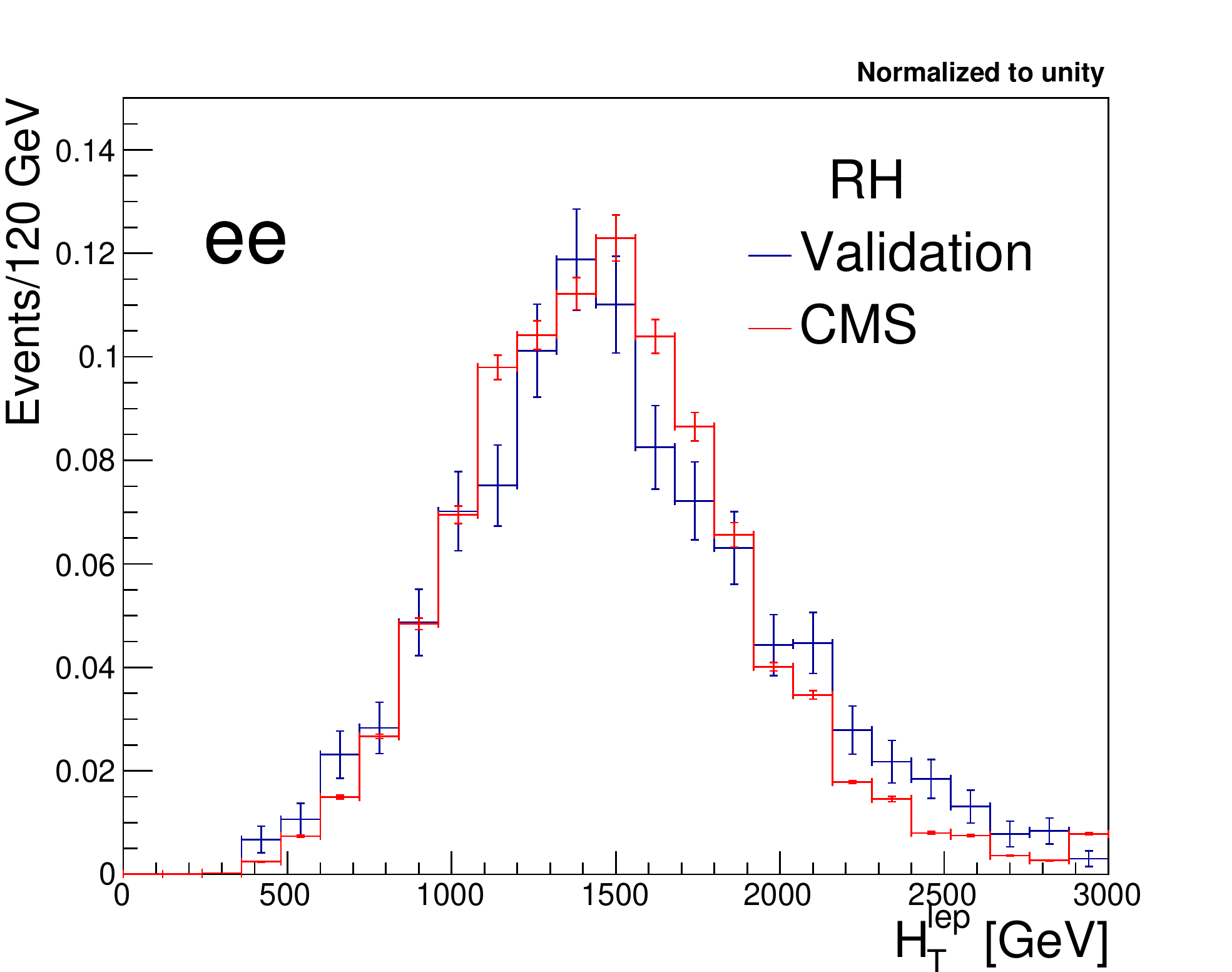}
\includegraphics[width=.325\textwidth]{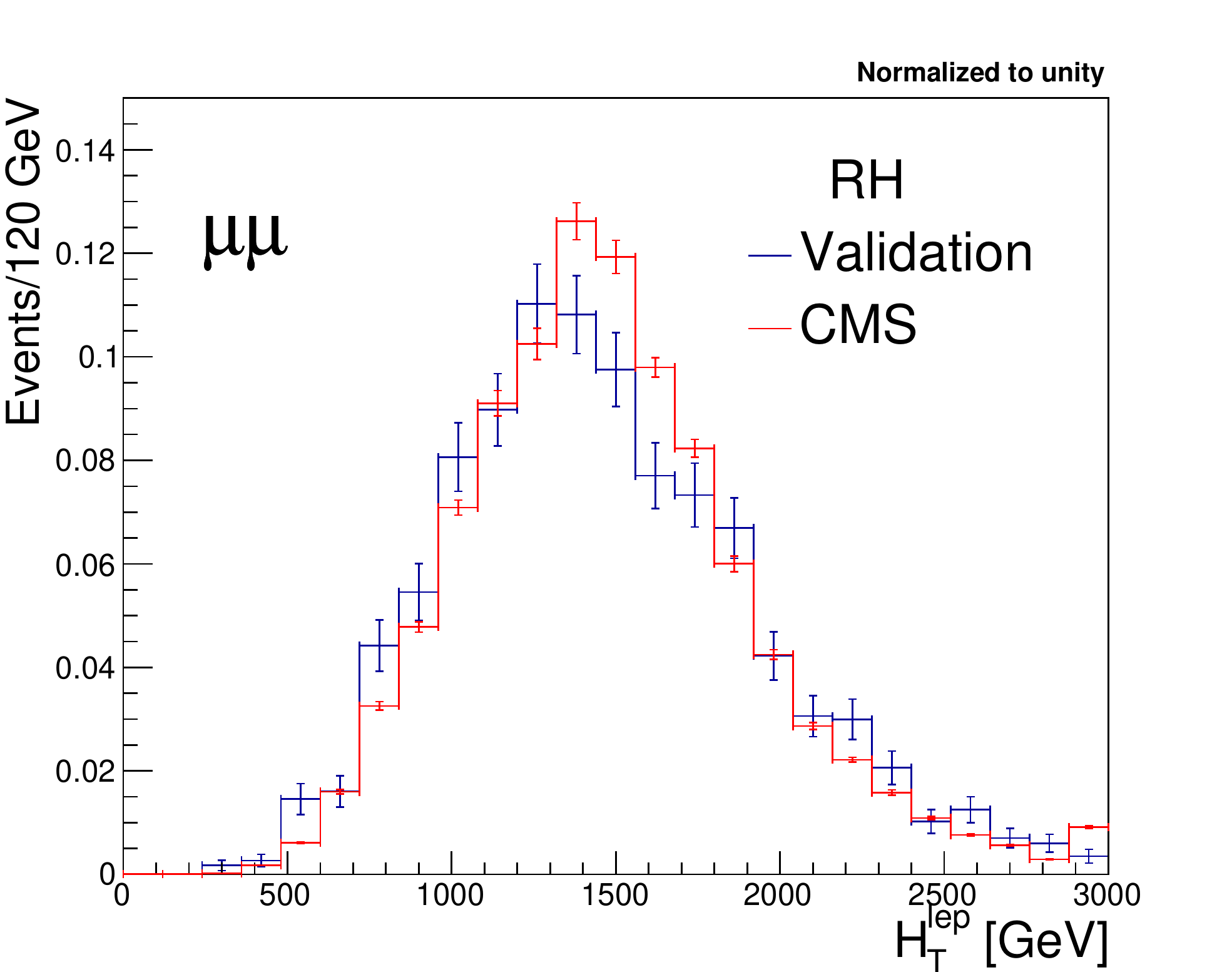}
\includegraphics[width=.325\textwidth]{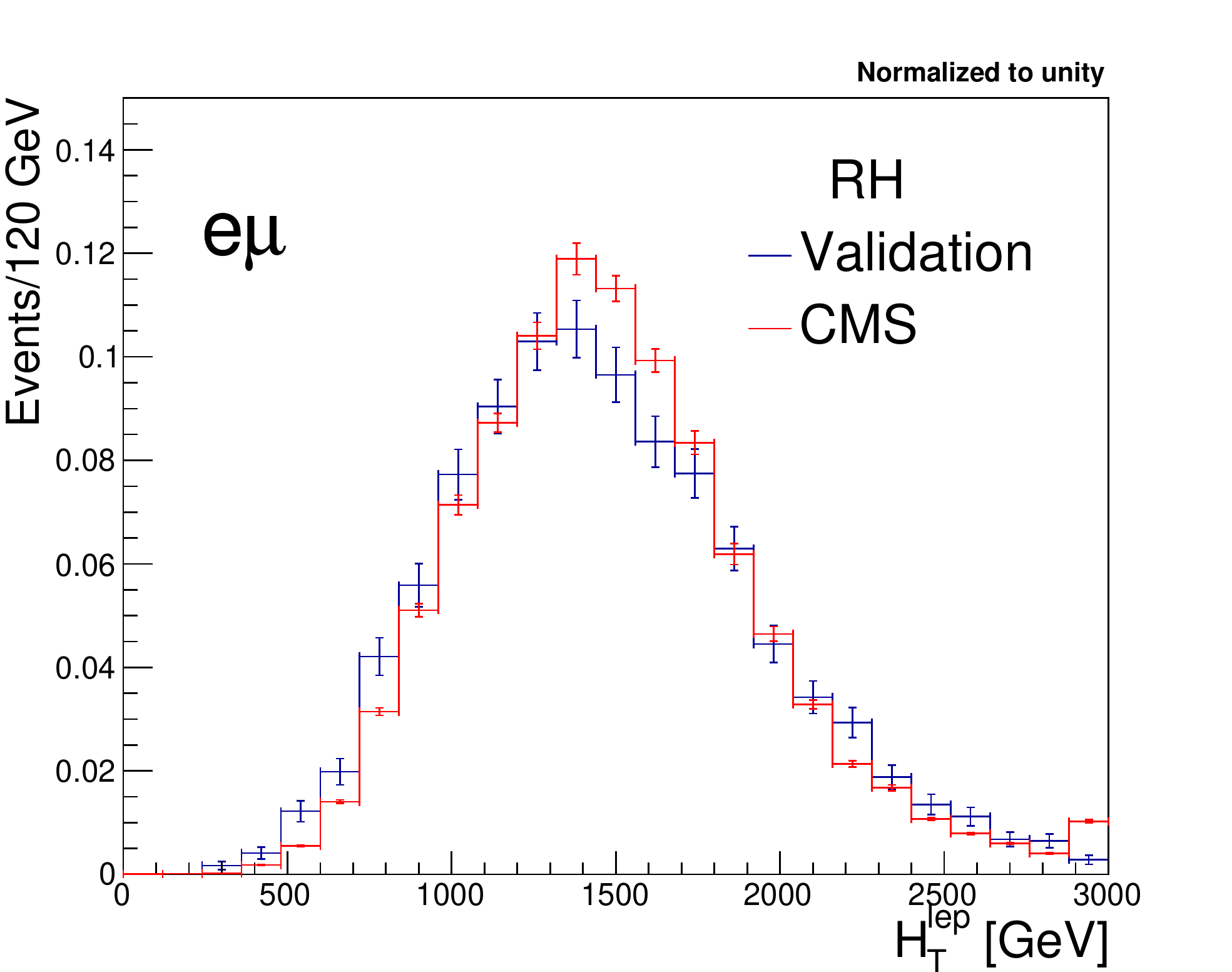}\\
\caption{\label{fig:HTlep} Expected $H_T^{\rm lep}$ distributions for a $X_{5/3}$ with mass $M_X = 1\TeV$  and with either left-handed \textbf{(top row)} or right-handed \textbf{(bottom row)} couplings. The distribution is plotted before the two final cuts $N_{\rm const}\geq 5$ and $H_T^{\rm lep} \geq 1200\GeV$. The blue line corresponds to the
results of our simulations while the red line corresponds to the results of~\cite{Sirunyan:2018yun}. All distributions are normalized to unity. The last bin includes the overflow events.}
\end{figure}

\begin{figure}[!htbp]
\centering
\includegraphics[width=.325\textwidth]{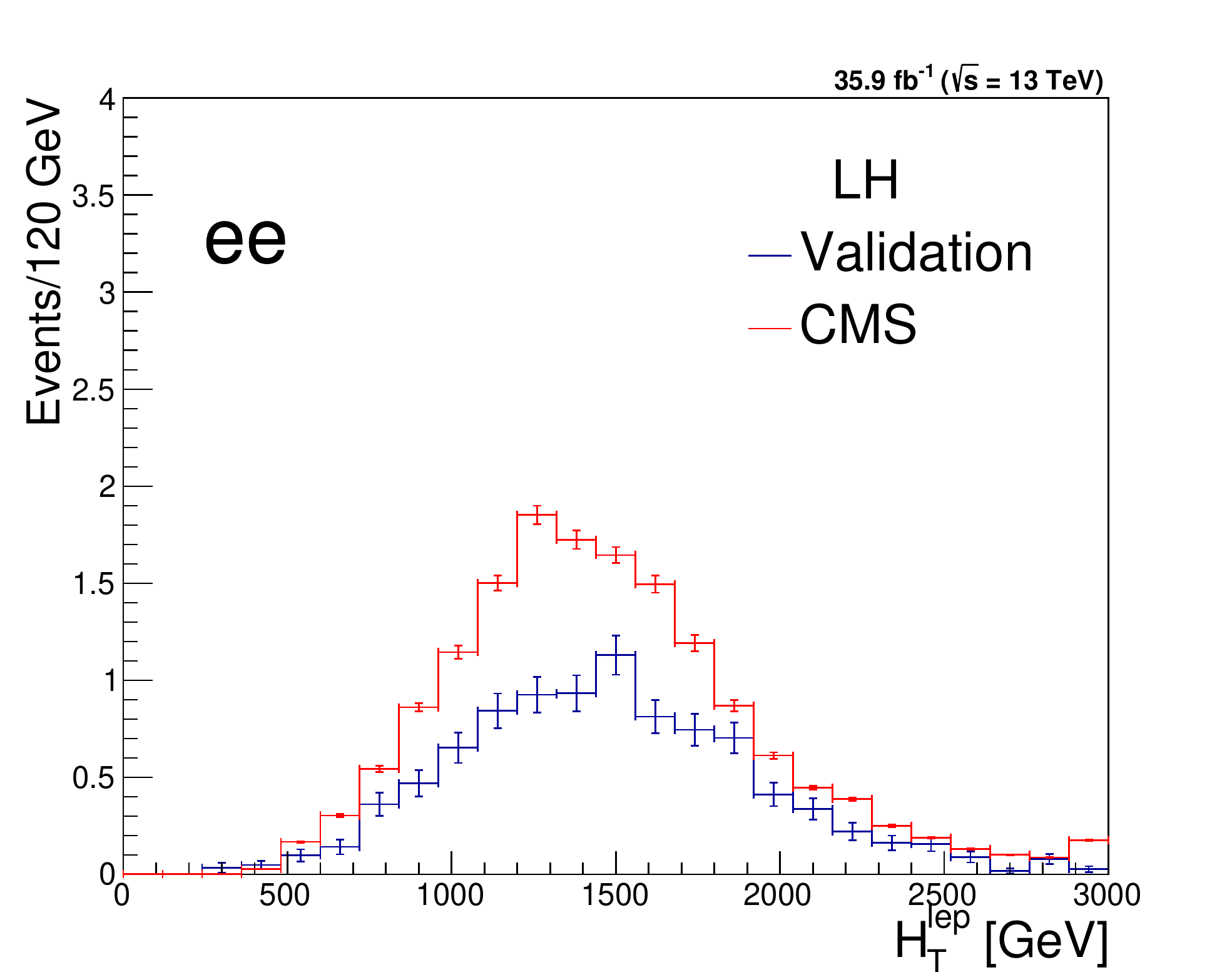}
\includegraphics[width=.325\textwidth]{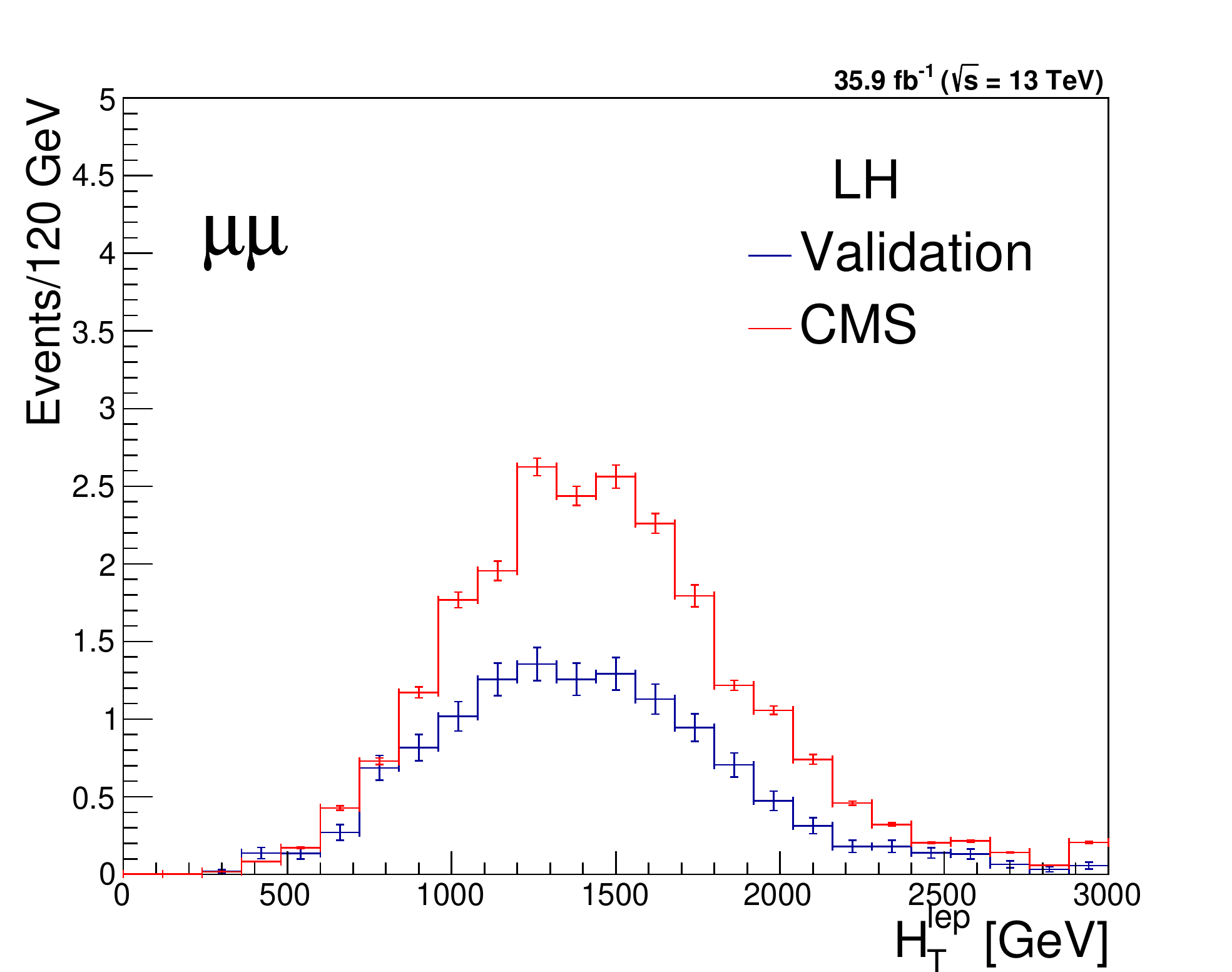}
\includegraphics[width=.325\textwidth]{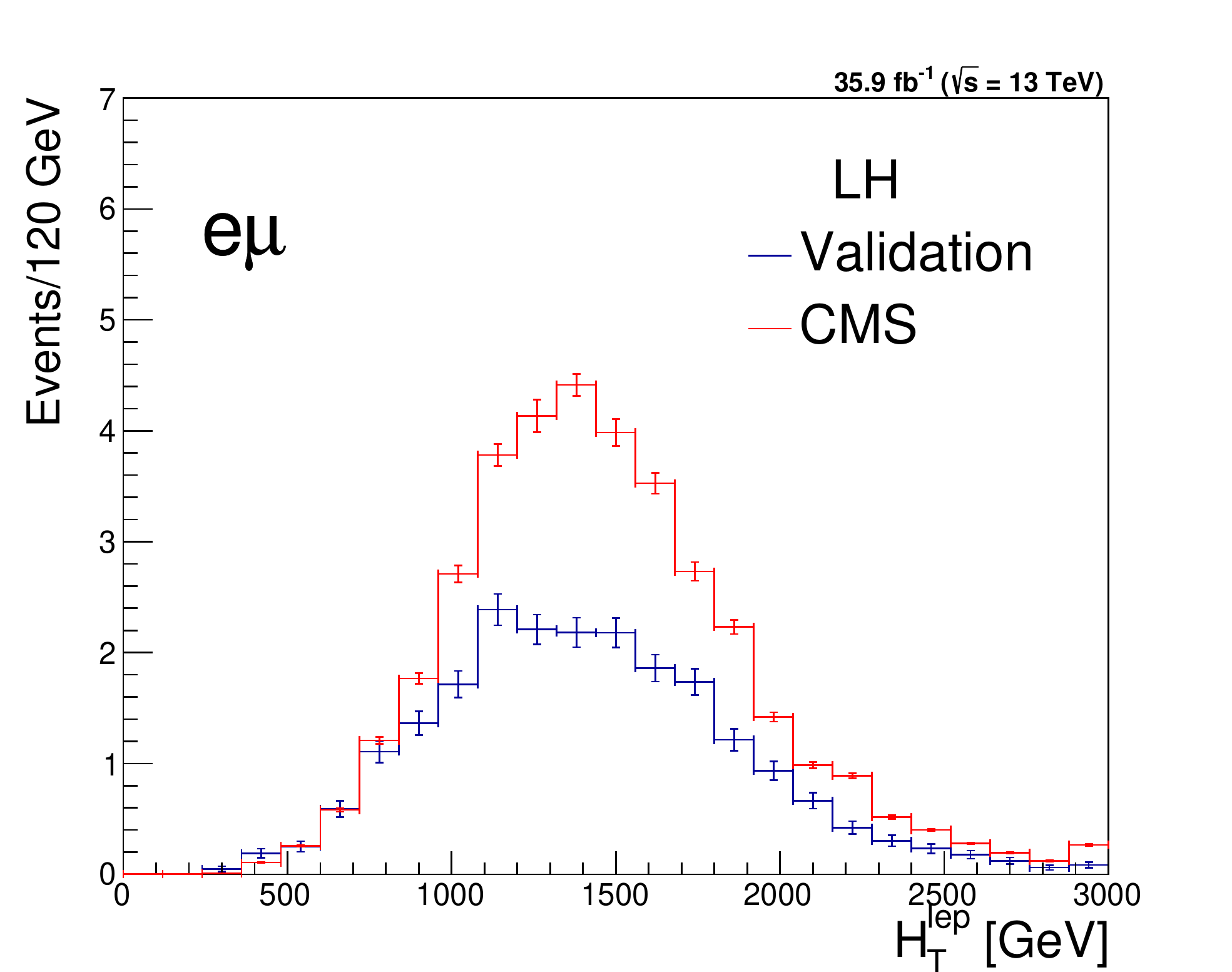}\\
\includegraphics[width=.325\textwidth]{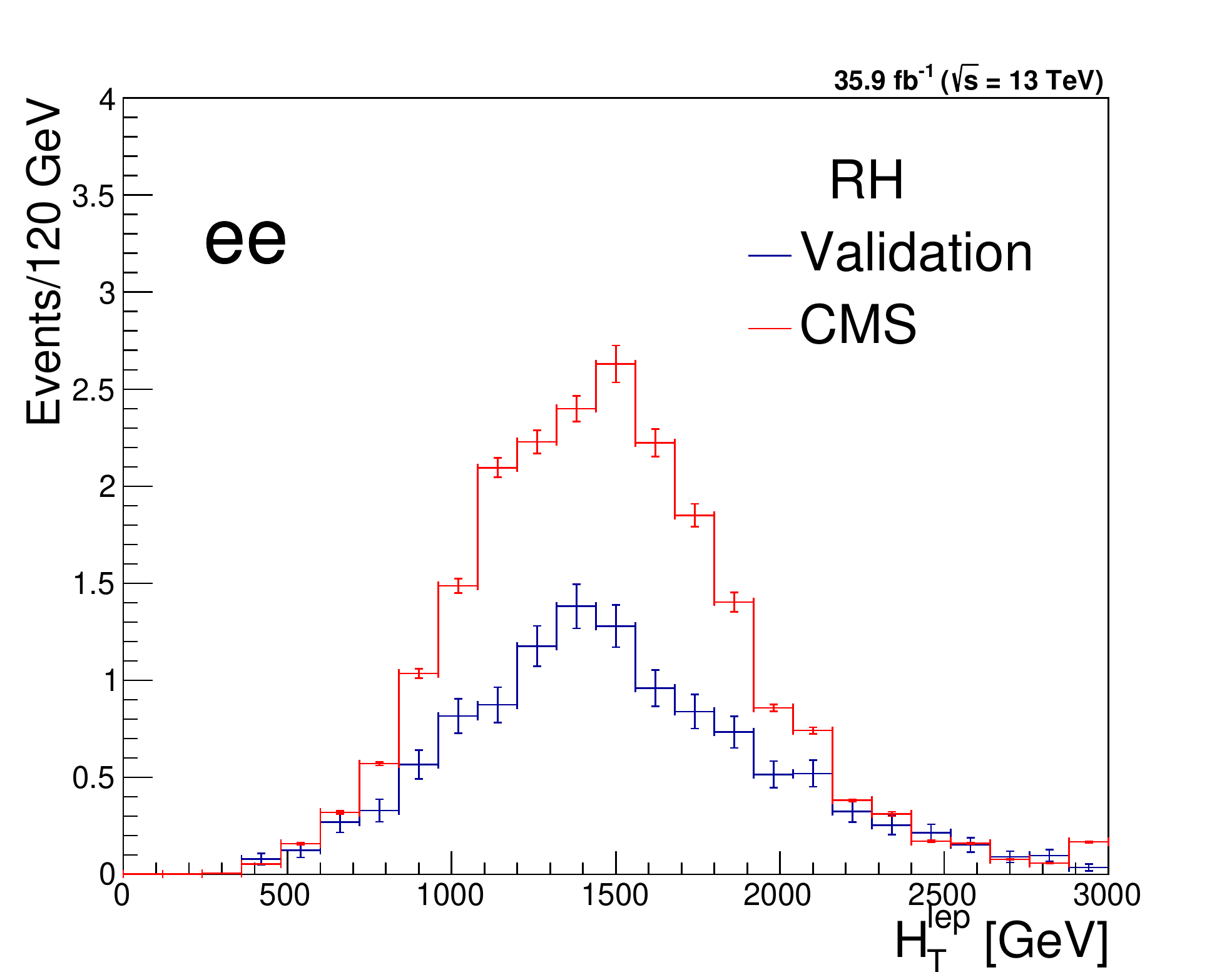}
\includegraphics[width=.325\textwidth]{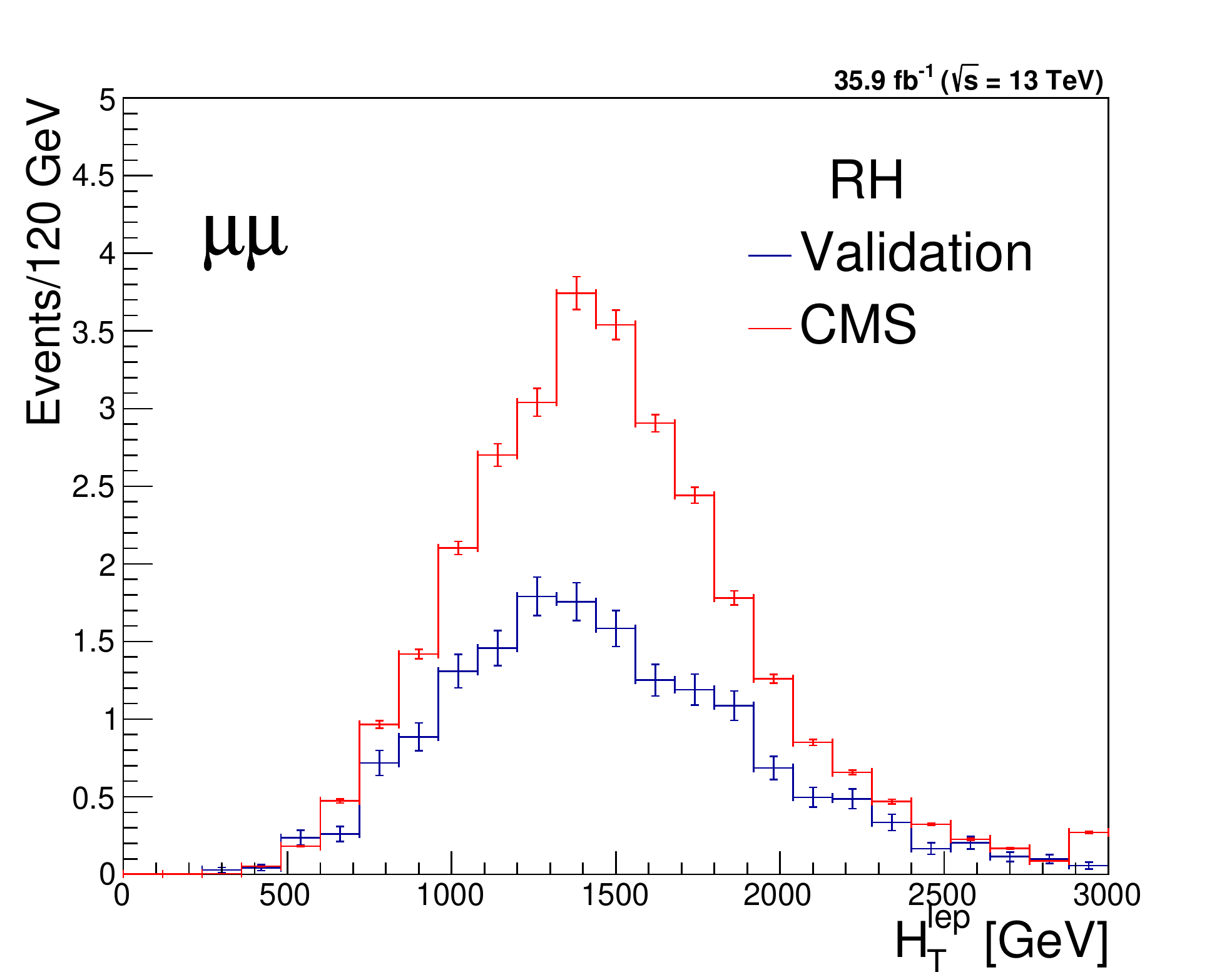}
\includegraphics[width=.325\textwidth]{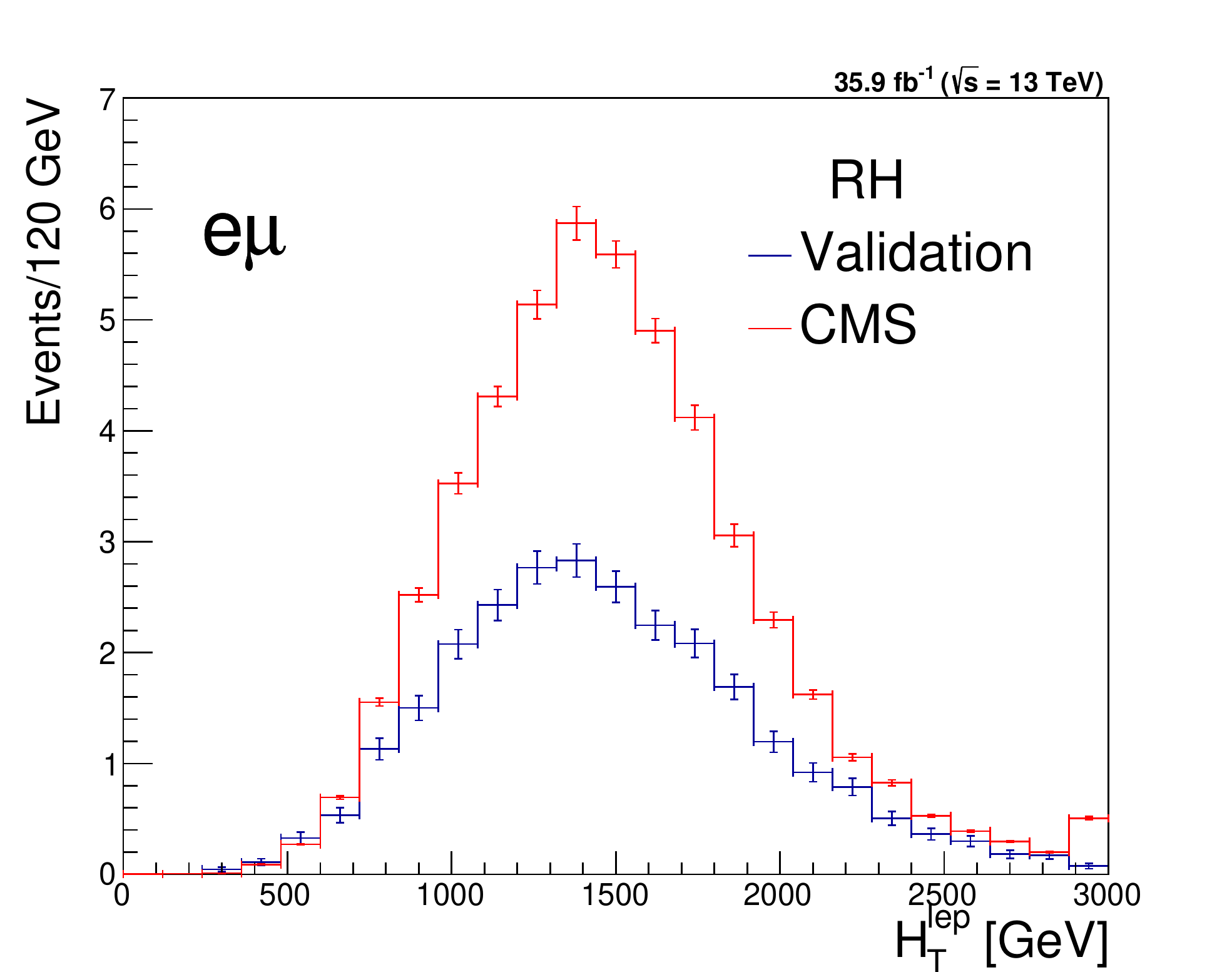}\\
\caption{\label{fig:HTleplumi} Same distributions as in Figure~\ref{fig:HTlep} but normalized to the number of events corresponding to an integrated luminosity of $\mathcal L = 35 \fb^{-1}$.}
\end{figure}

Our implementation has also been validated at differential level: for both coupling scenarios we compare the $H_T^{\rm lep}$ spectrum after all but the last two cuts, namely before applying 
$N_{\rm const}\geq 5$ and $H_T^{\rm lep} \geq 1200\GeV$, with the distributions from Figure 2 of the CMS analysis \cite{Sirunyan:2018yun}.
Our comparison is presented 
in Figure~\ref{fig:HTlep} for both left-handed and right-handed couplings, normalising all distributions to unity to show the agreement of the shapes.
In Figure~\ref{fig:HTleplumi}, all spectra are instead
normalised to the number of events corresponding to the luminosity of the
CMS search, namely  $\mathcal L = 35.9\fb^{-1}$.
The result is that all distributions obtained in our simulation
framework consistently lie below the CMS ones, which signals that at the corresponding stage of the cut-flow we get a lower efficiency than CMS, uniformly distributed along the bins of $H_T^{\rm lep}$.
However, given the excellent agreement on the expected number of events after all the cuts, as Table~\ref{tab:Svalidation} displays,
we can reasonably assume that a compensation of effects should emerge when applying the remaining two cuts, independent of $M_X$ and most probably originating from the difference between our fast detector simulation and the full detector description adopted by CMS. It is not possible to elaborate further on this issue, as the CMS analysis does not provide the step-by-step information about the cut-flow efficiencies.

Finally, we compare the upper limits on the signal cross-section as function of the mass of $X_{5/3}$ yielded by our implementation and by the CMS analysis. 
For the purpose of obtaining such limits, it is useful summarizing in
Table~\ref{tab:ex_VLQ_bckg_nobs} the number of observed and background events
reported by CMS in the different SRs.
\begin{table}
\centering
\begin{tabular}{c|ccc}
\toprule
SR                 & $ee$         & $\mu\mu$   & $e\mu$       \\
\midrule
$n^{obs}_{\rm SR}$ & 10           & 12         & 26 \\
$b_{\rm SR}$       & 10.9$\pm$1.9 & 11.2$\pm$2 & 23.2$\pm$3.7 \\
\bottomrule
\end{tabular}
\caption{\label{tab:ex_VLQ_bckg_nobs} The number of observed and of background events in the three SRs of the CMS search~\cite{Sirunyan:2018yun}.} 
\end{table}
Then, the only results we can exploit from~\cite{Sirunyan:2018yun} are the expected and observed 95\% CL upper limits  on the cross-section for the combination of the three same-sign di-lepton SRs (Figure 7a of~\cite{Sirunyan:2018yun}, reported in HEPData~\cite{1697570}). We combined our limits through the CLs method~\cite{Read:2000ru,Read:2002hq}, including a theoretical systematic uncertainty on the signal yield of 10\% (results are very weakly sensitive to this uncertainty though). The agreement with the CMS results is very good for both LH and RH scenarios, with a slightly better performance for the RH scenario. We thus conclude that the validation of the recast implementation is successful.
\begin{figure}[!htbp]
\centering
\includegraphics[width=.49\textwidth]{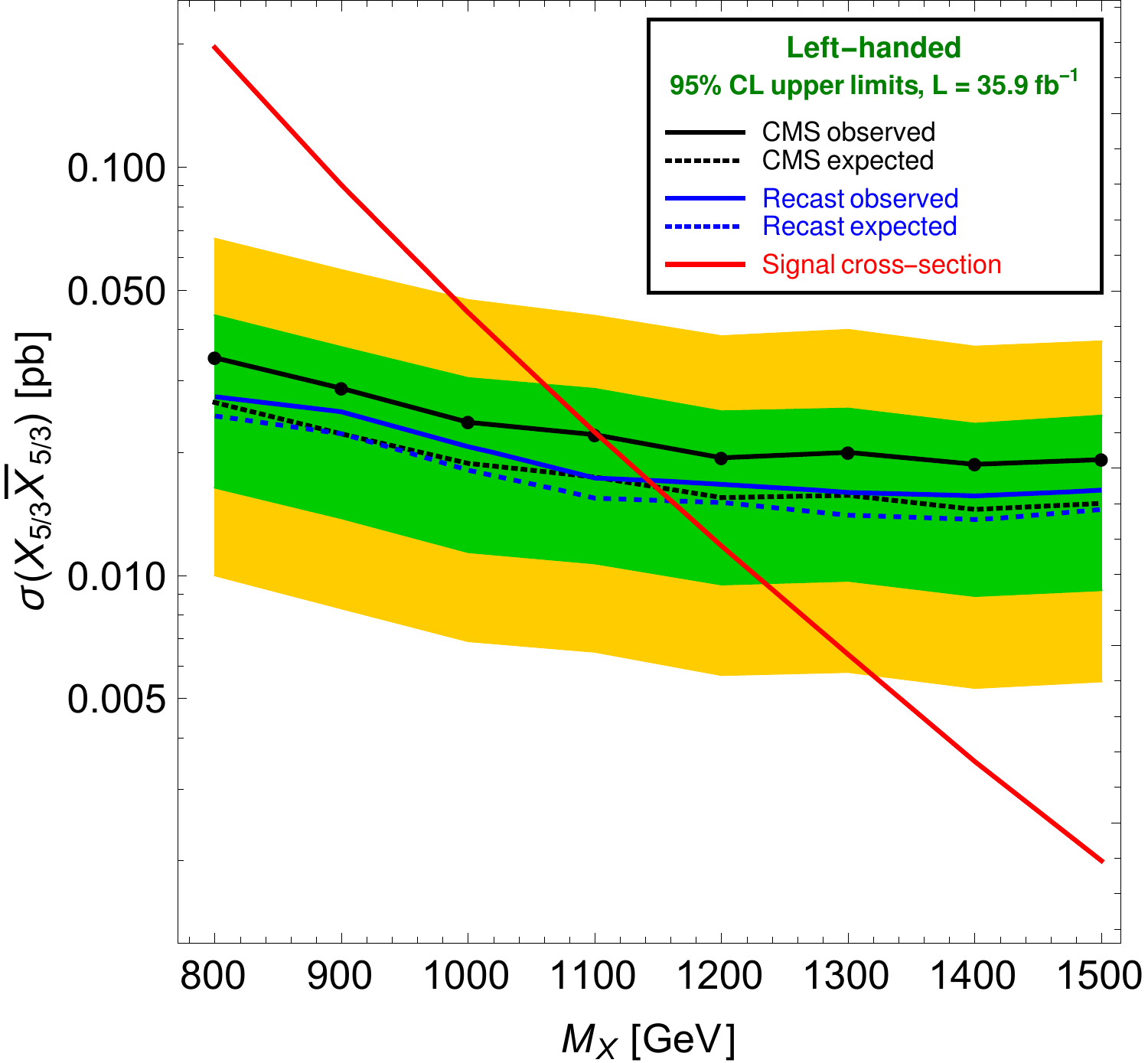}
\includegraphics[width=.49\textwidth]{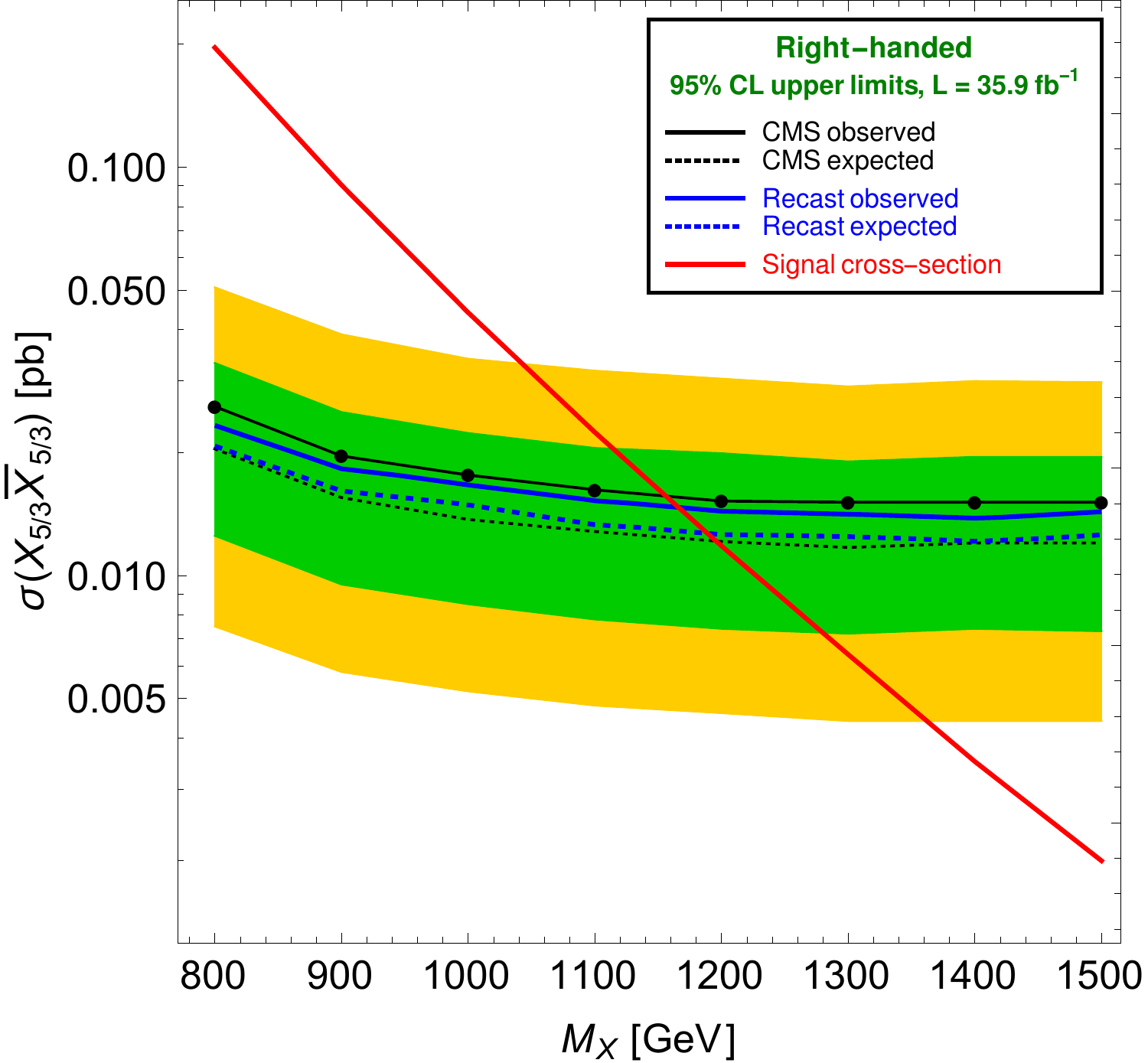}
\caption{\label{fig:upperXSvalidation} Observed and expected upper limits on the pair-production cross section of left-handed (left) and
right-handed (right) $X_{5/3}$ assuming $BR(X_{5/3} \to t W^+) = 1$. The results of our simulations (blue lines) are compared with the CMS results (black lines) from Figure 7 of~\cite{Sirunyan:2018yun}, reported in HEPData~\cite{1697570}. The red line represents the theoretical cross section used by the CMS collaboration.}
\end{figure}

\newpage

\section{Model-independent efficiencies, bounds and projections for all decay channels and signal regions}\label{app:otherbounds}
This Appendix focuses on the model-independent efficiencies, bounds and projections of the remaining decay channels of the exotic bosons, combined with all possible SRs. Along the same lines of Section~\ref{sec:LHCpheno}, the full set of relevant plots is collected in the next pages.
In this Appendix, exclusive interactions of both $X_{5/3}$ and the charged bosons are always assumed: this implies the presence of unphysical scenarios, in which it is assumed that the two $X_{5/3}$ or the two charged bosons propagating in the diagrams decay to different particles, always with 100\% BRs. Such results are only functional for a reconstruction of generic signals where the BRs are not 100\% and multiple combinations of final states are allowed. If the scenarios are physical and the sensitivity of the CMS search in a specific final state is not negligible, bounds and projections are provided as well.

A further simplifying assumption is that the masses of singly and doubly charged bosons are considered degenerate in the calculation of recast efficiencies for processes where one $X_{5/3}$ decays into a singly charged boson and the
other one into a doubly charged boson.

\subsection{Pair production of $X_{5/3}$ with both branches decaying into $V^\pm$}

\begin{figure}[t!]
  \centering
  \includegraphics[width=.325\textwidth]{figures/Efficiencies_XVP_eve_SRee}
  \includegraphics[width=.325\textwidth]{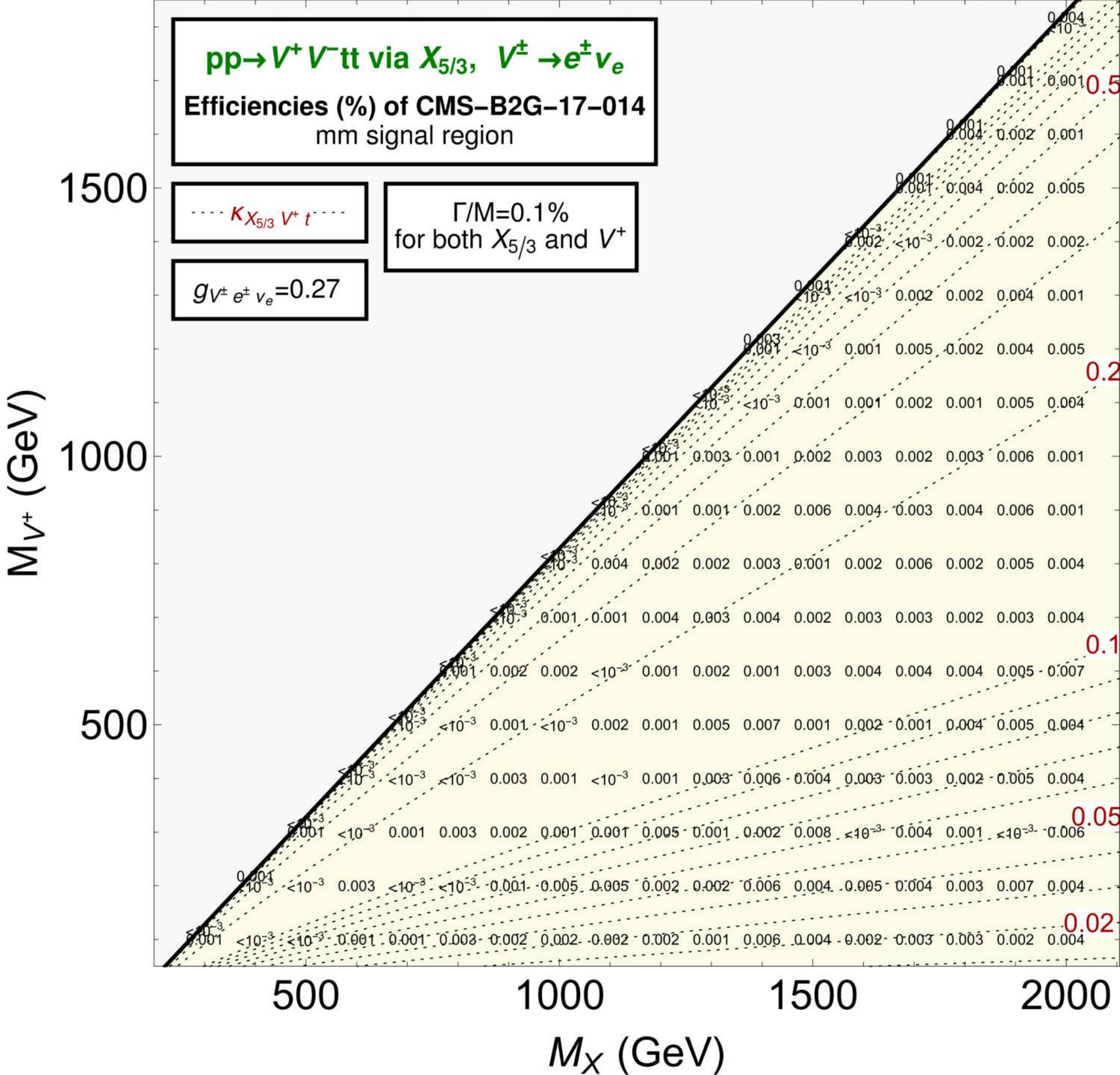}
  \includegraphics[width=.325\textwidth]{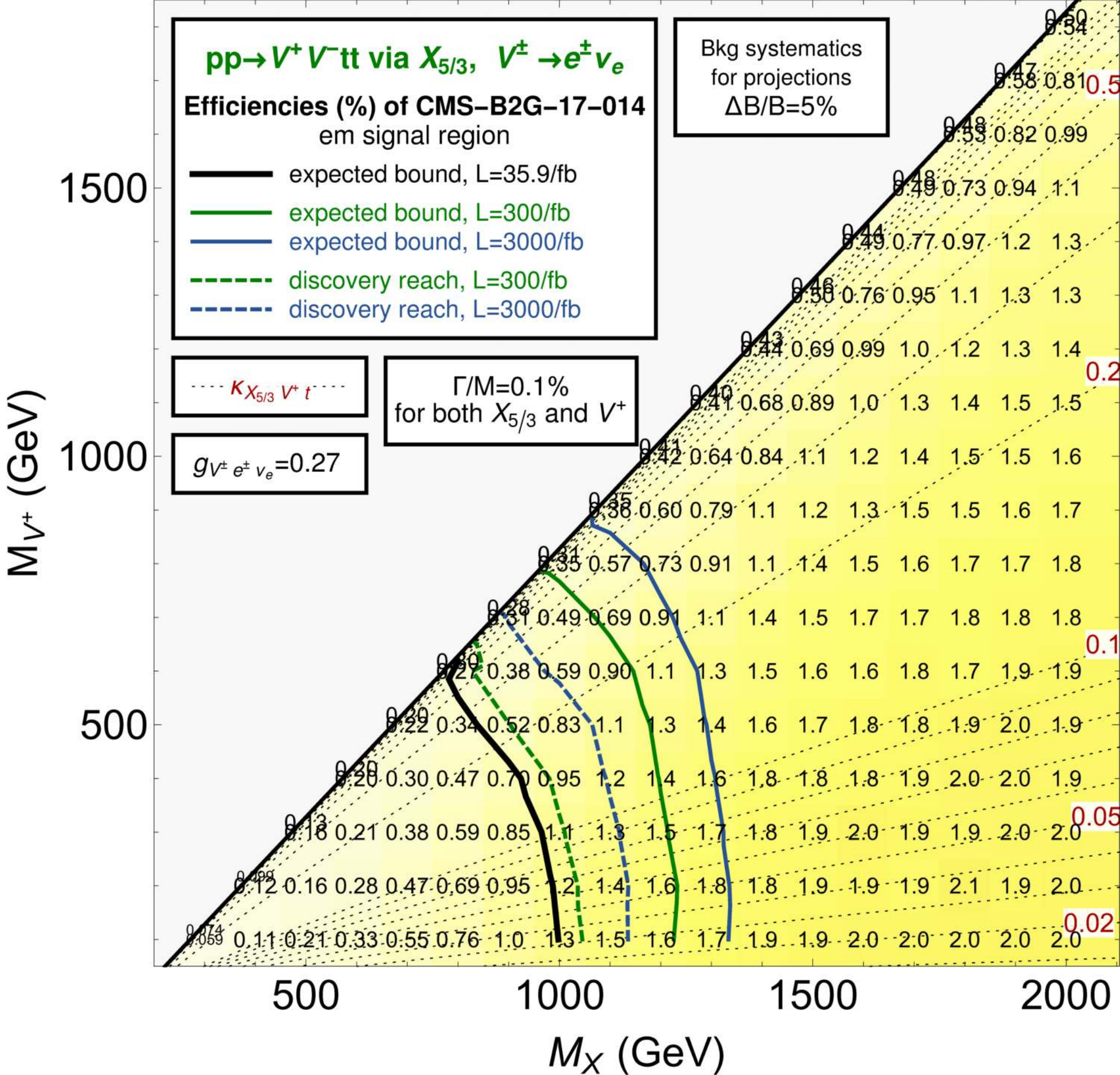}\\
  \includegraphics[width=.325\textwidth]{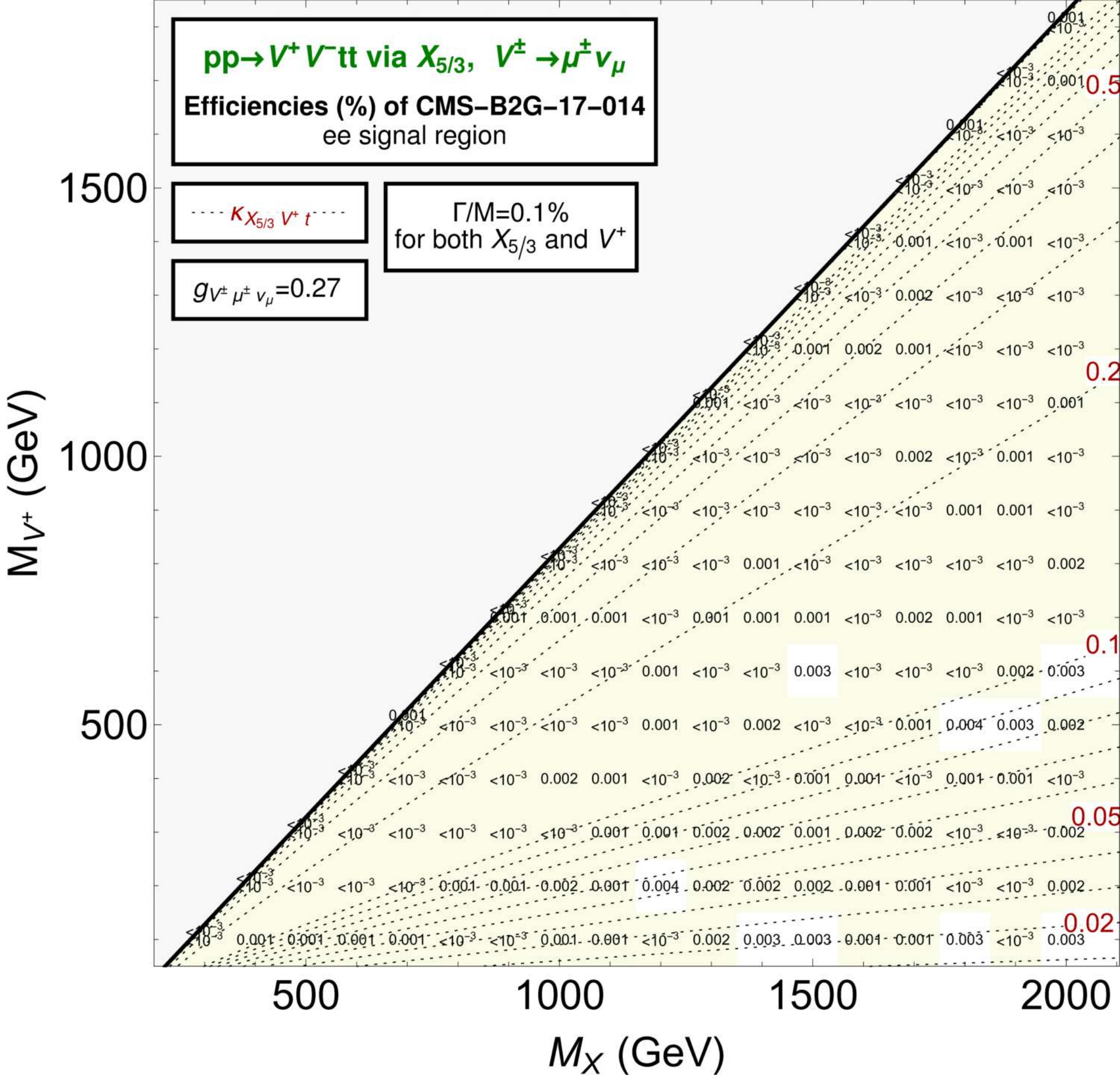}
  \includegraphics[width=.325\textwidth]{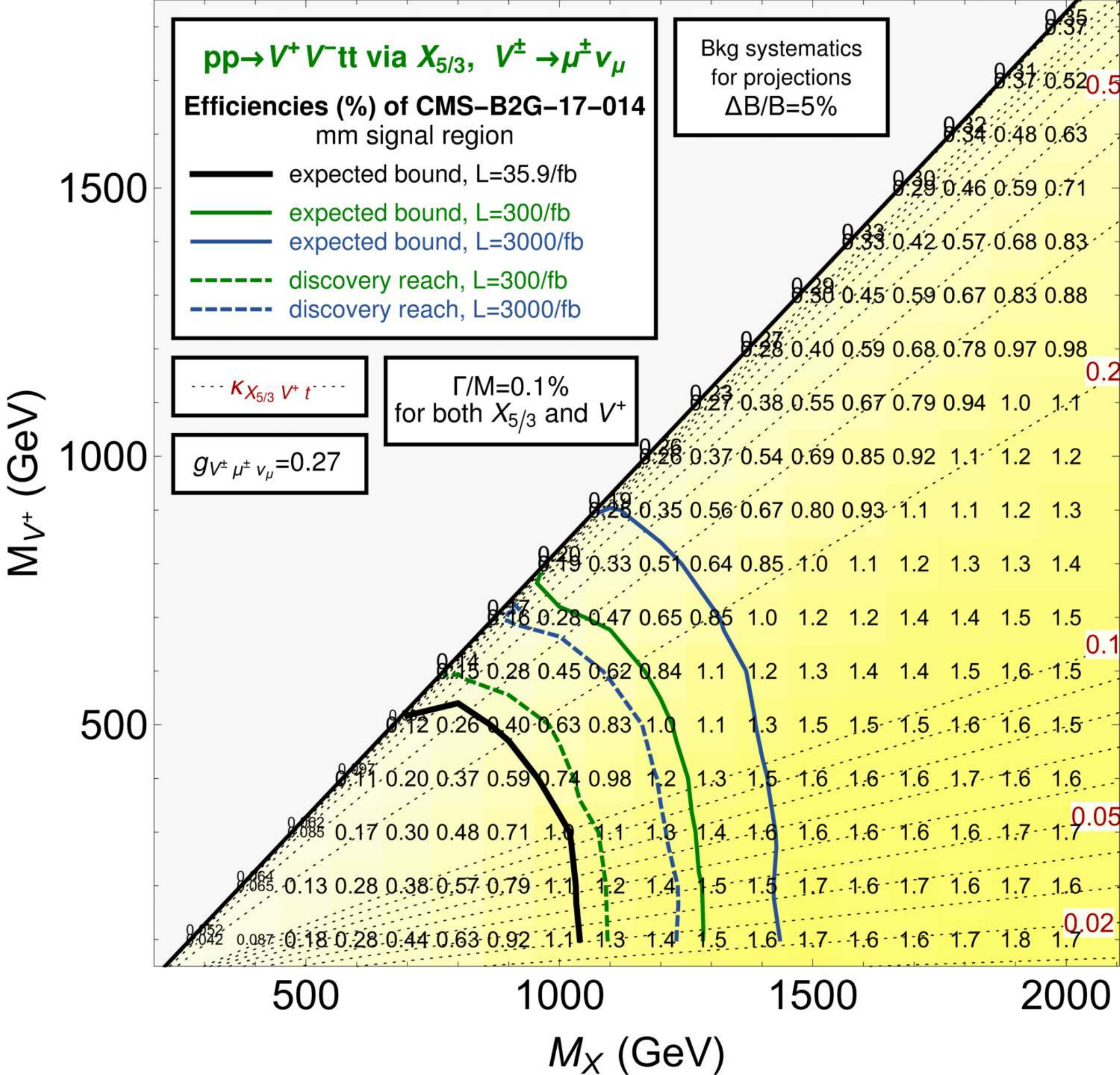}
  \includegraphics[width=.325\textwidth]{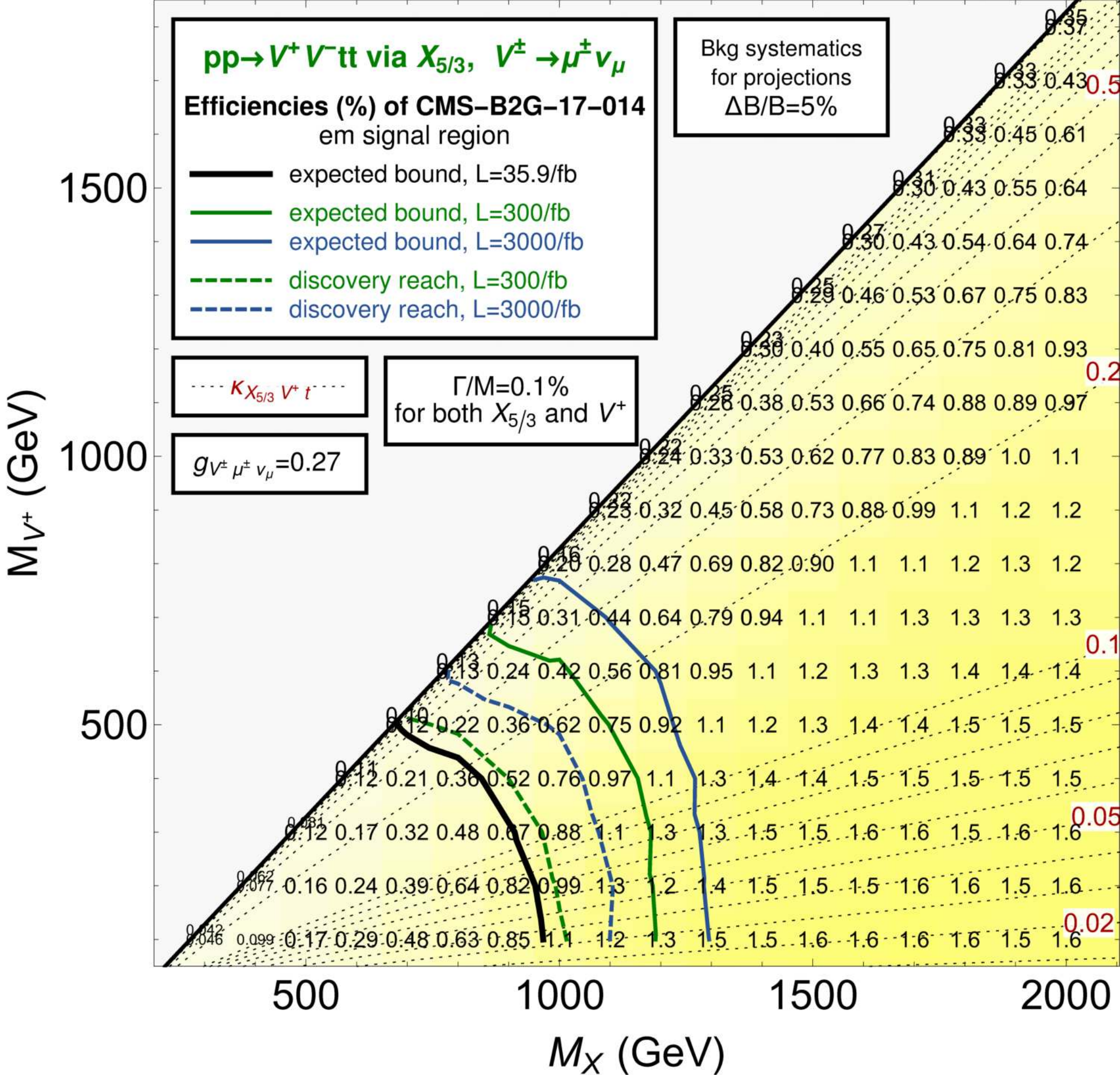}
  \caption{\label{fig:boundVP} Efficiencies for scenarios in which $V^{+}$ decays to the conjugate final state of $V^{-}$: $V^\pm\to e^\pm \nu_e$ {\bf (top row)} $V^\pm\to \mu^\pm \nu_\mu$ {\bf (bottom row)}, in the $ee$ {\bf (left column)}, $\mu\mu$ {\bf (central column)} and $e\mu$ {\bf (right column)} SRs of~\cite{Sirunyan:2018yun}. The meaning of curves, colours and symbols is the same as in Figure~\ref{fig:boundseSRee}. Due to the small efficiencies in SRs not targeting the leptons with same sign arising from the different signal hypotheses, the bounds and projections corresponding to 100\% decays of the vector bosons are shown only for the most sensitive SRs.}
\end{figure}

In Figure~\ref{fig:boundVP}, efficiencies for the cases involving the singly charged vector $V$ against the three signal hypotheses of the CMS analysis (same-sign $ee$, $\mu\mu$ and $e\mu$) are shown. In these plots, the decays of the $V$ boson and its conjugate have been assumed to involve same lepton flavours with $100\%$ saturated $BR$s. Top-central and bottom-left panels represent the analysis of the decay $V^{\pm}\to l^{\pm}\nu_l$ against the signal hypothesis containing same-sign $l'l'$ with $l\neq l'$. In these cases, the efficiency drops and it is not possible to extract relevant phenomenological information.

\begin{figure}[t!]
  \centering
  \includegraphics[width=.325\textwidth]{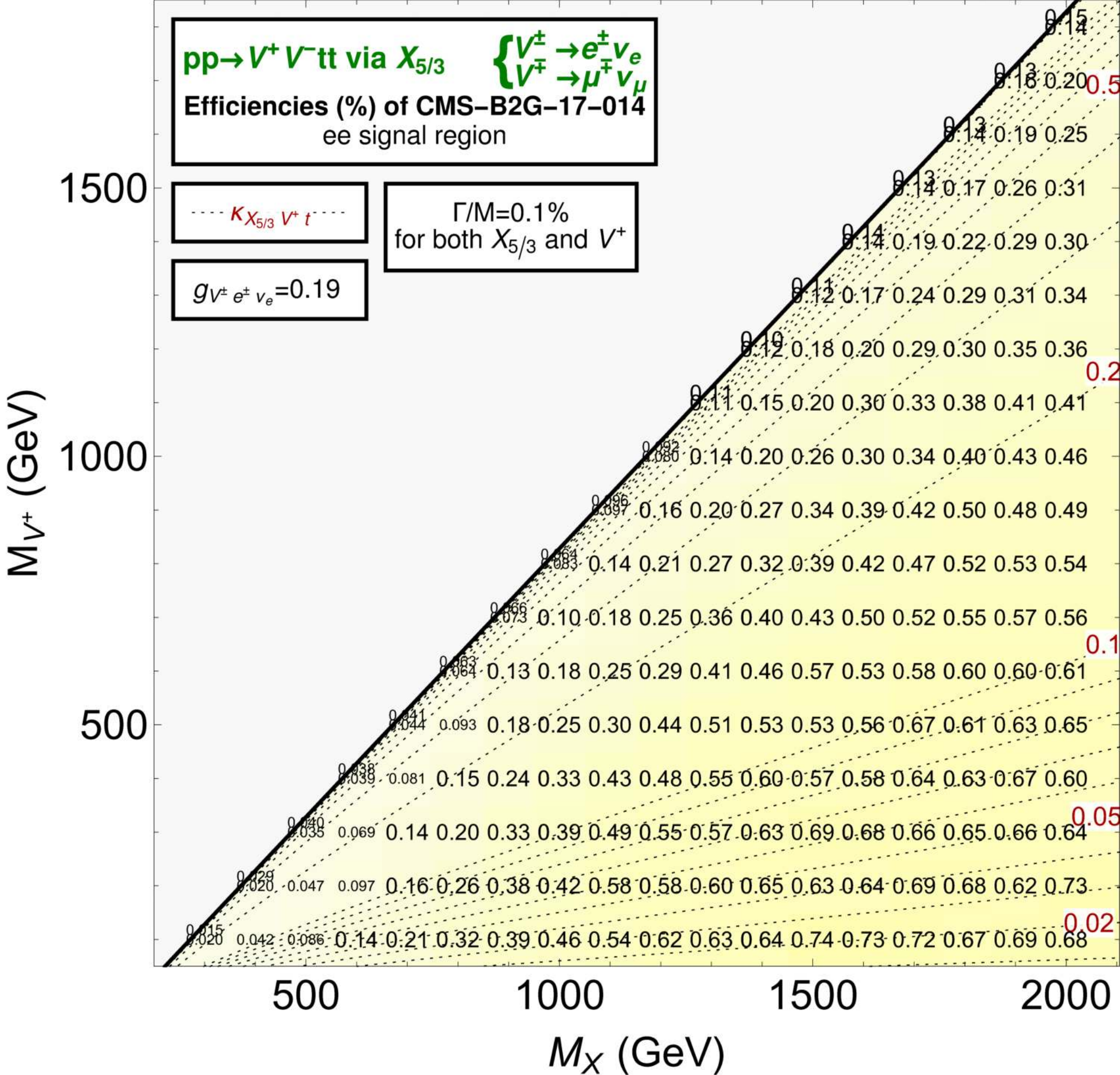}
  \includegraphics[width=.325\textwidth]{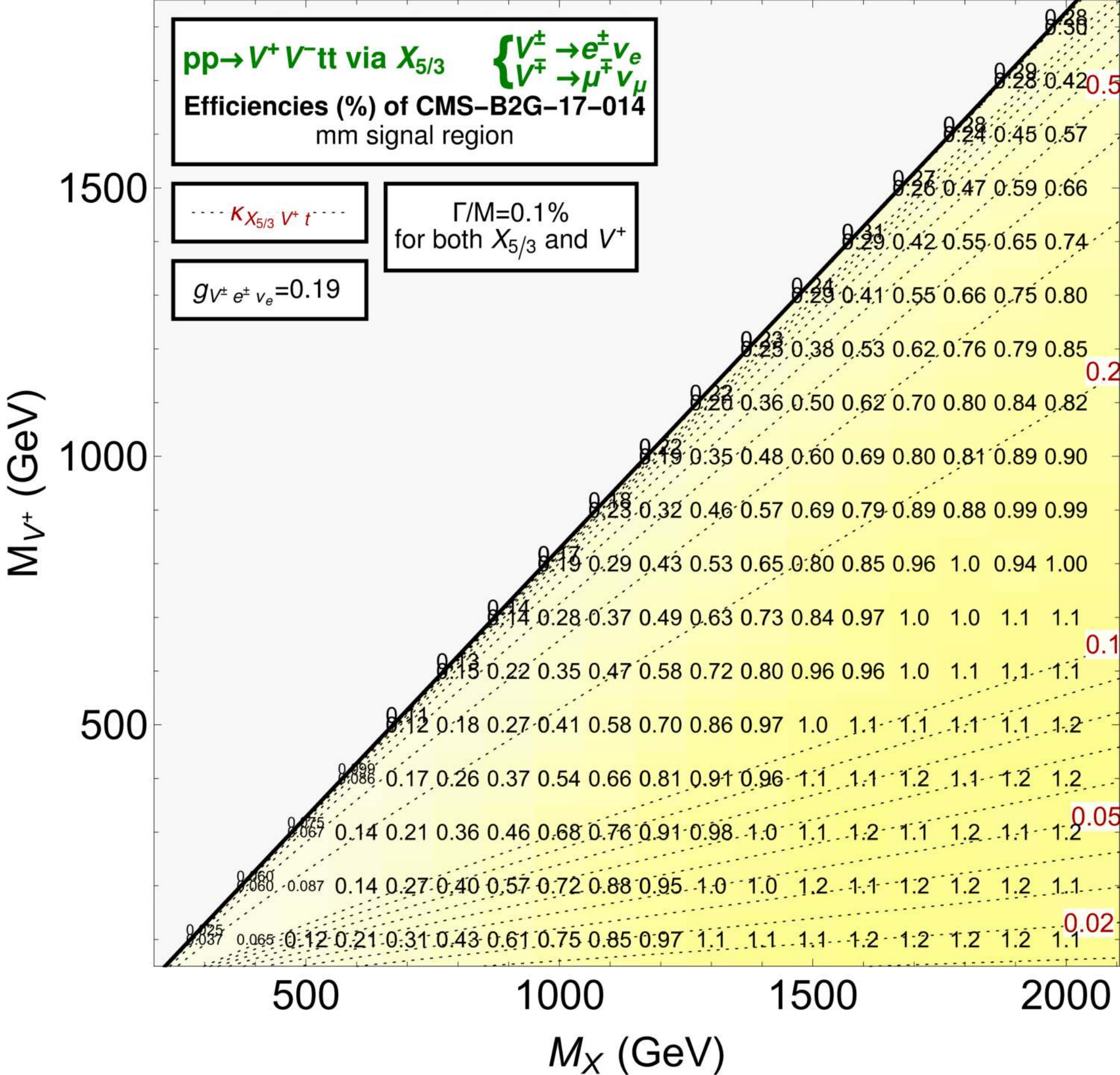}
  \includegraphics[width=.325\textwidth]{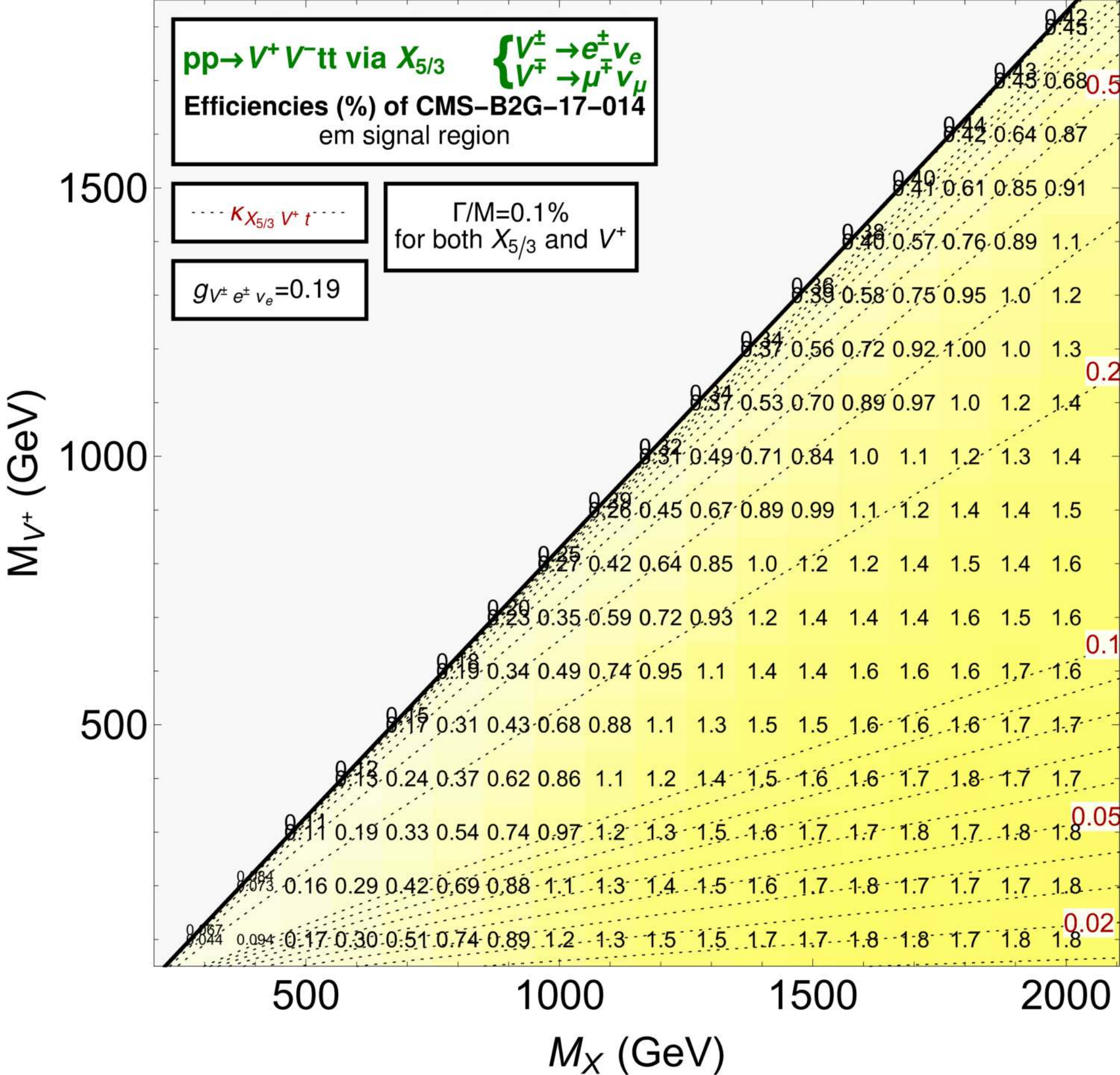}
  \caption{\label{fig:boundVP2} Efficiencies for scenarios in which $V^{\pm}$ decays to $e^\pm \nu_e$ and $V^{\mp}$ to $\mu^\mp \nu_\mu$, in the $ee$ {\bf (left column)}, $\mu\mu$ {\bf (central column)} and $e\mu$ {\bf (right column)} SRs of~\cite{Sirunyan:2018yun}. The meaning of colours and symbols is the same as in Figure~\ref{fig:boundseSRee}. Provided that each $BR$ is set to $100\%$, these scenarios are unphysical when considered individually, thus no bounds or projections are shown.}
\end{figure}

Figure~\ref{fig:boundVP2} is analogous to Figure~\ref{fig:boundVP}, but
the decays of $V$ and its conjugate have been assumed to involve different lepton flavours, e.g. $V^+\to e^+\nu_e$ and $V^-\to \mu^+\nu_\mu$,  with $100\%$ saturated $BR$s.

\subsection{Pair production of $X_{5/3}$ with both branches decaying into $V^{\pm\pm}$}
 
\begin{figure}[thbp]
  \centering
  \includegraphics[width=.325\textwidth]{figures/Efficiencies_XVPP_ee_SRee}
  \includegraphics[width=.325\textwidth]{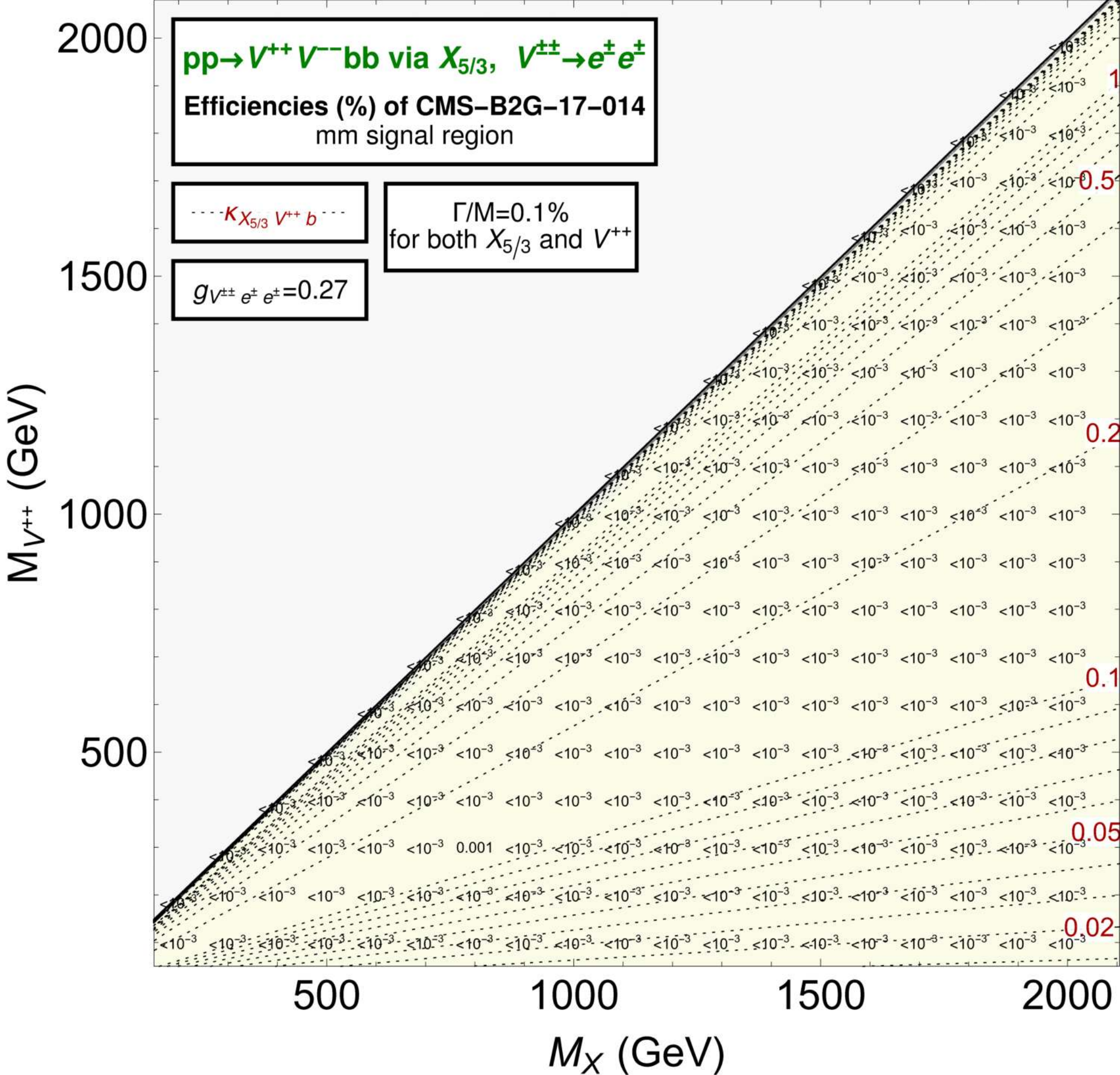}
  \includegraphics[width=.325\textwidth]{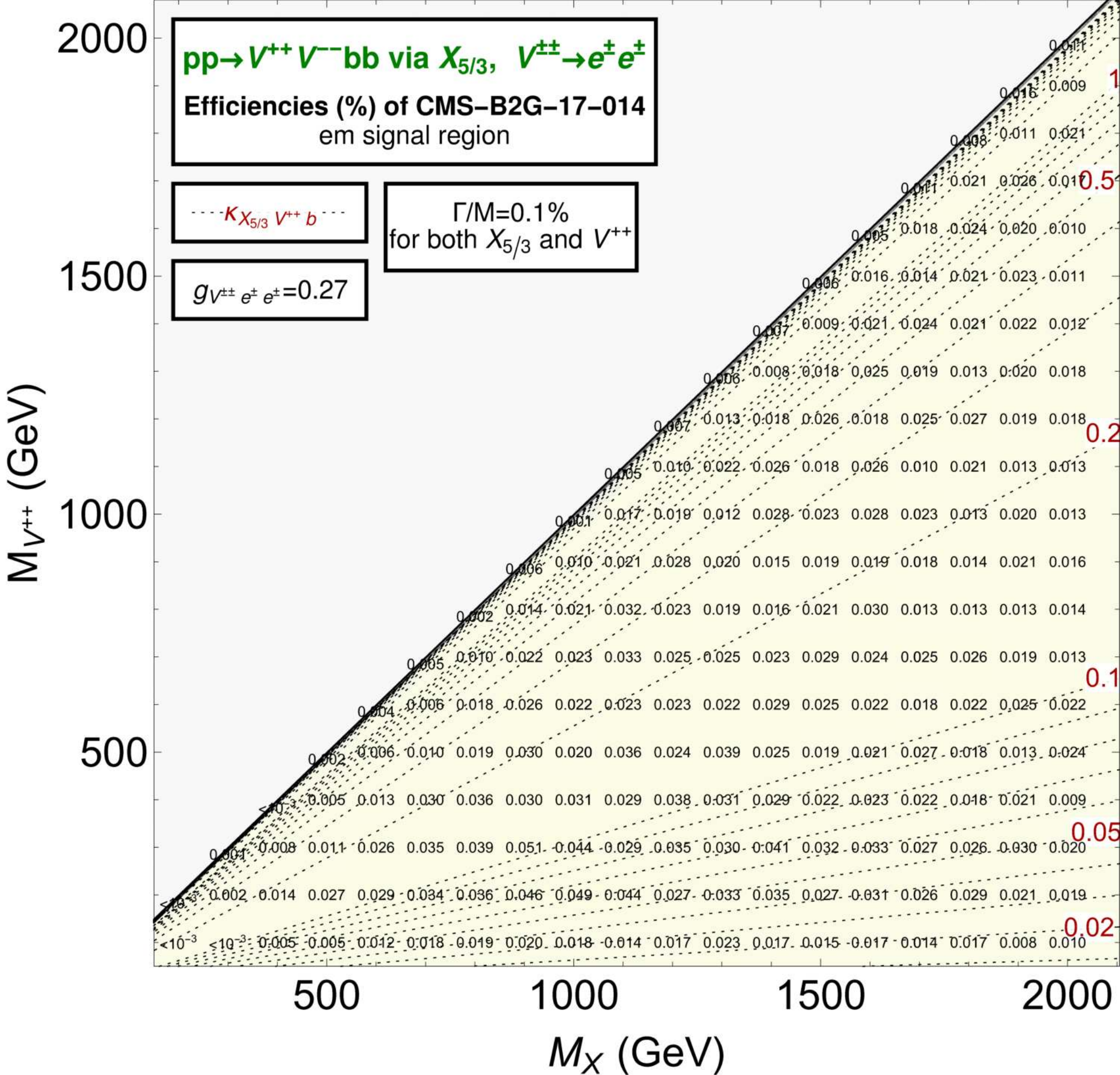}\\
  \includegraphics[width=.325\textwidth]{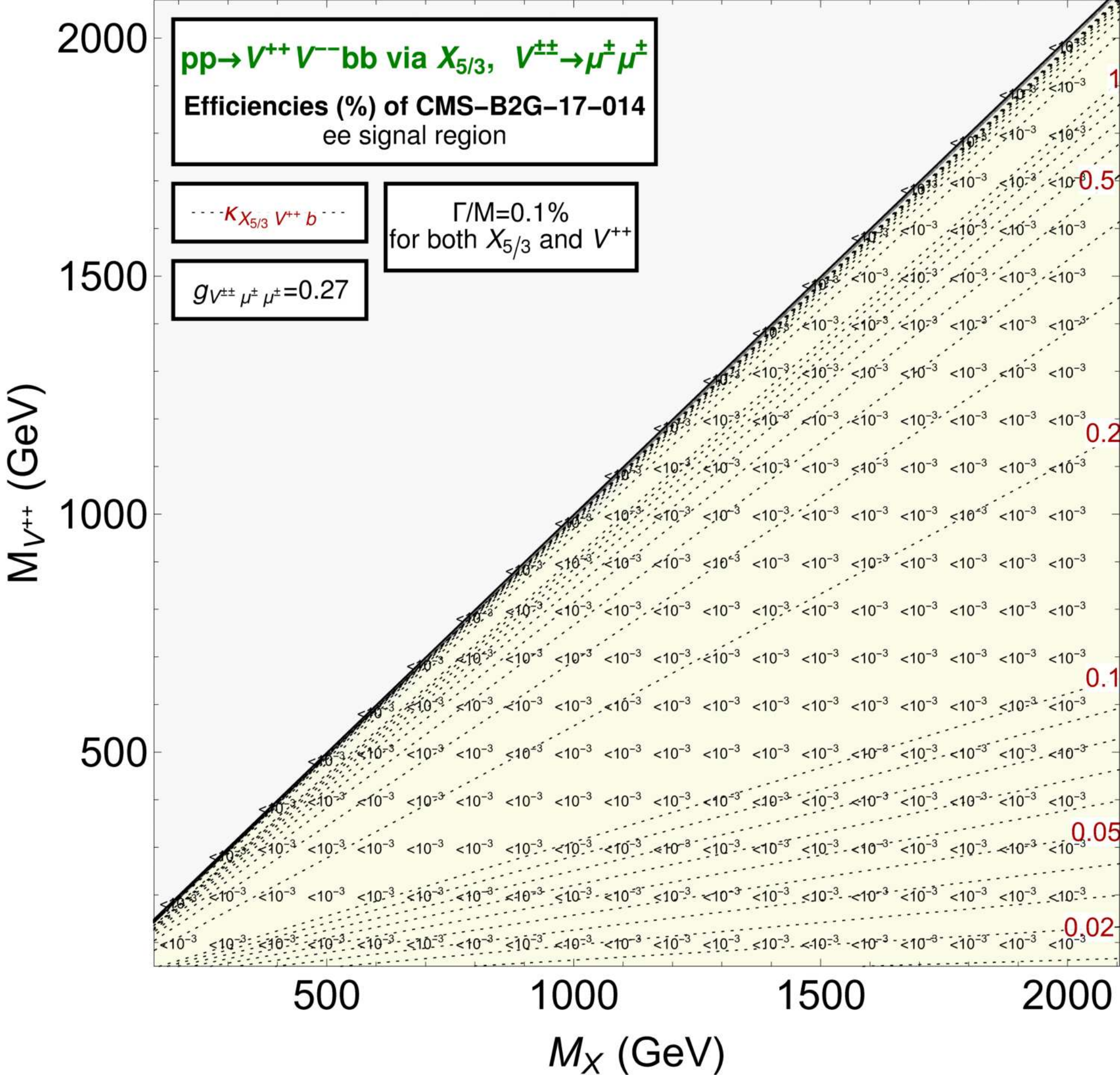}
  \includegraphics[width=.325\textwidth]{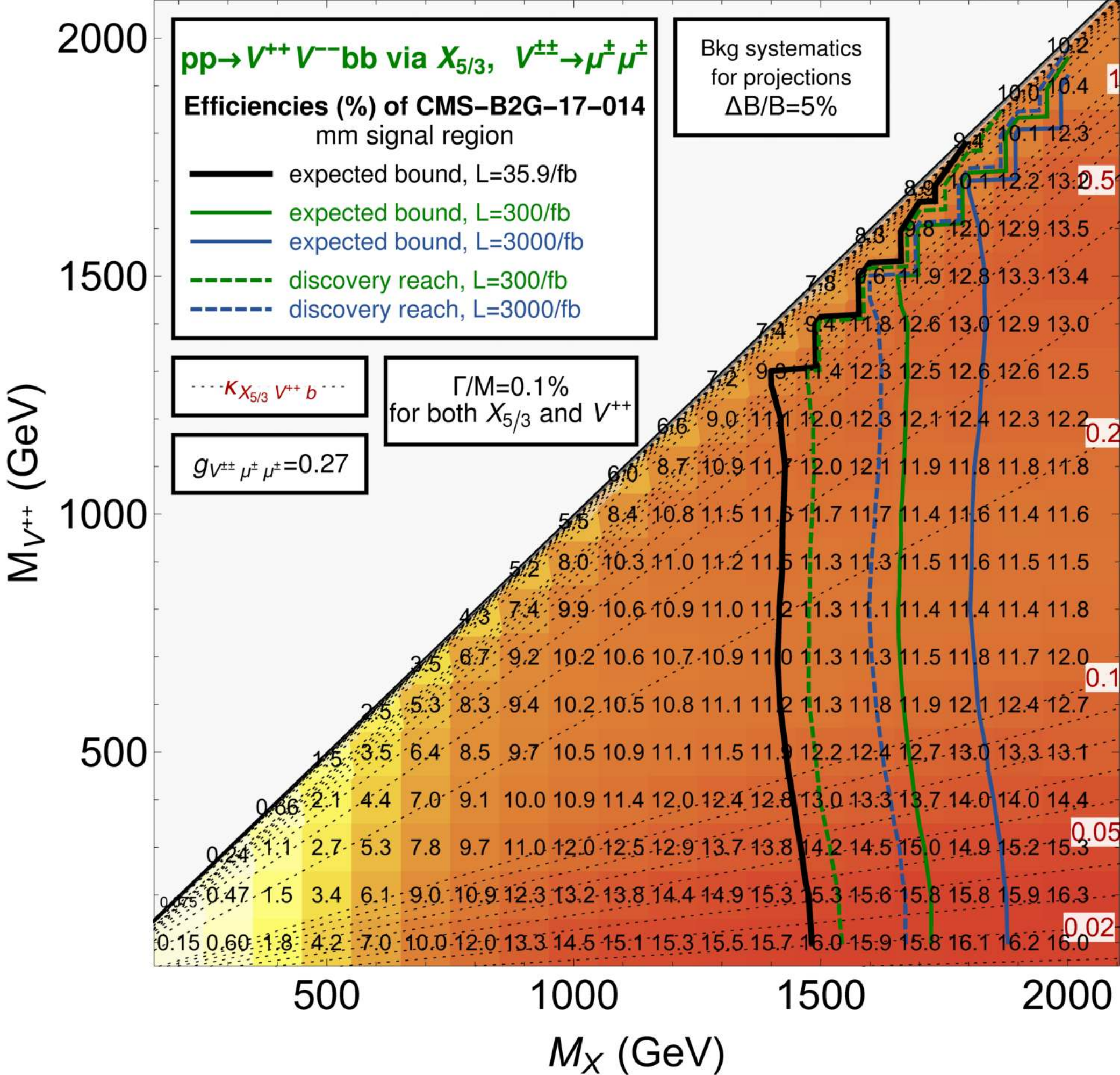}
  \includegraphics[width=.325\textwidth]{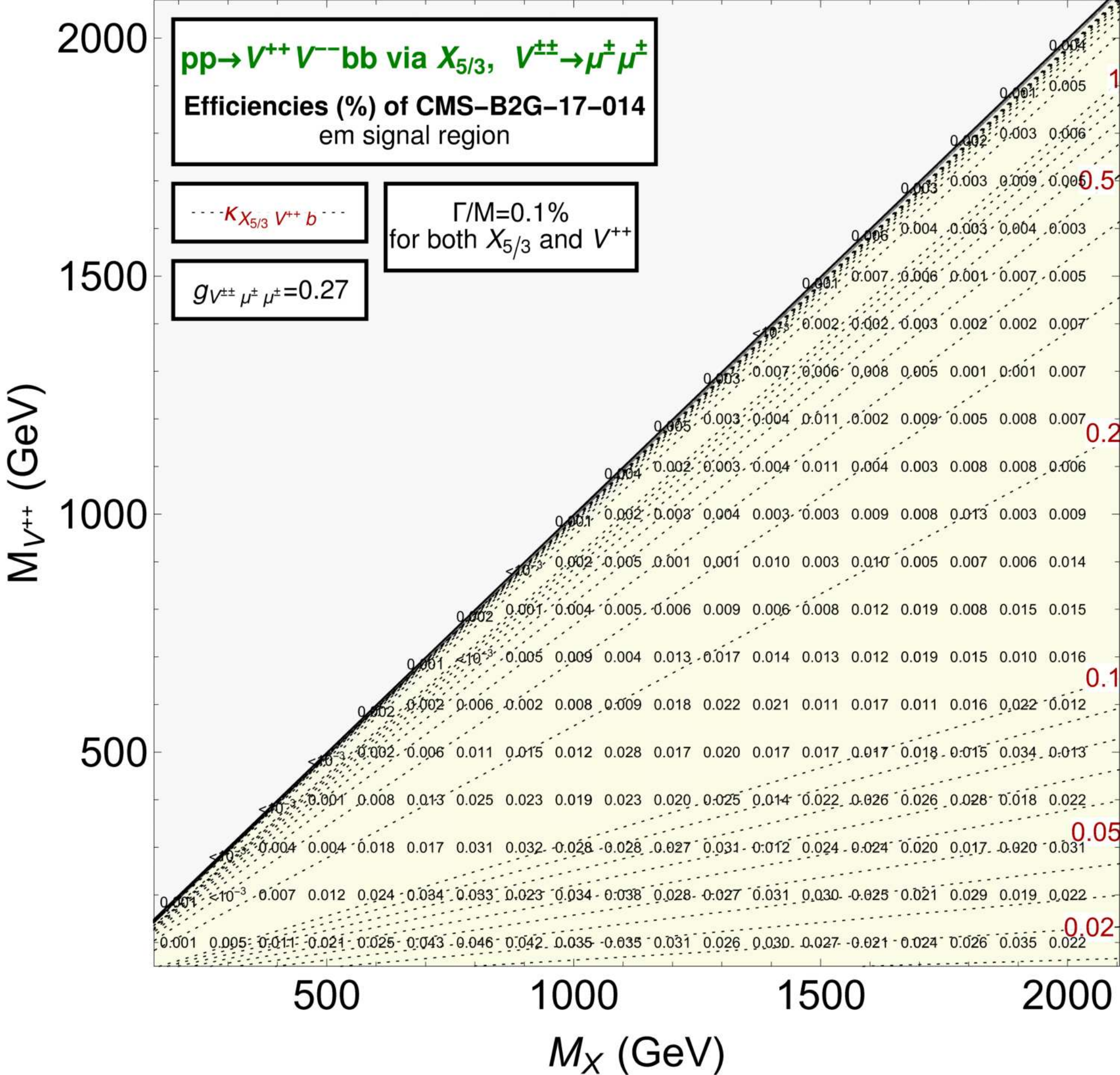}\\
  \includegraphics[width=.325\textwidth]{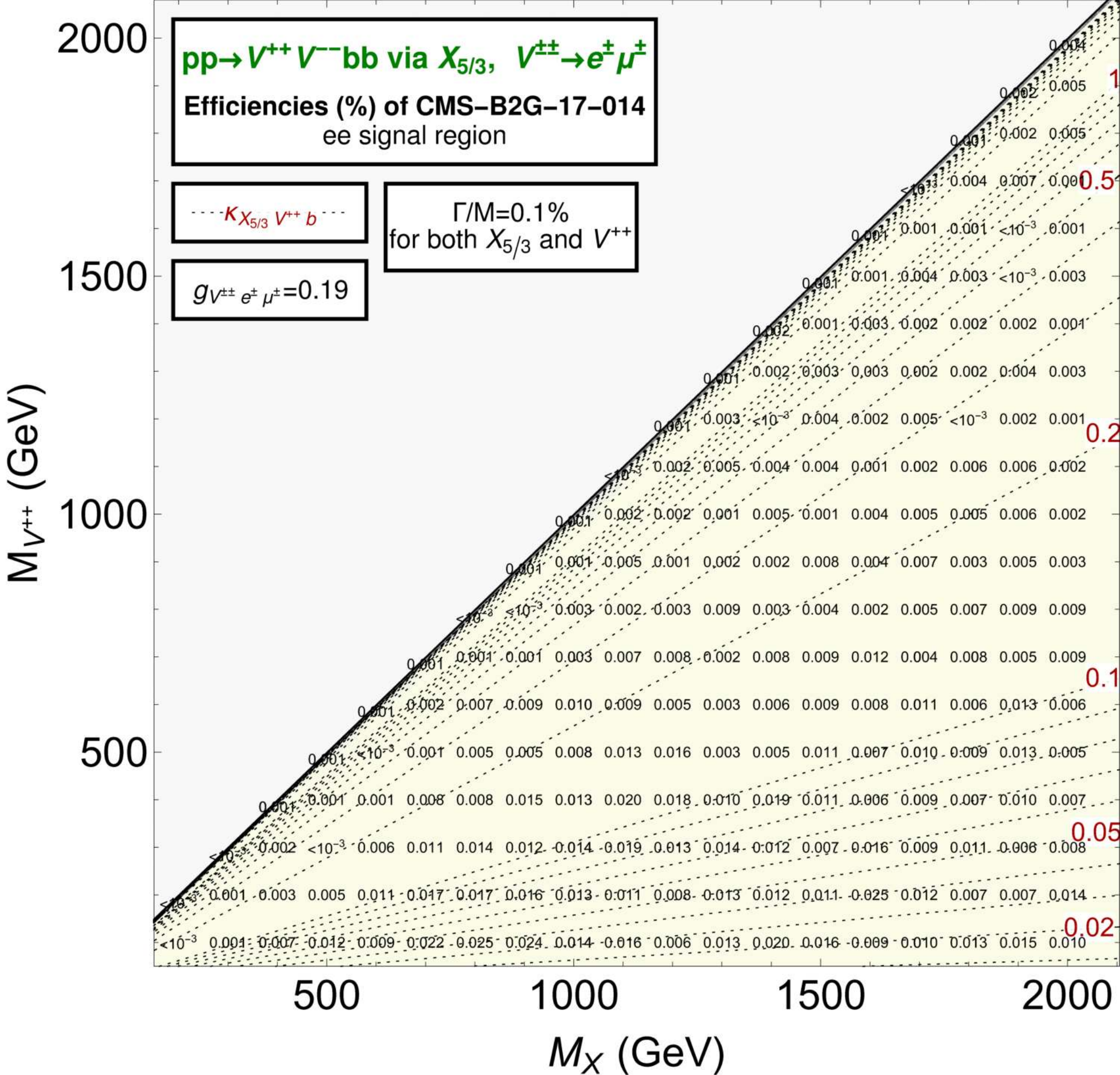}
  \includegraphics[width=.325\textwidth]{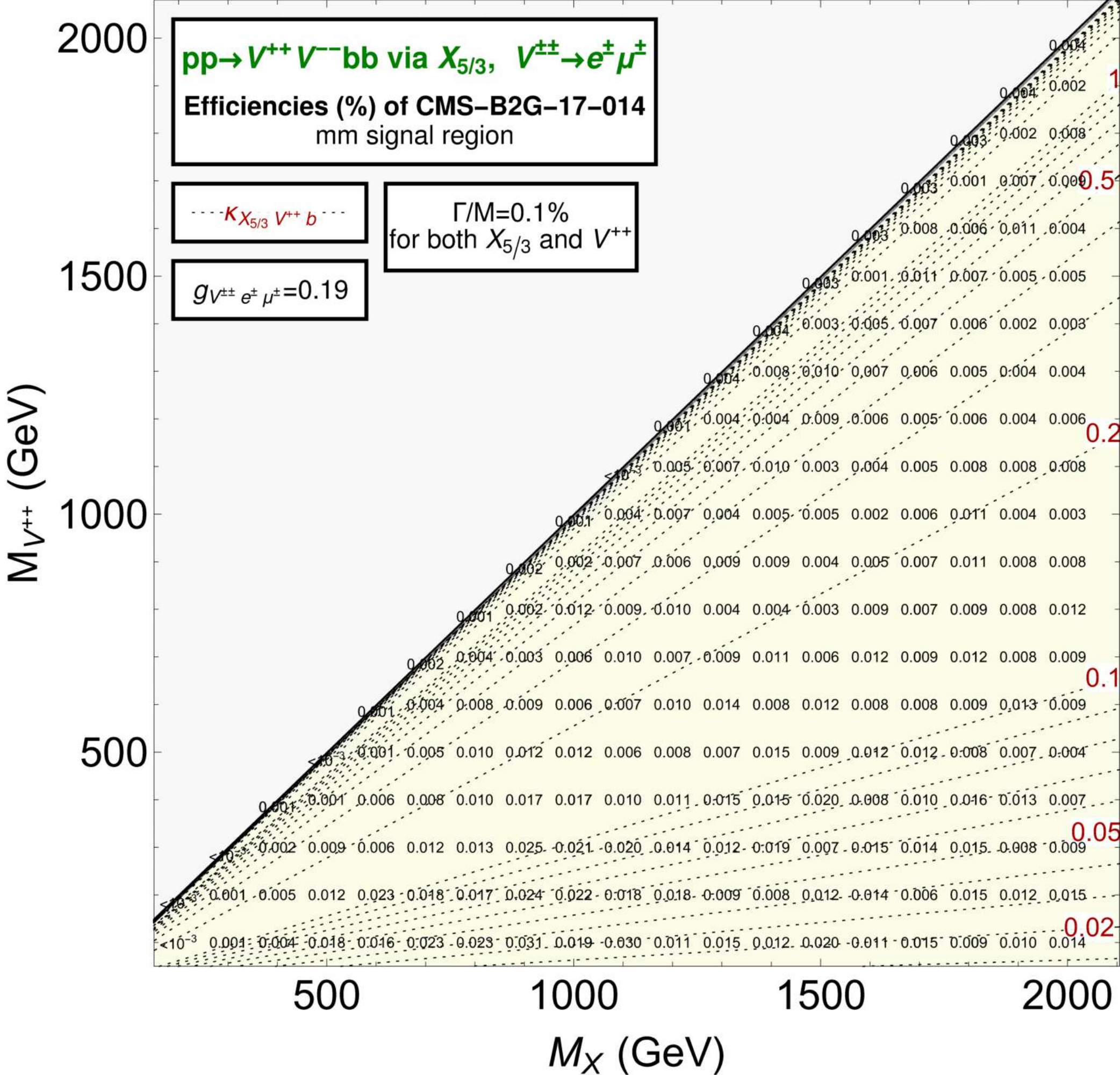}
  \includegraphics[width=.325\textwidth]{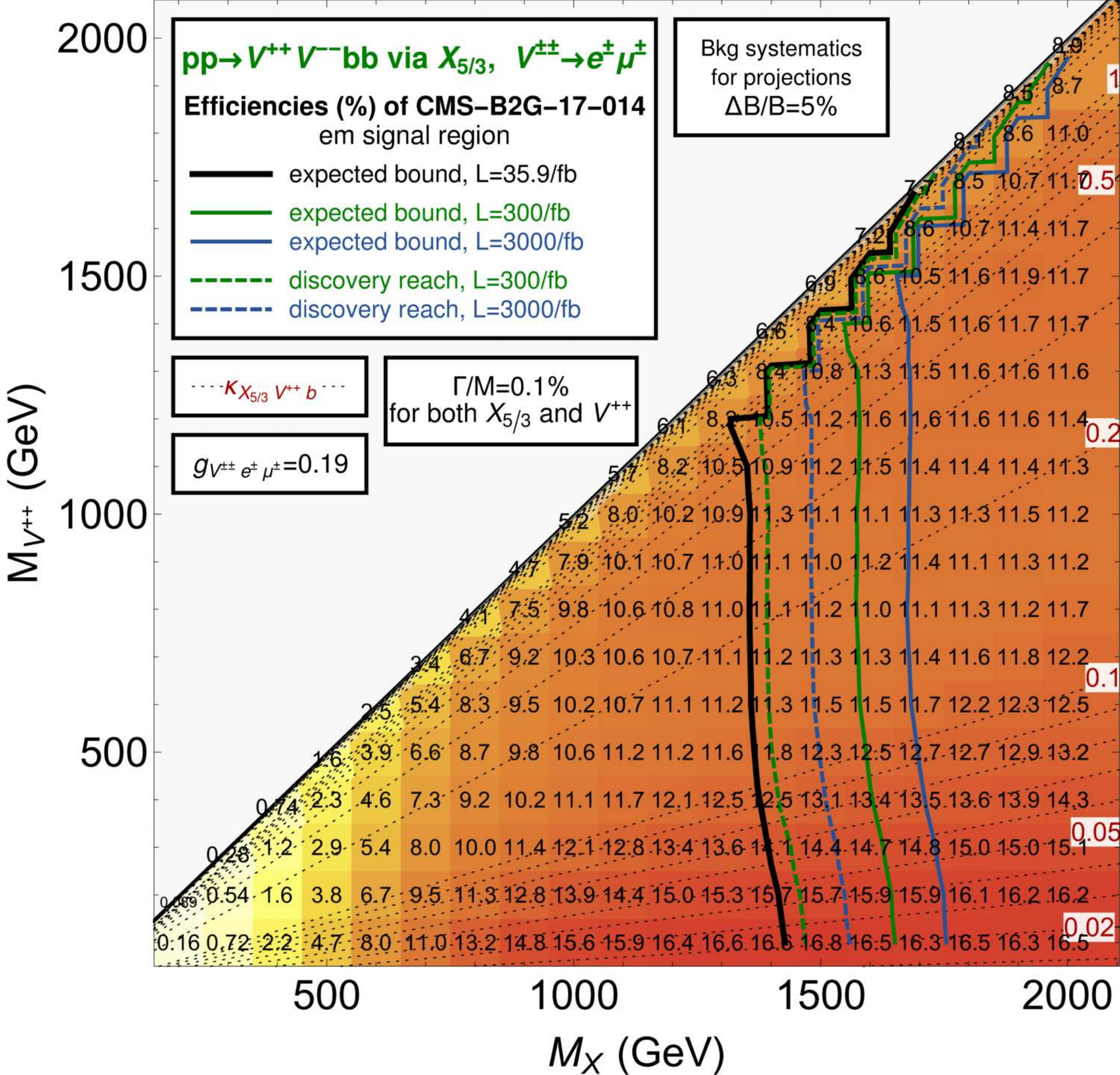}
  \caption{\label{fig:boundVPP} Efficiencies for scenarios in which $V^{++}$ decays to the conjugate final state of $V^{--}$: $V^{\pm\pm} \to e^\pm e^\pm$ {\bf (top row)}, $V^{\pm\pm} \to \mu^\pm \mu^\pm$ {\bf (central row)} and $V^{\pm\pm} \to e^\pm \mu^\pm$ {\bf (bottom row)} in the $ee$ {\bf (left column)}, $\mu\mu$ {\bf (central column)} and $e\mu$ SRs {\bf (right column)} of~\cite{Sirunyan:2018yun}. The meaning of curves, colours and symbols is the same as in Figure~\ref{fig:boundseSRee}. Due to the small efficiencies in SRs not targeting the leptons with same sign arising from the different signal hypotheses, the bounds and projections corresponding to 100\% decays of the vector bosons are shown only for the most sensitive SRs.}
\end{figure}

Figure~\ref{fig:boundVPP} is the counterpart of Figure~\ref{fig:boundVP}, but
for doubly charged vectors: one has the three CMS SRs and both $V^{++}$ and $V^{--}$ decay into leptons with the same flavour  with $100\%$ saturated $BR$s.
Only for top-left, central-central and bottom-right panels the same sign leptons coming from the vector $V$ match the different SRs, giving rise to relevant phenomenological information.
 
\begin{figure}[thbp]
  \centering
  \includegraphics[width=.325\textwidth]{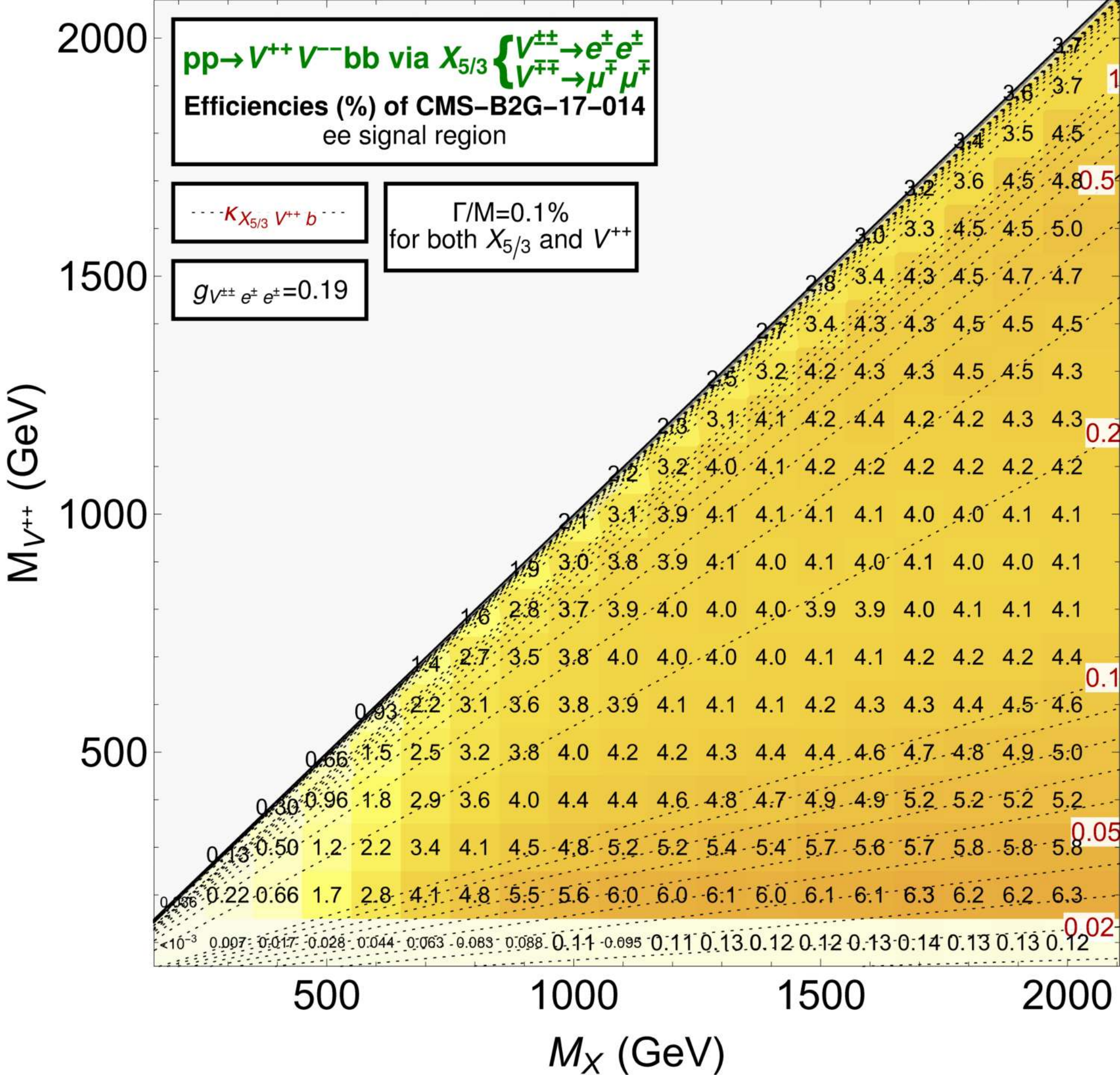}
  \includegraphics[width=.325\textwidth]{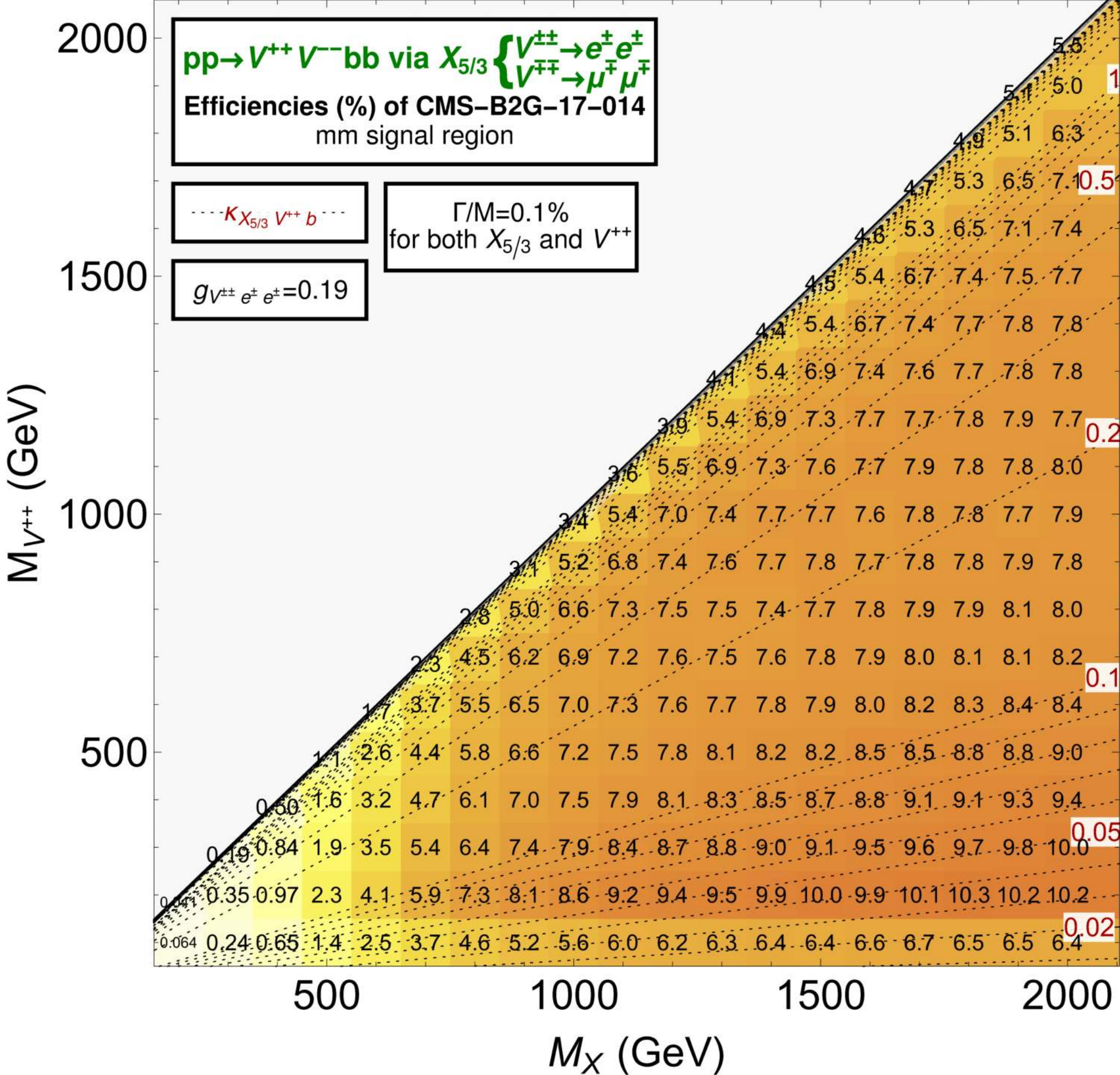}
  \includegraphics[width=.325\textwidth]{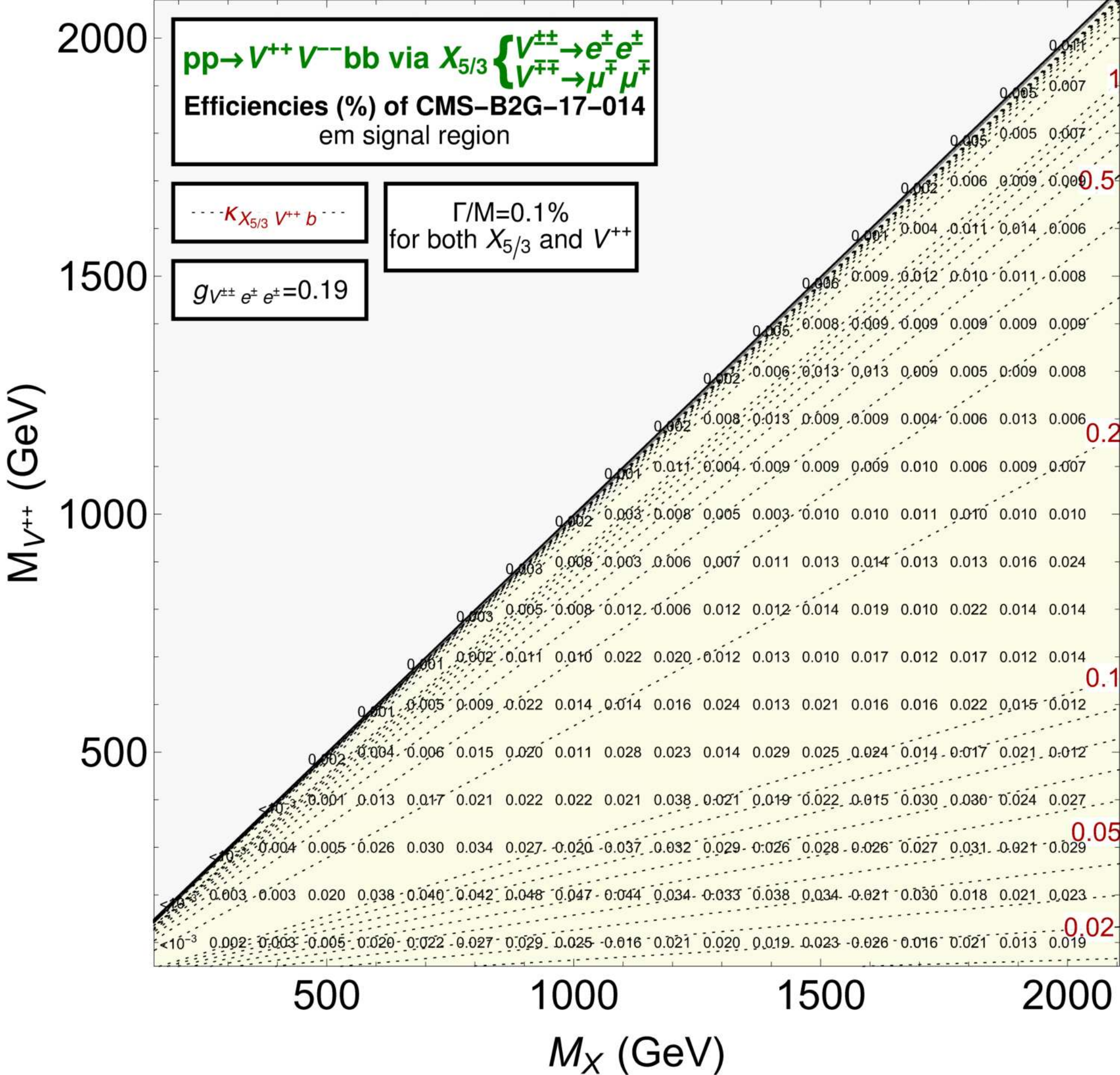}\\
  \includegraphics[width=.325\textwidth]{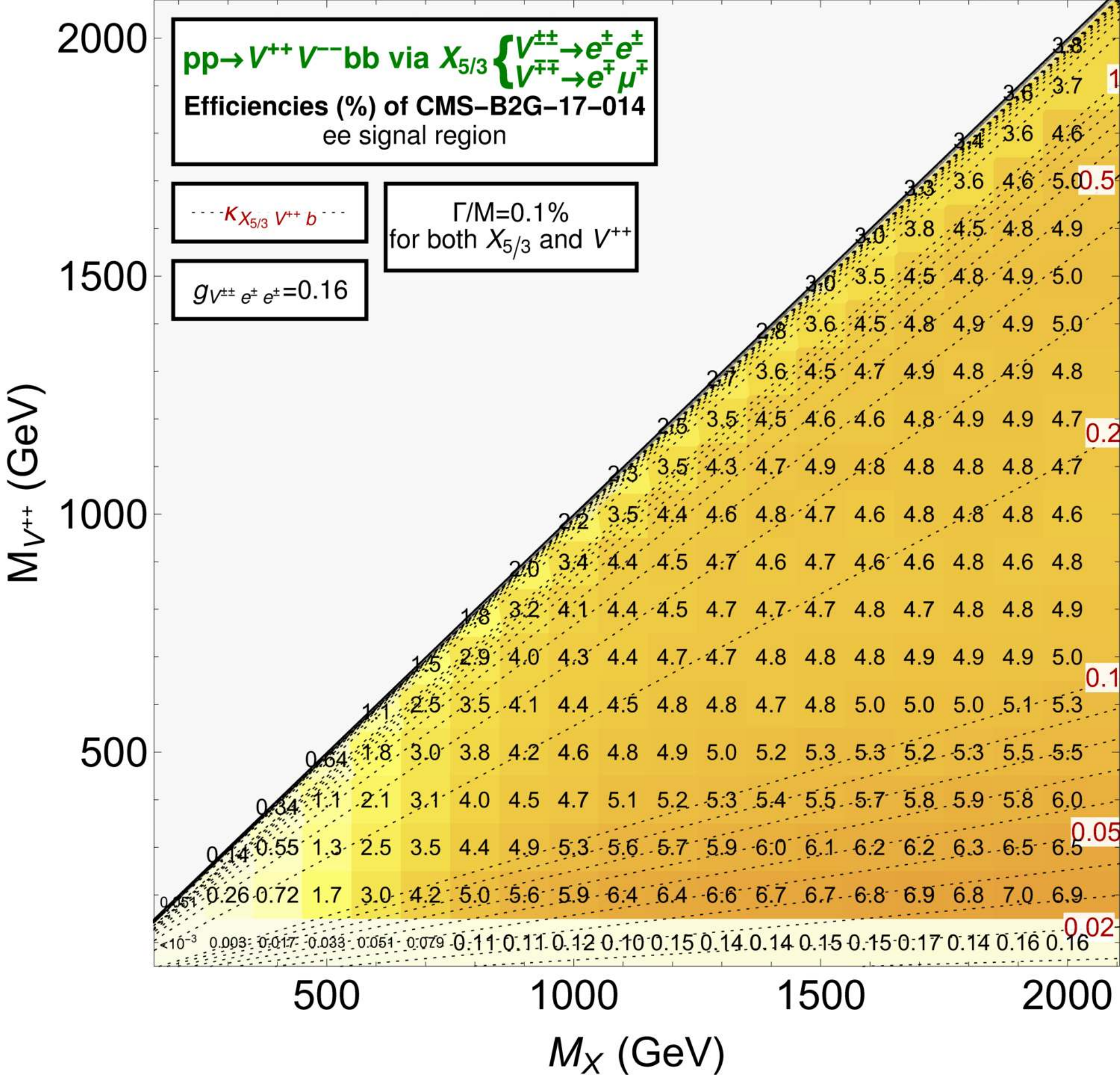}
  \includegraphics[width=.325\textwidth]{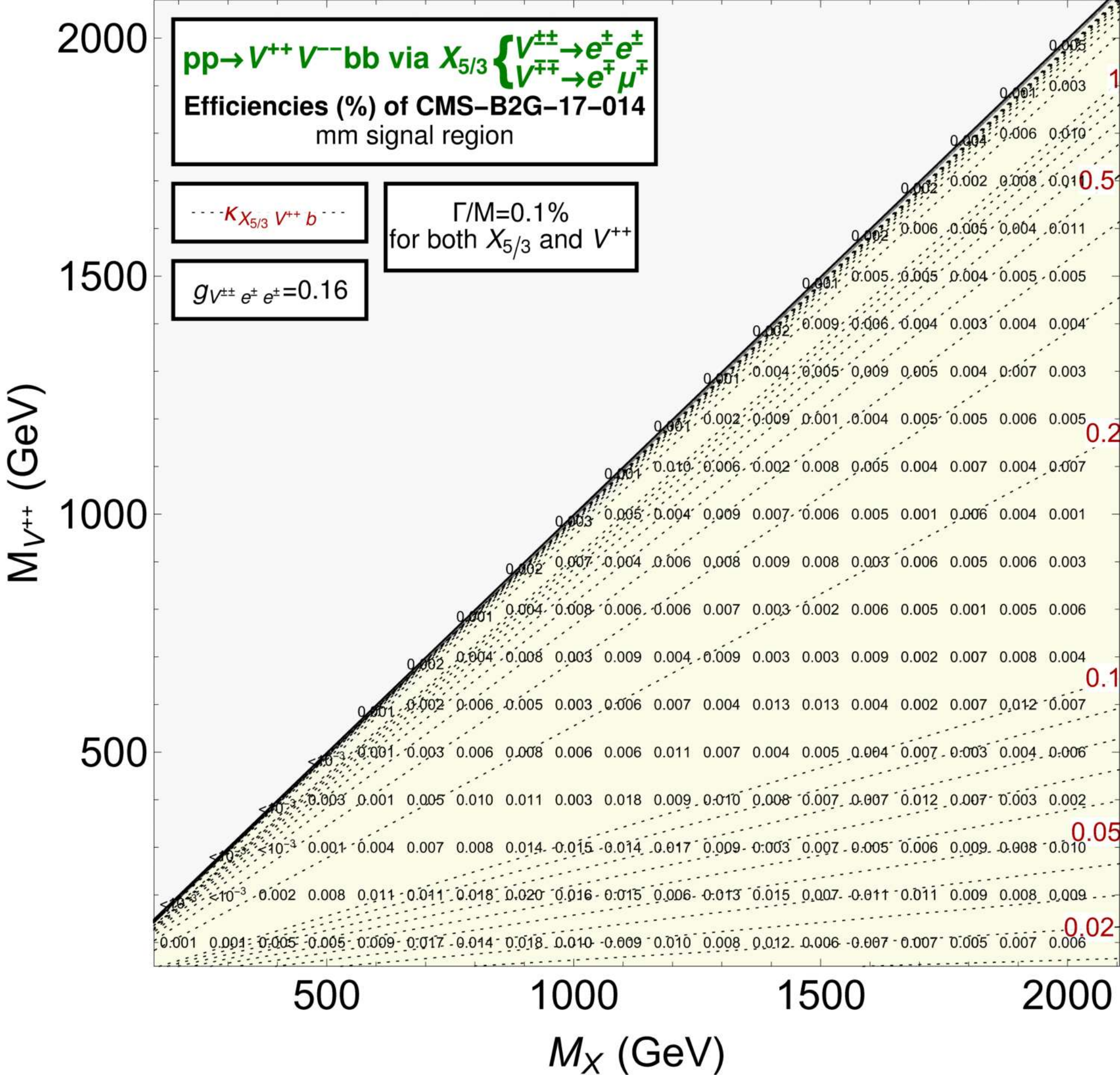}
  \includegraphics[width=.325\textwidth]{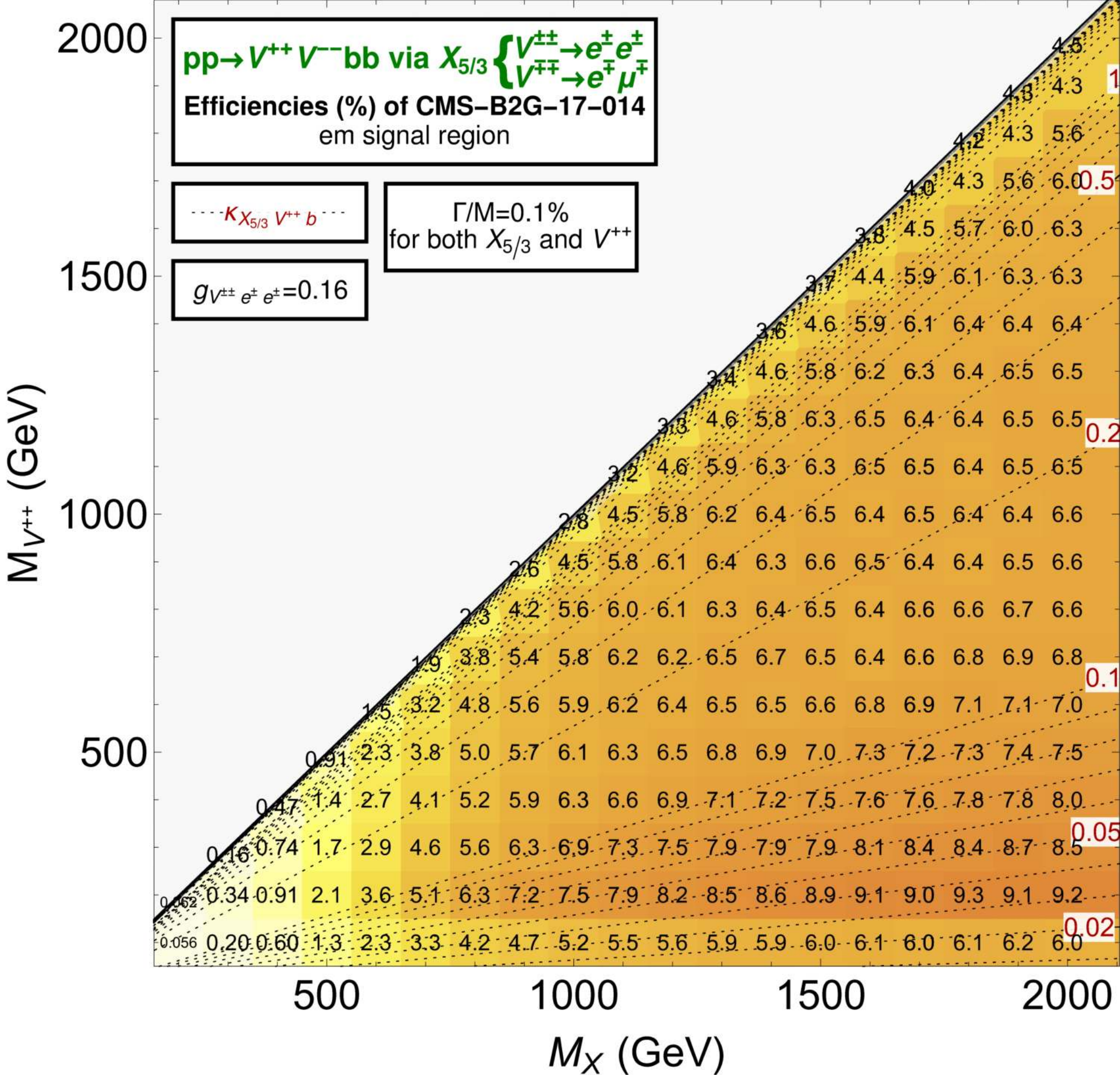}\\
  \includegraphics[width=.325\textwidth]{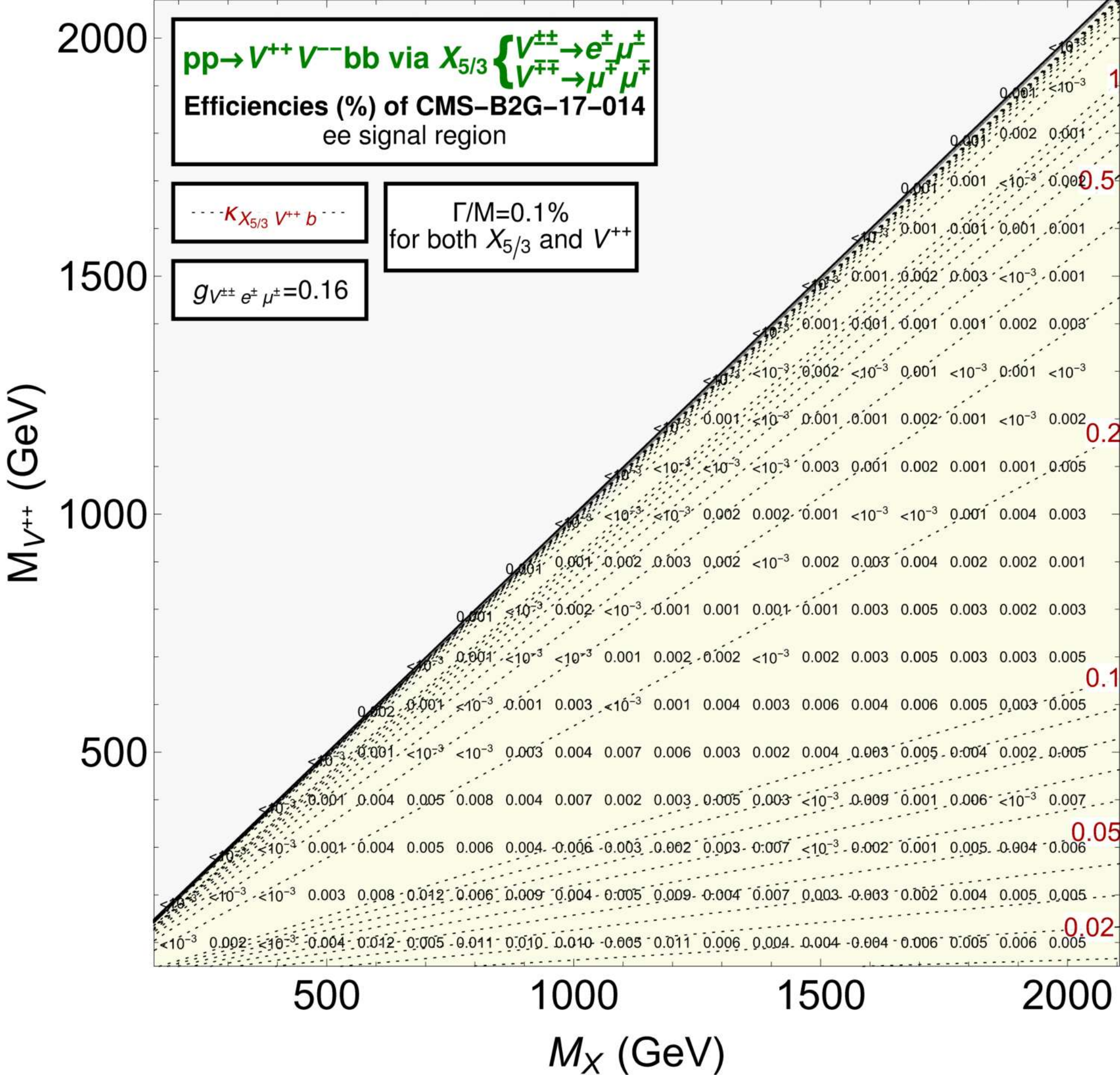}
  \includegraphics[width=.325\textwidth]{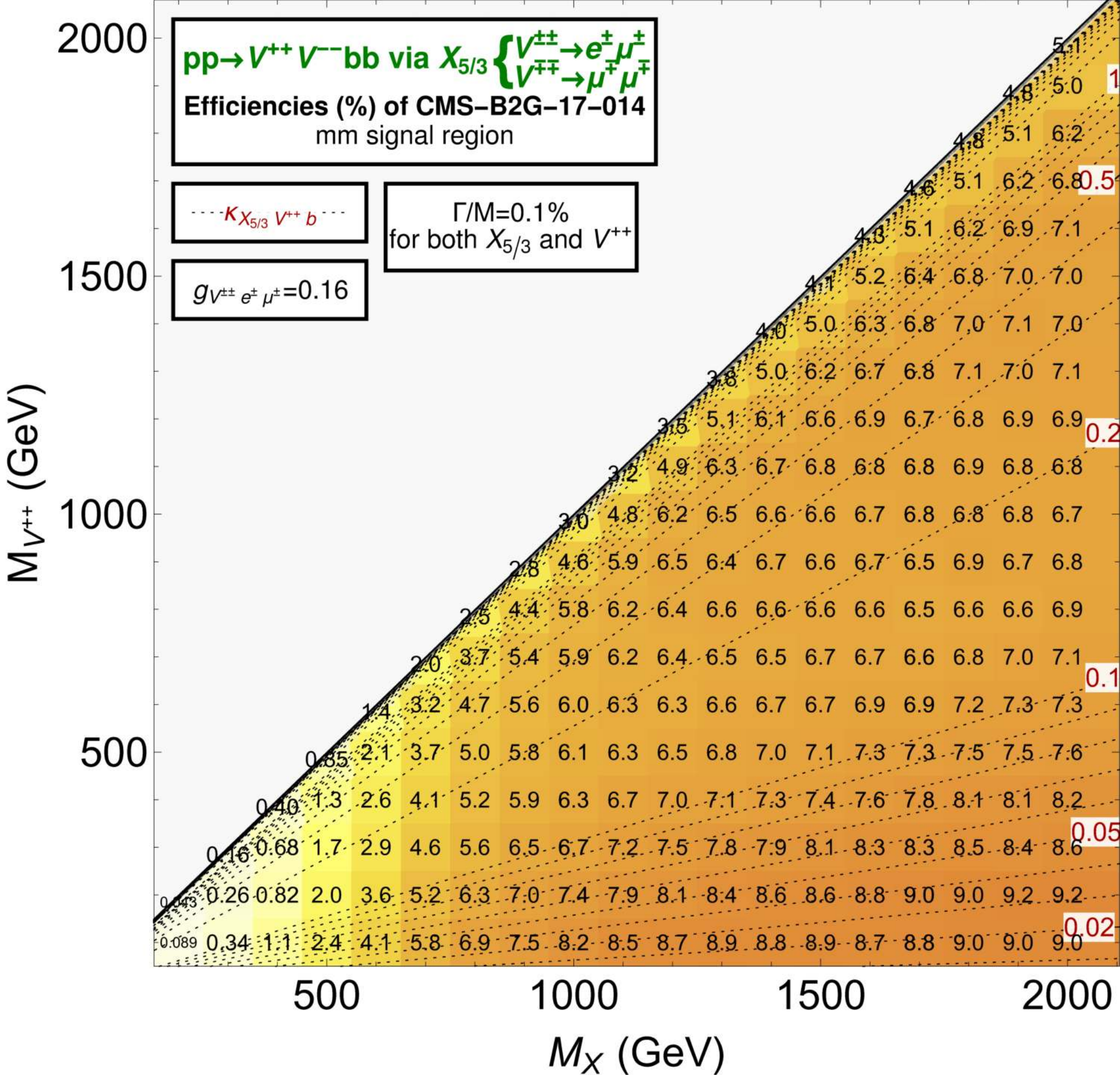}
  \includegraphics[width=.325\textwidth]{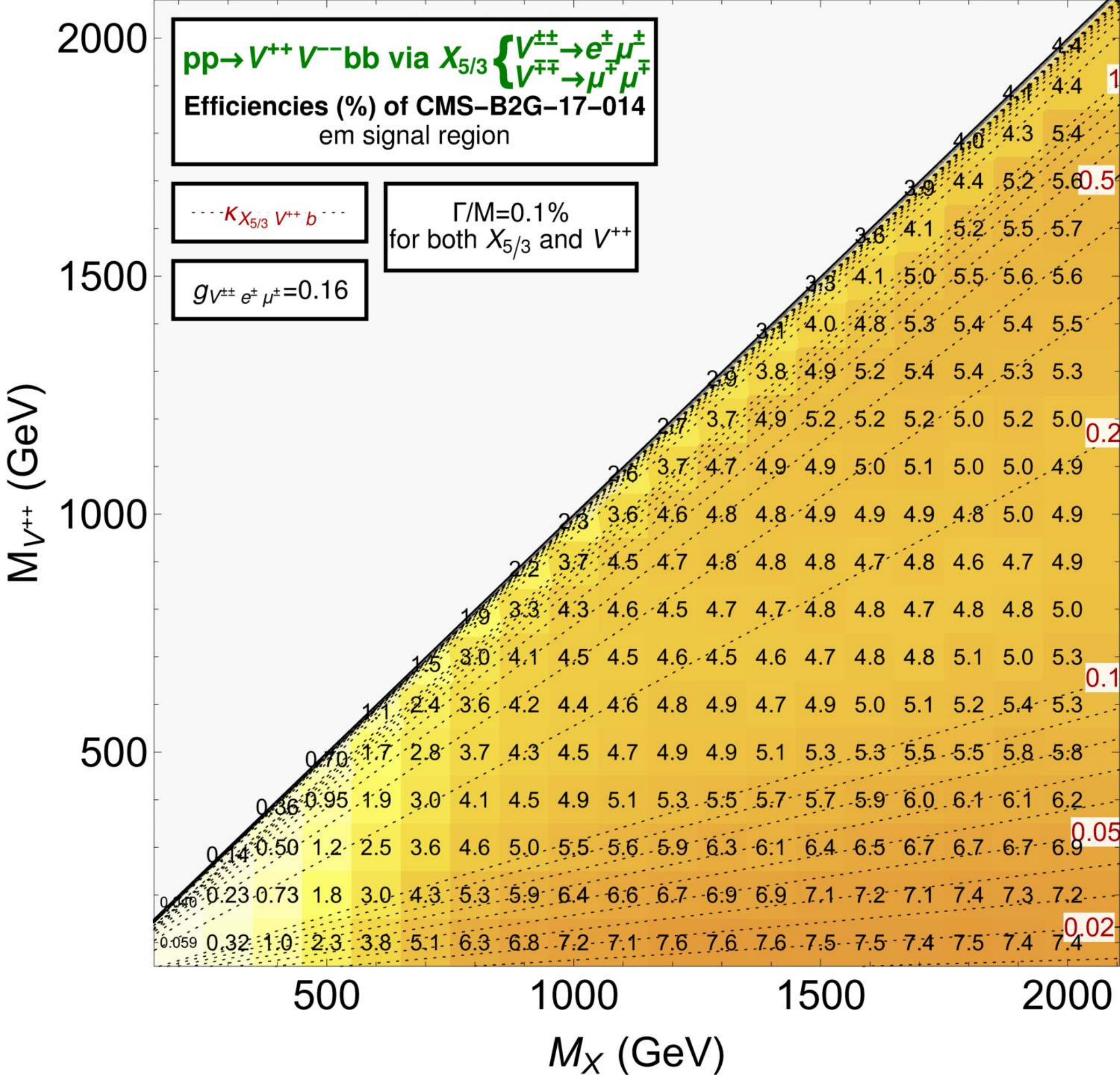}
  \caption{\label{fig:boundVPP2} Efficiencies for scenarios in which $V^{++}$ decays to a different final state than $V^{--}$. The meaning of colours and symbols is the same as in Figure~\ref{fig:boundseSRee}. These scenarios are unphysical if considered individually, thus no bounds or projections are shown.}
\end{figure}

Likewise, Figure~\ref{fig:boundVPP2} is analogous to Figure~\ref{fig:boundVPP},
but assumes that $V^{++}$ and $V^{--}$ decay into leptons with different flavour,
e.g. $V^{++}\to e^+e^+$ and $V^{--}\to \mu^-\mu^-$.

\subsection{Pair production of $X_{5/3}$ where one branch decays into $V^{\pm\pm}$ and the other into $V^\mp$ with same mass}

\begin{figure}[thbp]
  \centering
  \includegraphics[width=.325\textwidth]{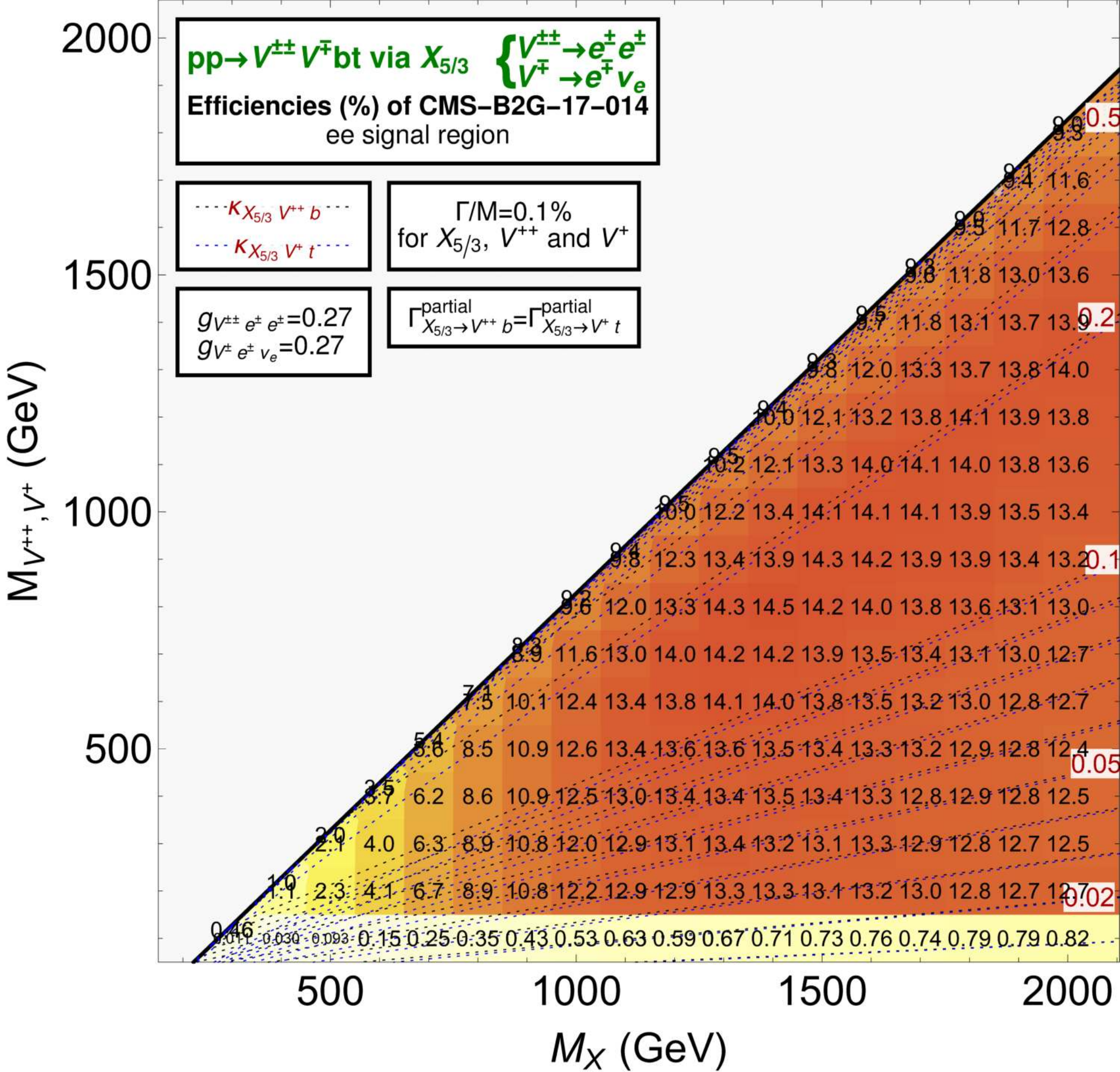}
  \includegraphics[width=.325\textwidth]{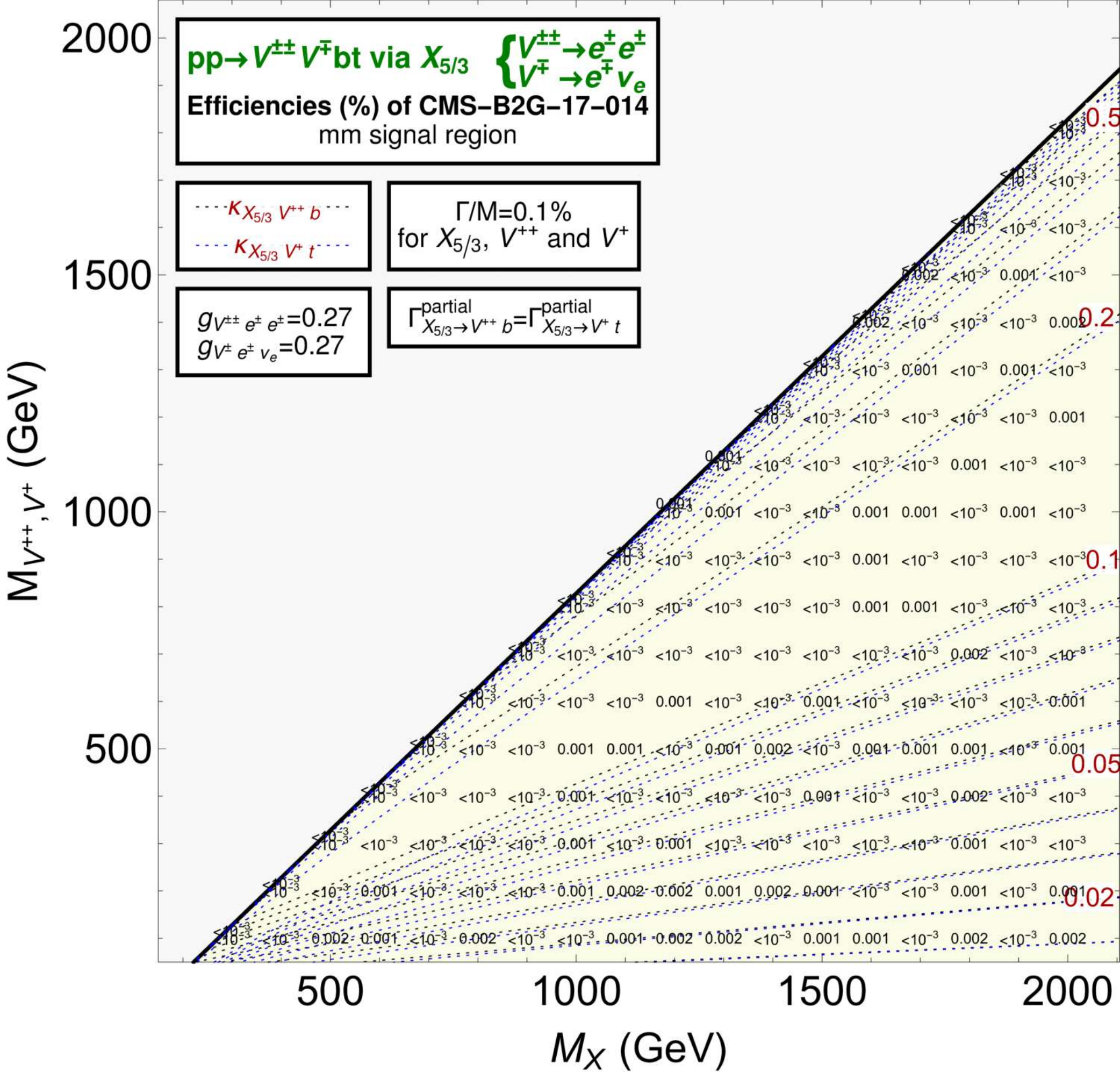}
  \includegraphics[width=.325\textwidth]{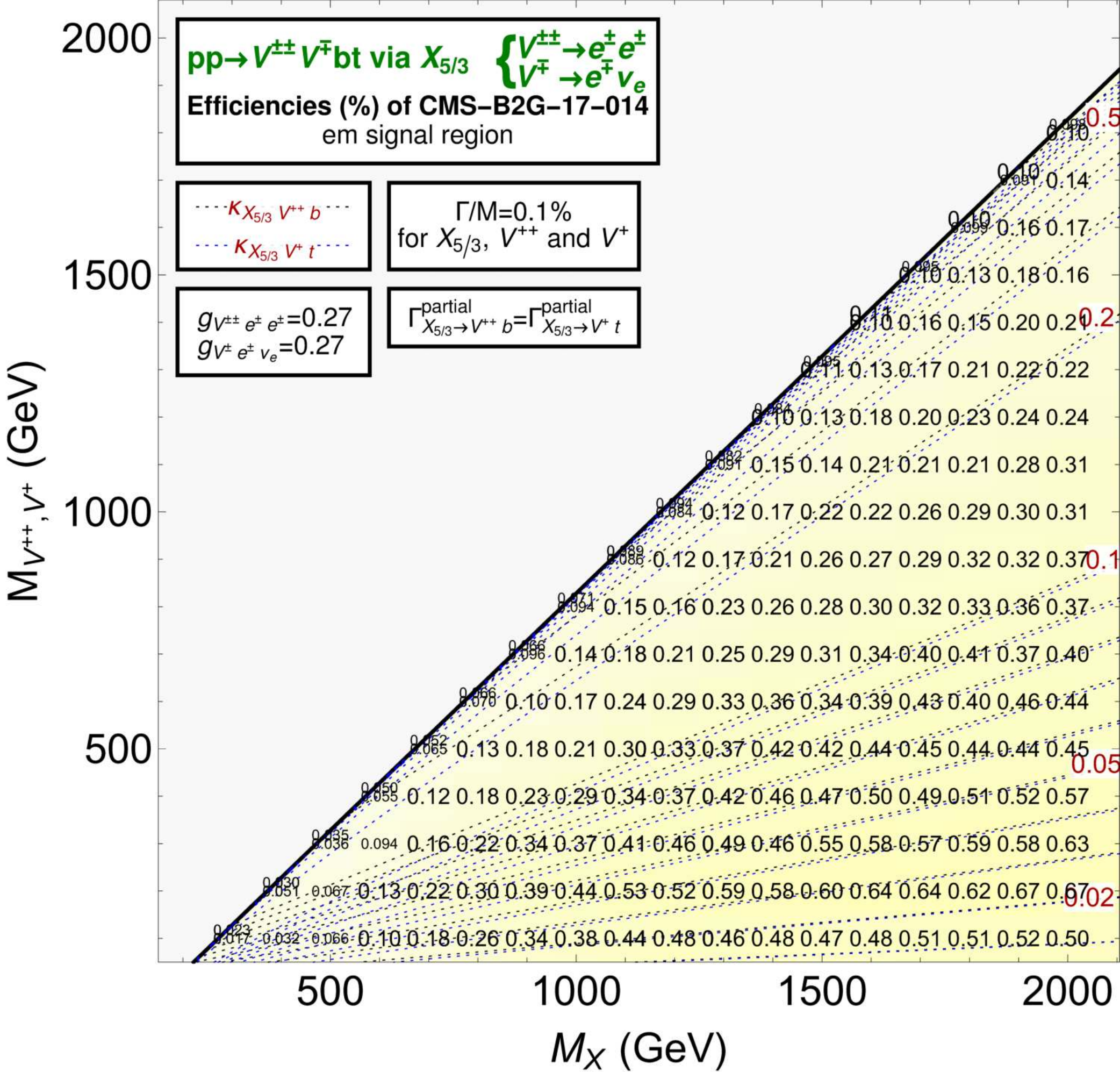}\\
  \includegraphics[width=.325\textwidth]{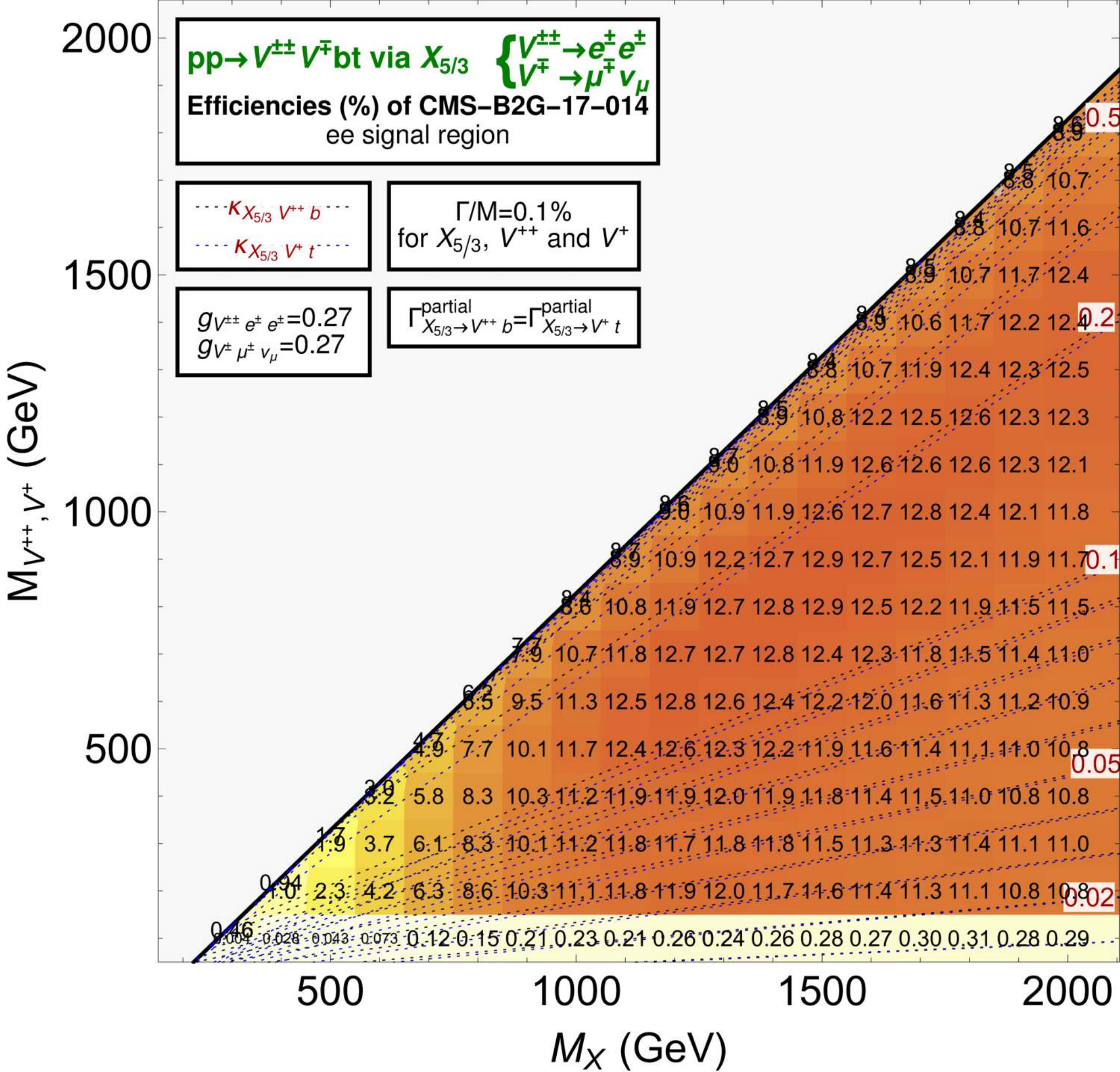}
  \includegraphics[width=.325\textwidth]{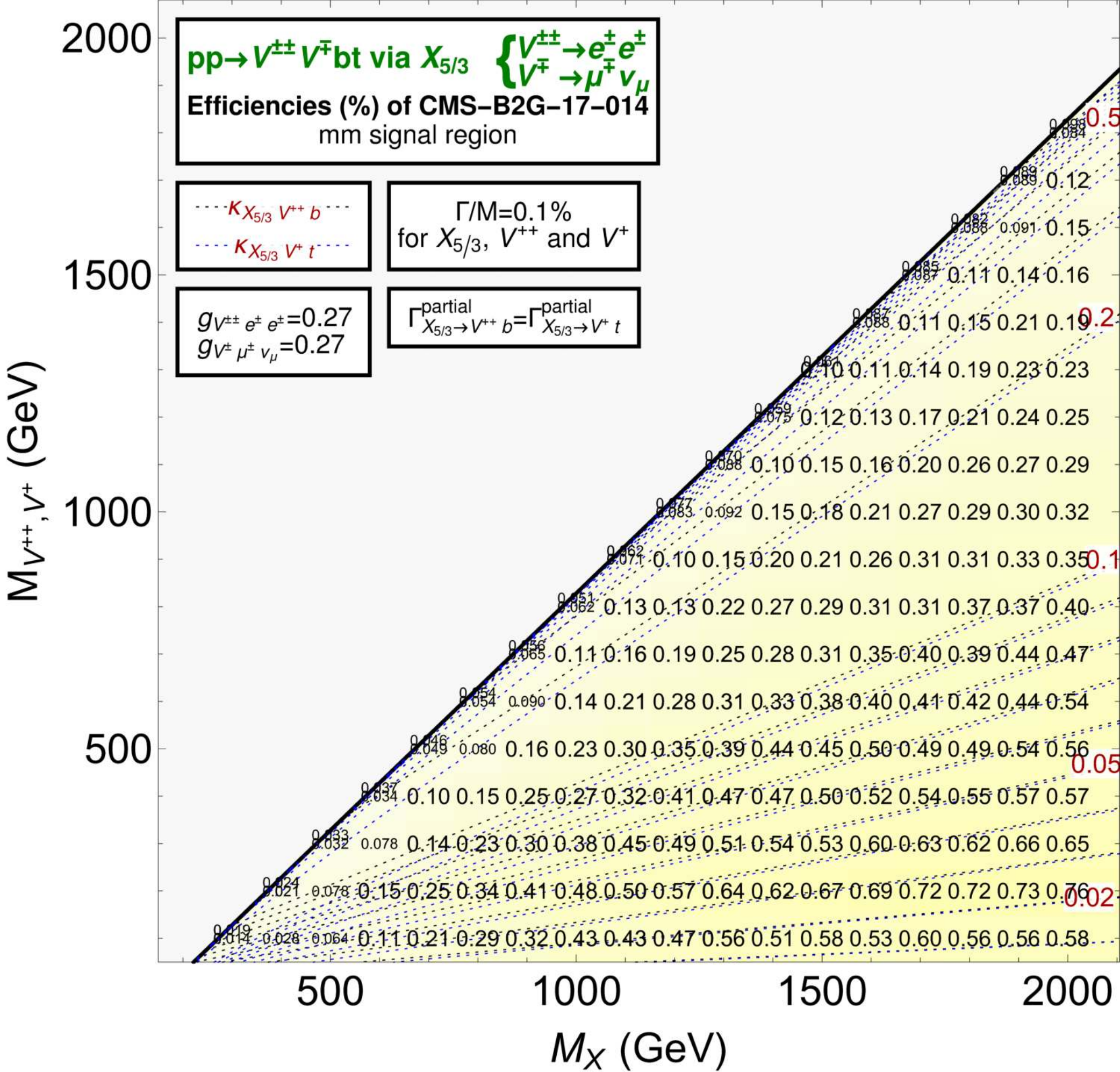}
  \includegraphics[width=.325\textwidth]{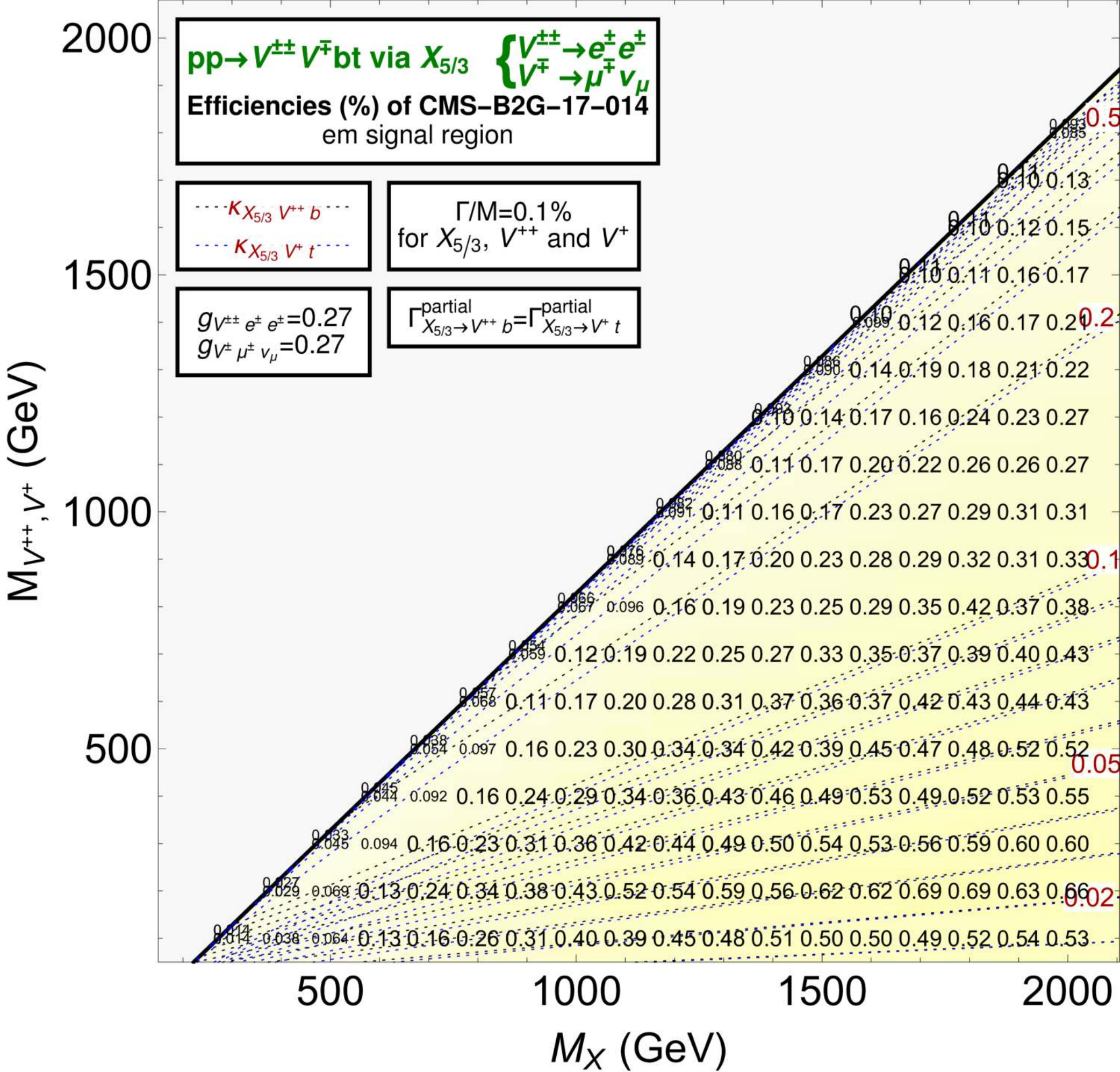}\\
  \caption{\label{fig:XVPP_XVP_ee} Efficiencies for scenarios in which $V^{\pm\pm}$ decays to $e^\pm e^\pm$ and $V^\mp$ decays to $e^\mp \nu_e$ {\bf (top row)} or to $\mu^\mp \nu_\mu$ {\bf (bottom row)} in the $ee$ {\bf (left column)}, $\mu\mu$ {\bf (central column)} and $e\mu$ SRs {\bf (right column)} of~\cite{Sirunyan:2018yun}. The meaning of colours and symbols is the same as in Figure~\ref{fig:boundseSRee}. The different curves corresponding to the couplings of the $X_{5/3}$ to $V^{++}$ and $V^+$ are represented in black and blue dotted lines respectively, and labelled together as curves with the same coupling value are very close in the plane. These scenarios are unphysical if considered individually, thus no bounds or projections are shown.}
\end{figure}

\begin{figure}[thbp]
  \centering
  \includegraphics[width=.325\textwidth]{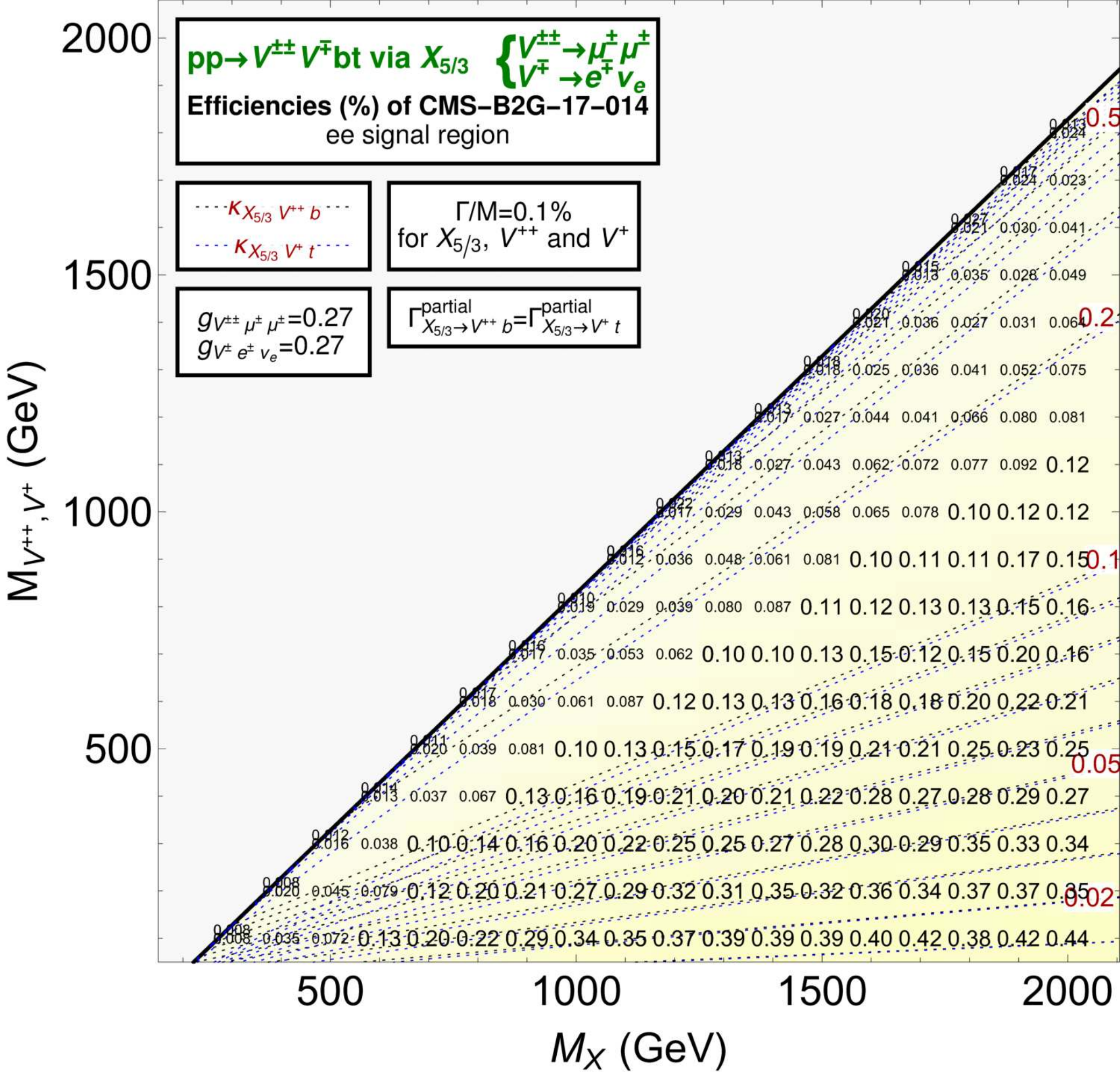}
  \includegraphics[width=.325\textwidth]{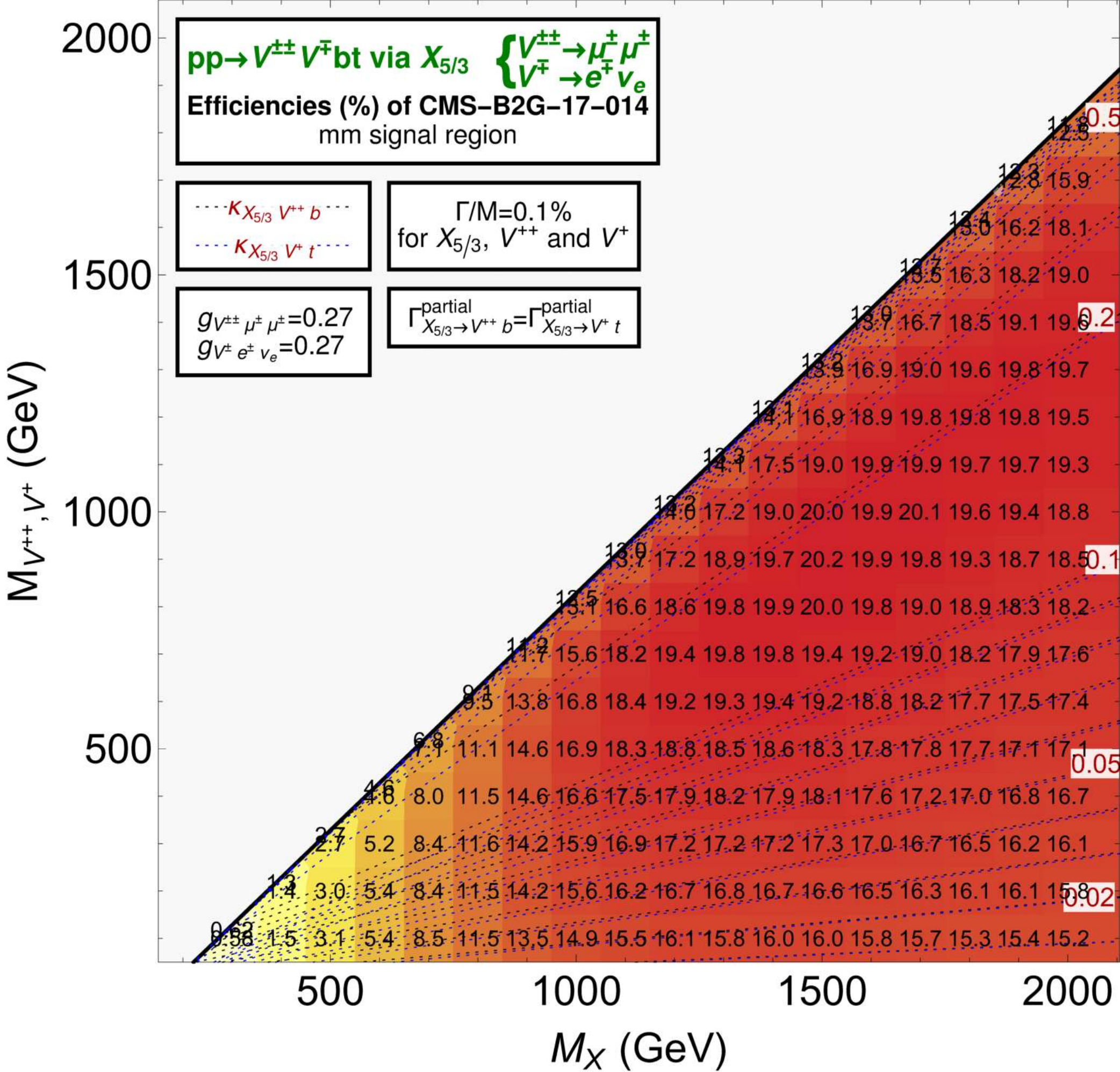}
  \includegraphics[width=.325\textwidth]{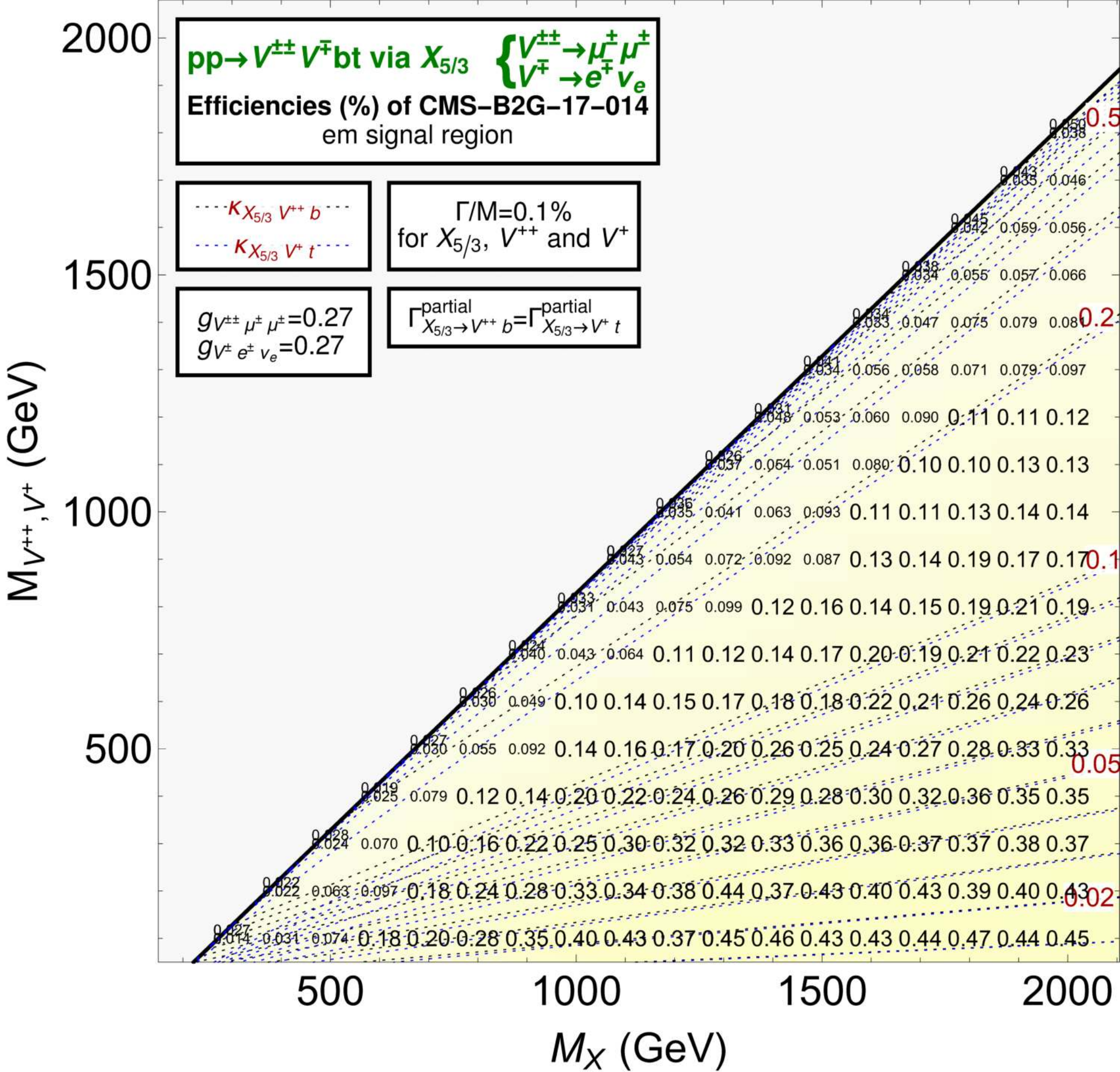}\\
  \includegraphics[width=.325\textwidth]{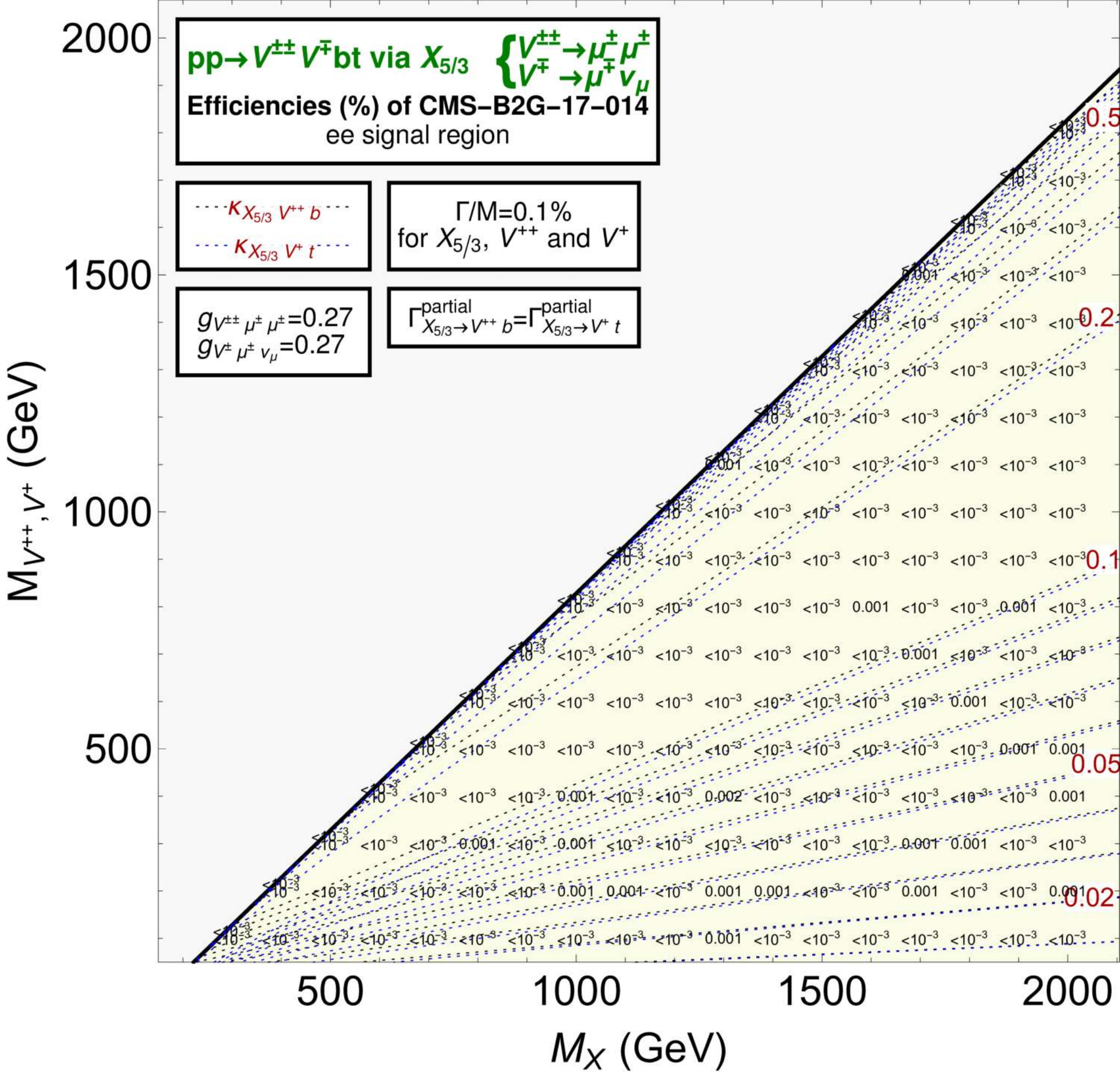}
  \includegraphics[width=.325\textwidth]{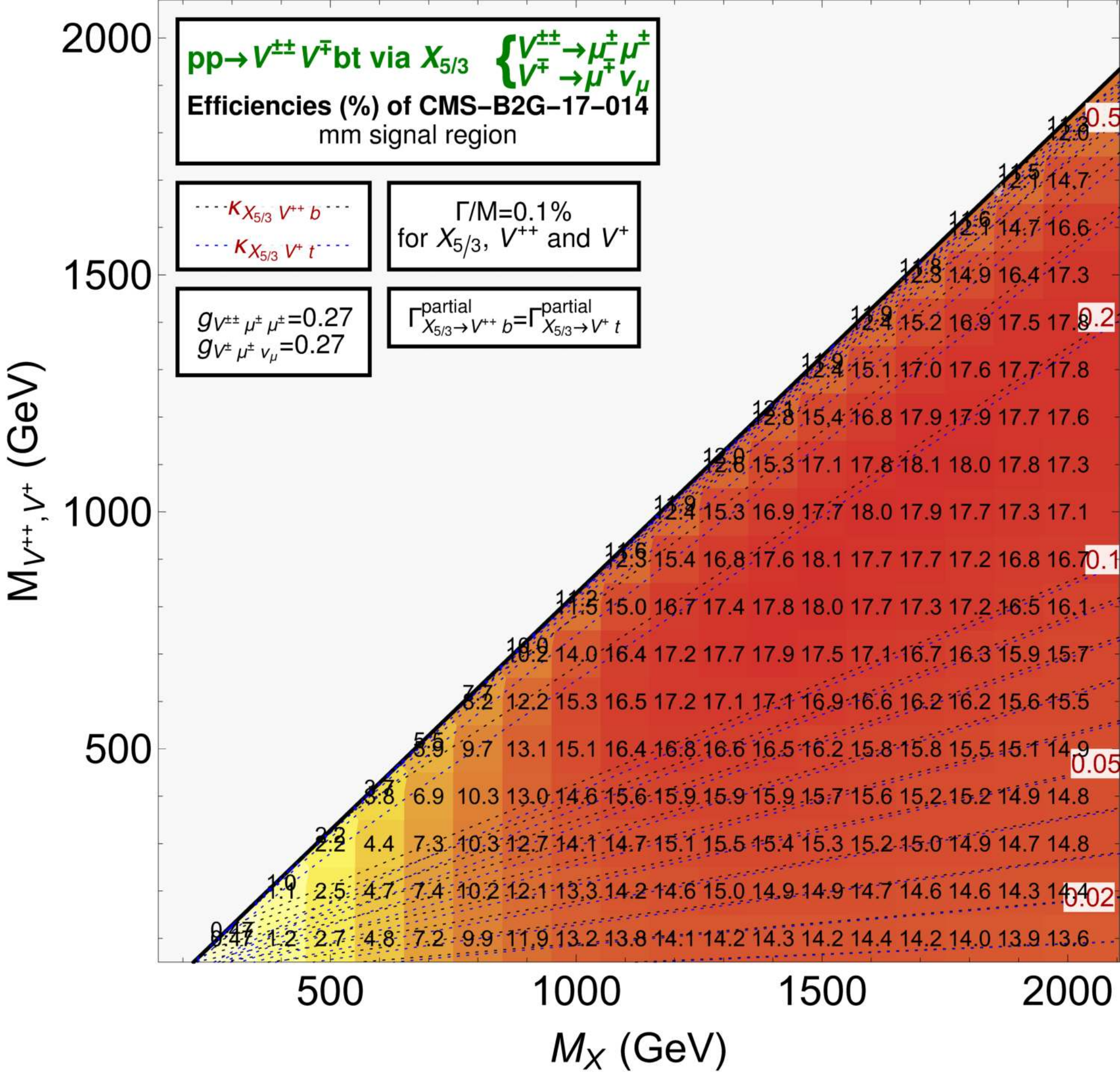}
  \includegraphics[width=.325\textwidth]{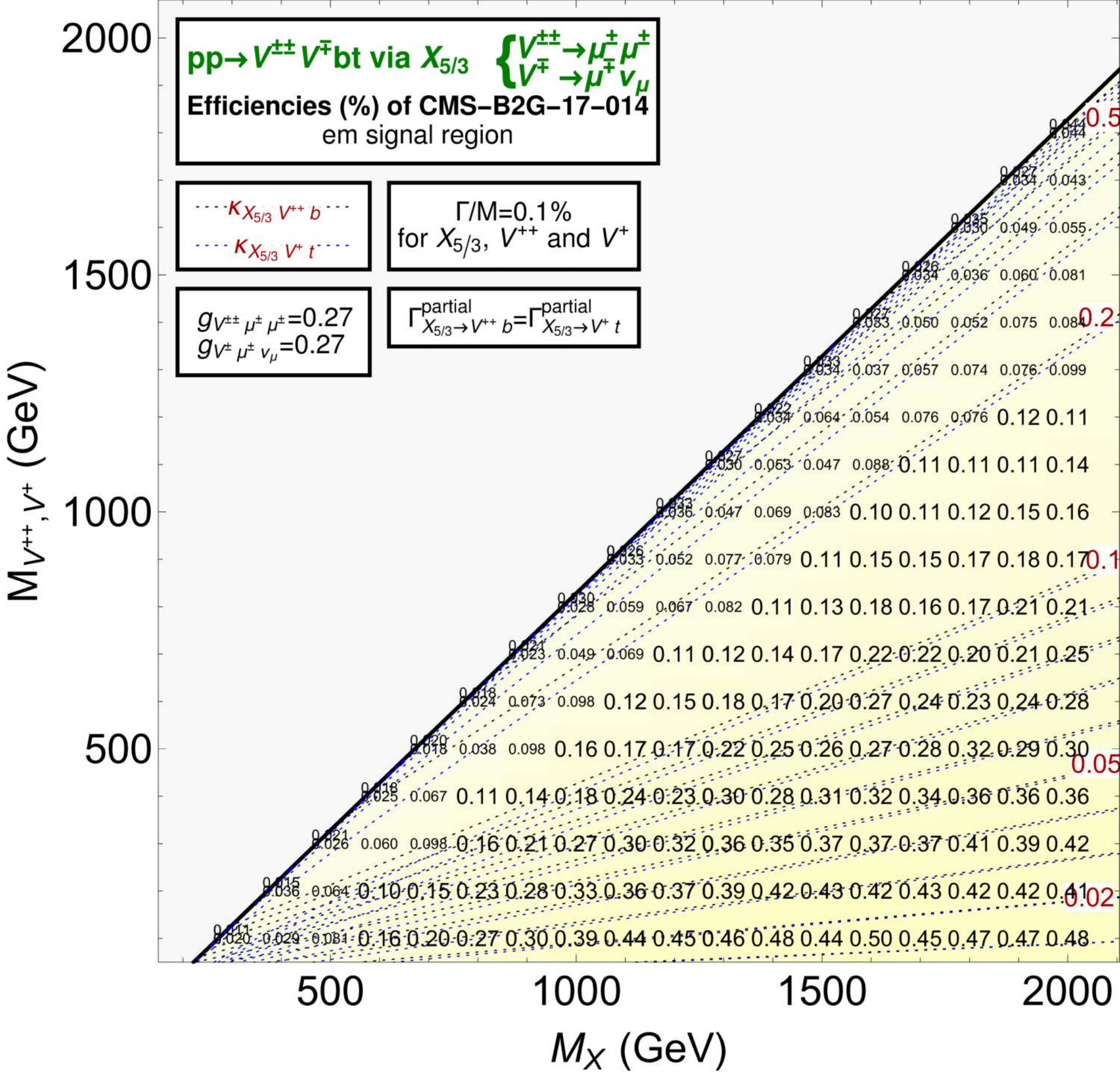}\\
  \caption{\label{fig:XVPP_XVP_mm} Same as Figure~\ref{fig:XVPP_XVP_ee}  but for $V^{\pm\pm}\to\mu^\pm \mu^\pm$ and $V^\mp\to e^\mp\nu_e$ or $V^\mp\to \mu^\mp\nu_\mu$.}
\end{figure}

\begin{figure}[thbp]
  \centering
  \includegraphics[width=.325\textwidth]{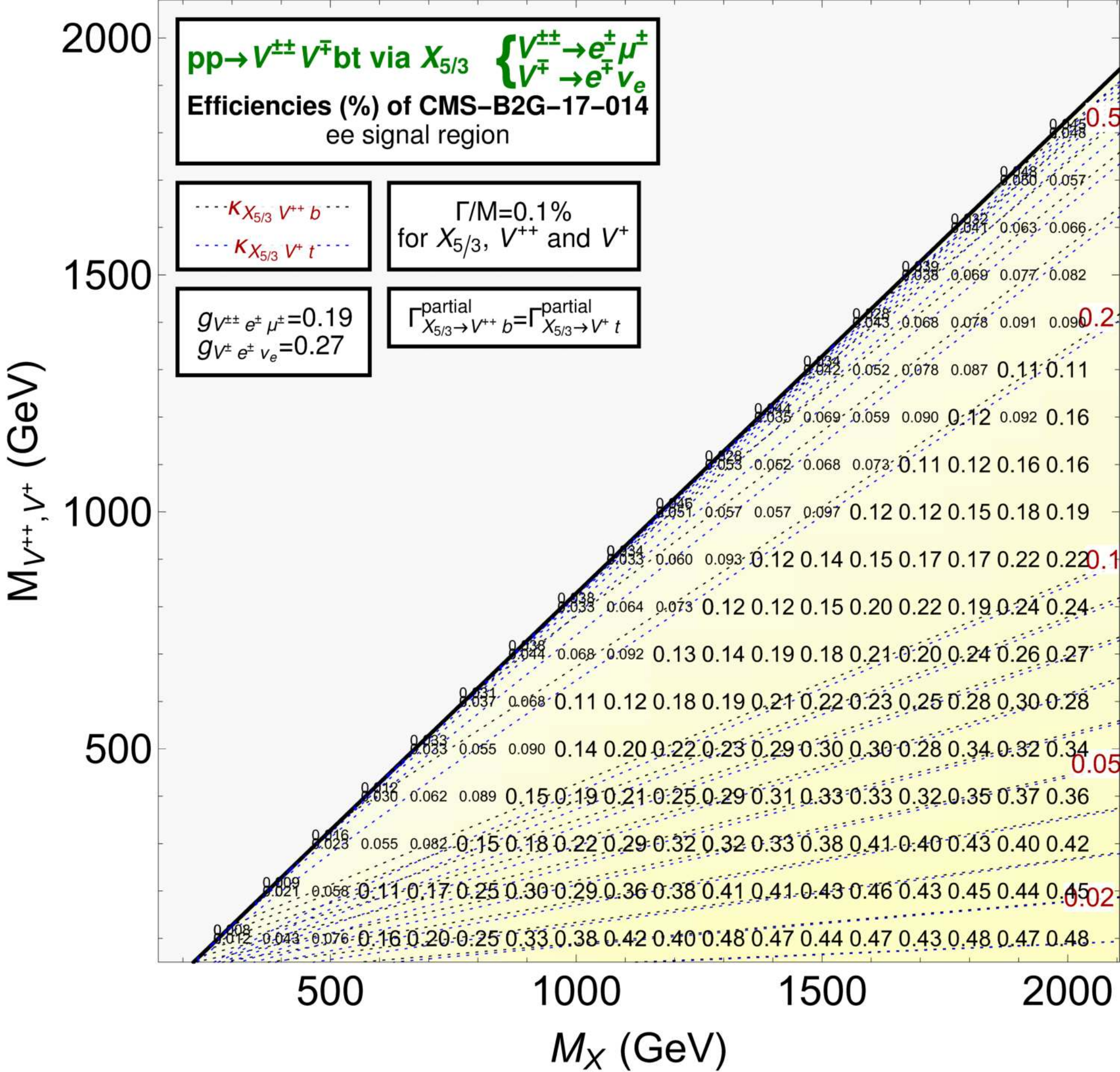}
  \includegraphics[width=.325\textwidth]{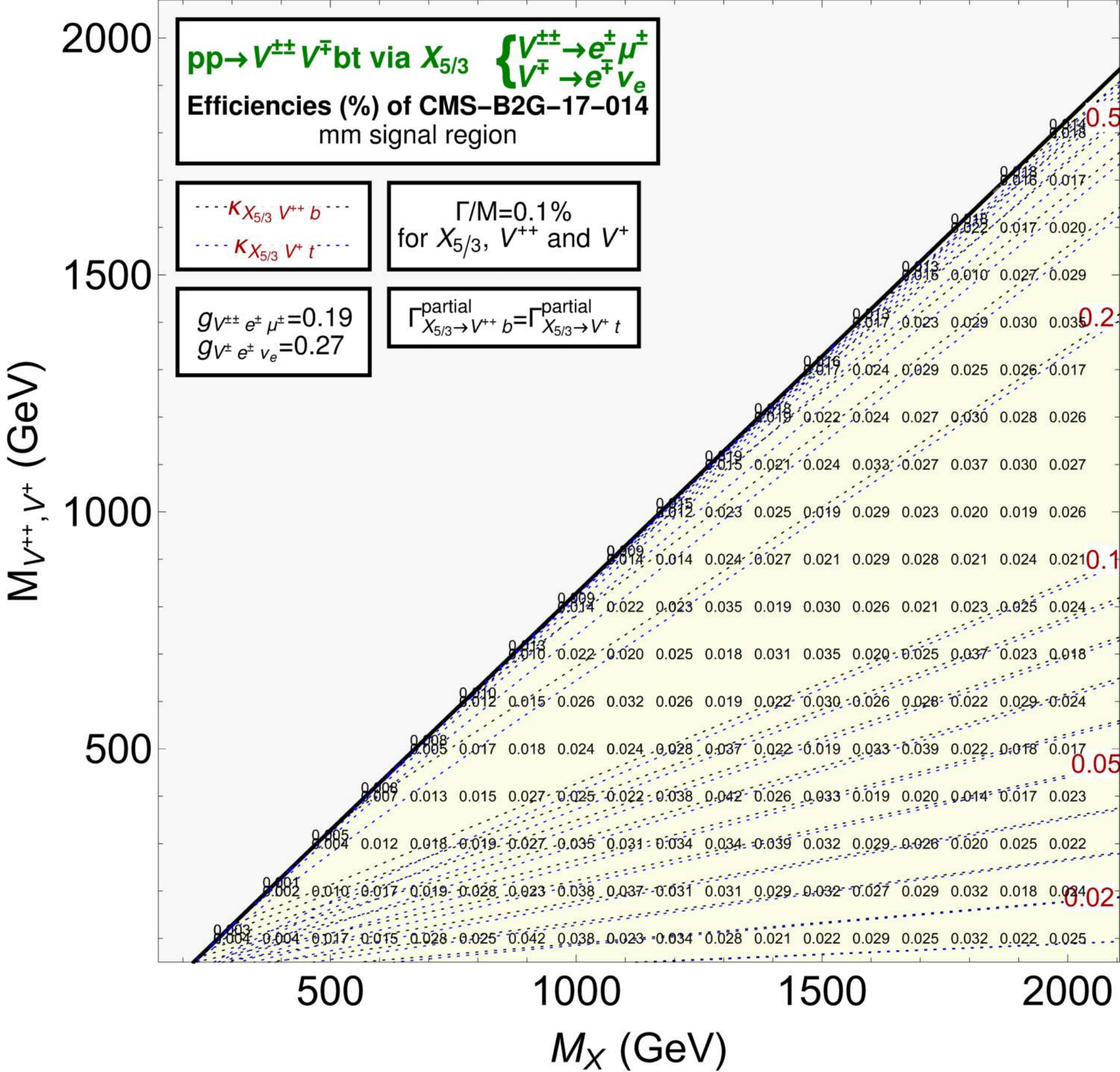}
  \includegraphics[width=.325\textwidth]{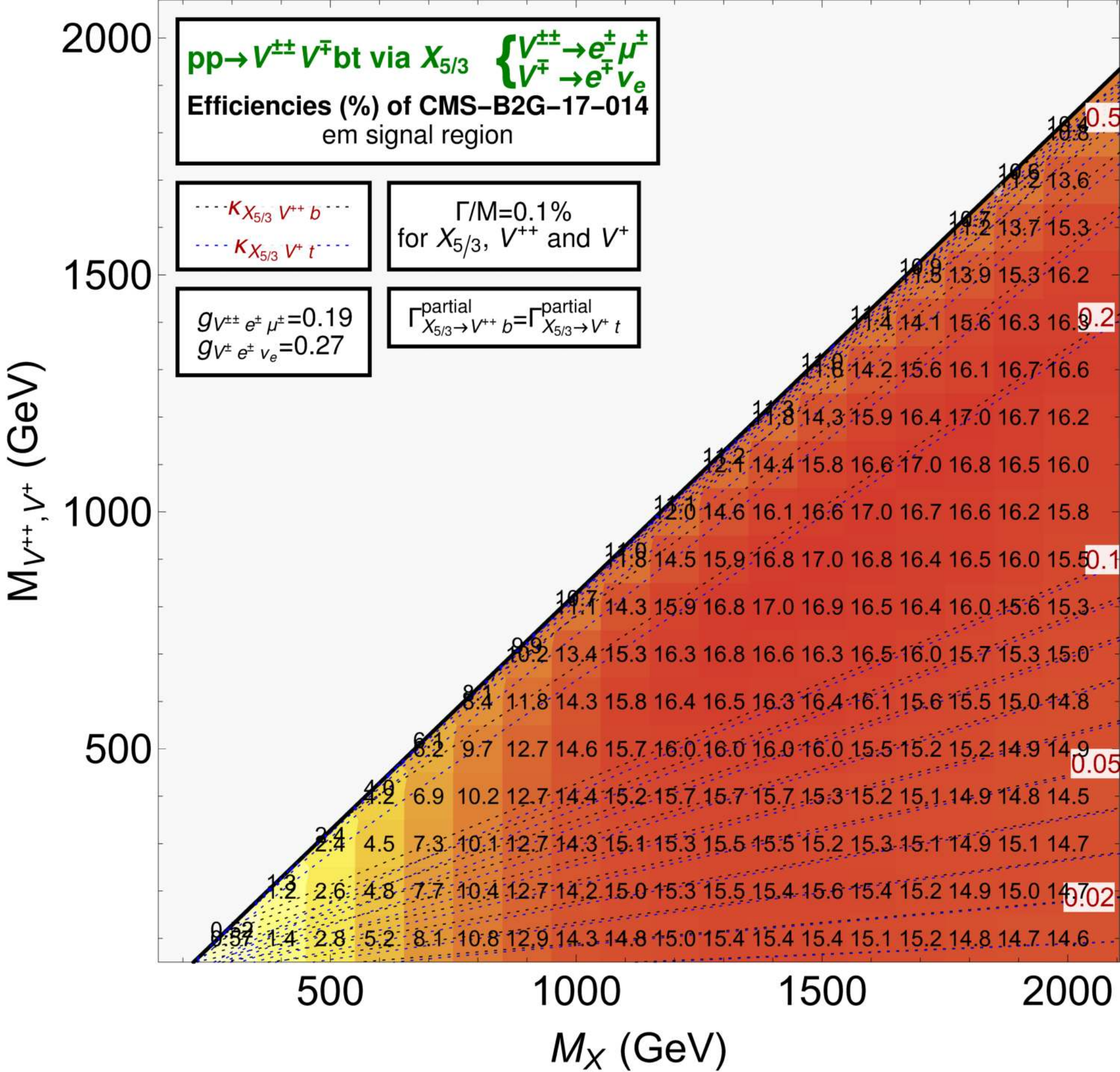}\\
  \includegraphics[width=.325\textwidth]{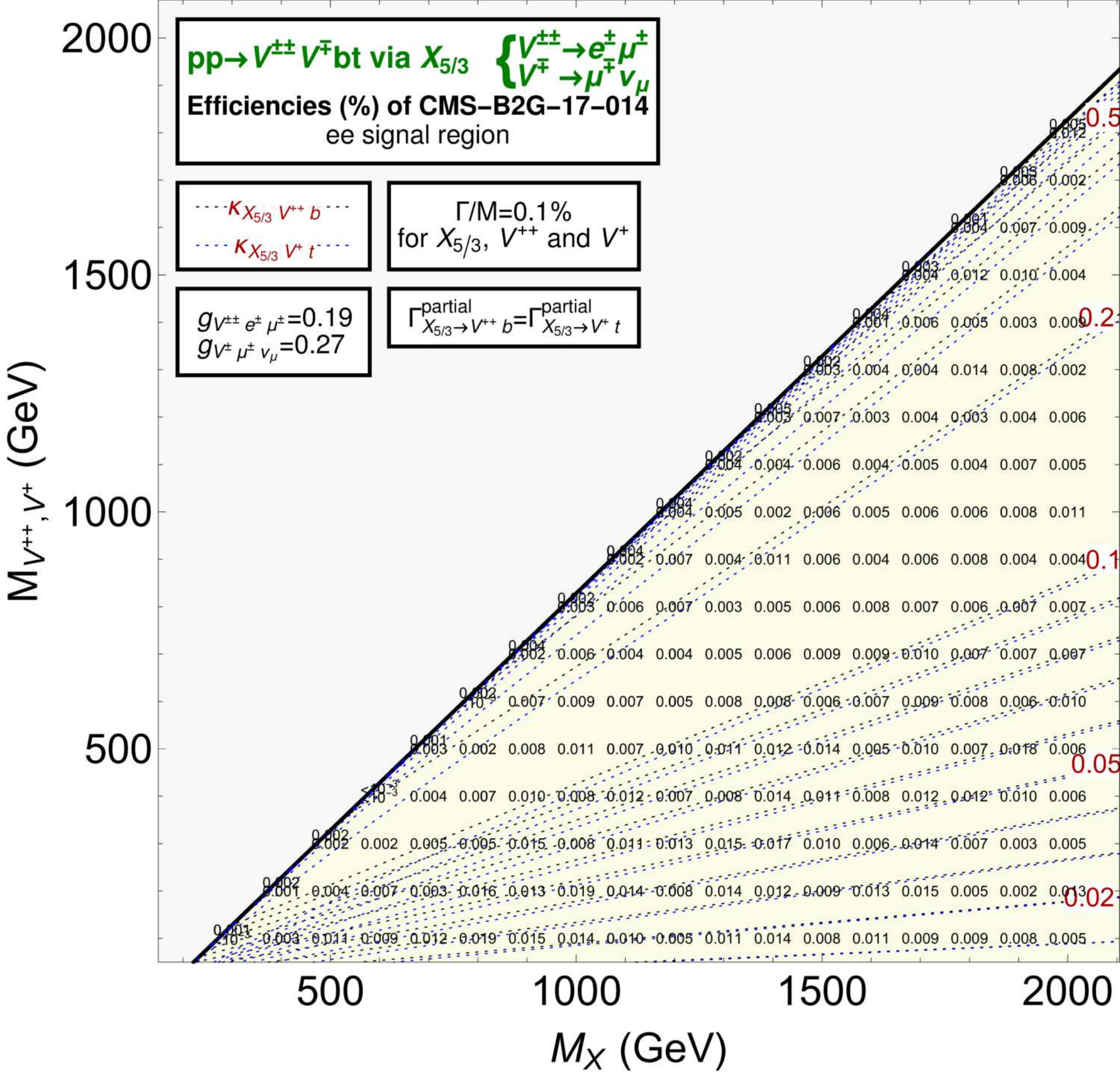}
  \includegraphics[width=.325\textwidth]{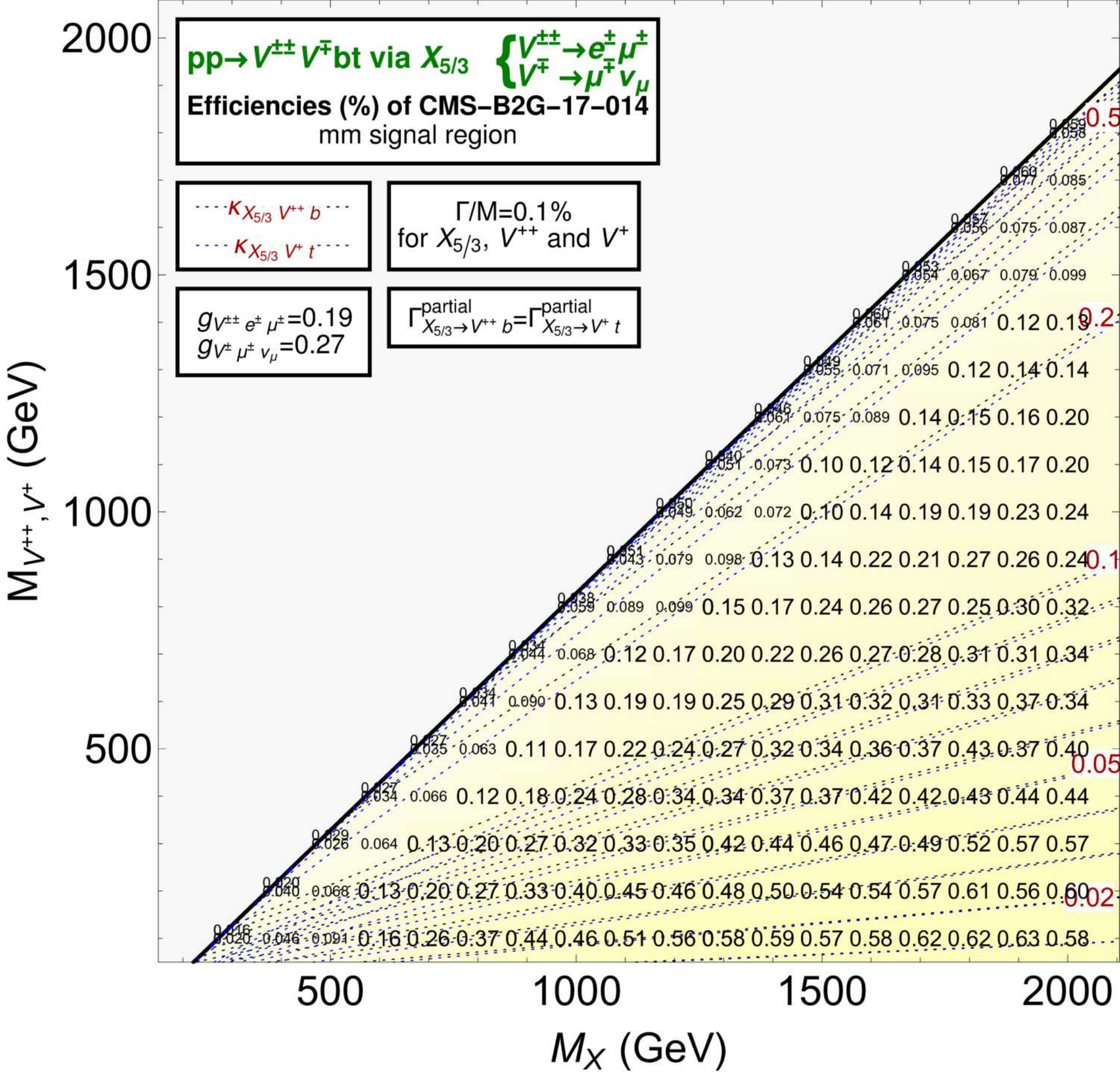}
  \includegraphics[width=.325\textwidth]{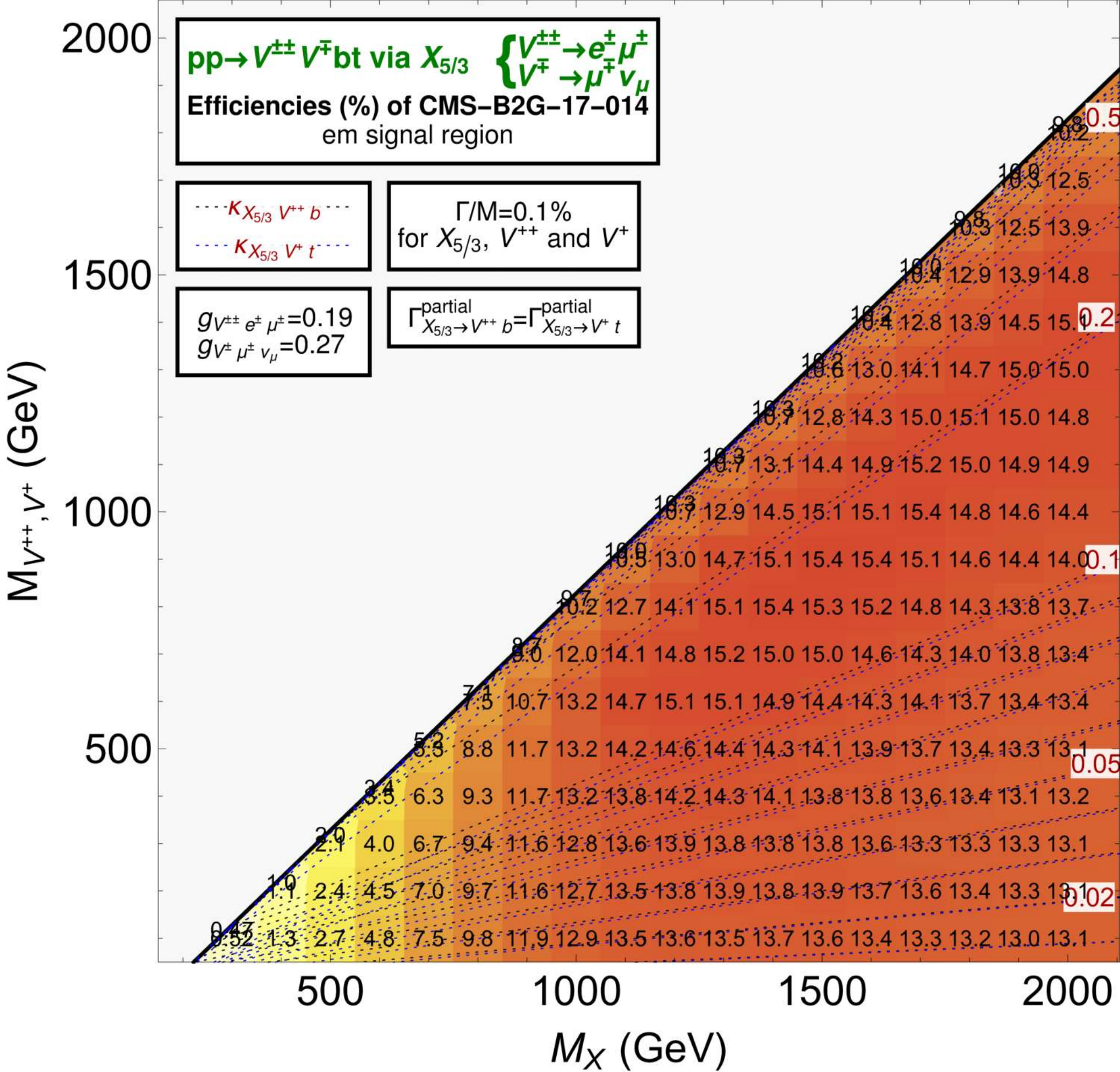}\\
  \caption{\label{fig:XVPP_XVP_em} Same as Figure~\ref{fig:XVPP_XVP_ee}  but for $V^{\pm\pm}\to e^\pm \mu^\pm$ and $V^\mp\to e^\mp\nu_e$ or $V^\mp\to \mu^\mp\nu_\mu$.}
\end{figure}

In Figure~\ref{fig:XVPP_XVP_ee}, \ref{fig:XVPP_XVP_mm} and \ref{fig:XVPP_XVP_em} the efficiencies for the cases involving a doubly charged and a singly charged vector $V$ with same mass against the three signal hypotheses of the CMS analysis (same-sign $ee$, $\mu\mu$ and $e\mu$) are presented. Here the decay of the exotic quark $X_{5/3}$ and its conjugate have been assumed to involve different vector bosons with $100\%$ saturated $BR$s. 
 
\clearpage

\bibliographystyle{JHEP}
\bibliography{BIB}

\providecommand{\href}[2]{#2}\begingroup\raggedright\begin{thebibliography}{10}

\bibitem{Aad:2012tfa}
{\scshape ATLAS} collaboration, \emph{{Observation of a new particle in the
  search for the Standard Model Higgs boson with the ATLAS detector at the
  LHC}}, \href{https://doi.org/10.1016/j.physletb.2012.08.020}{\emph{Phys.
  Lett. B} {\bfseries 716} (2012) 1}
  [\href{https://arxiv.org/abs/1207.7214}{{\ttfamily 1207.7214}}].

\bibitem{Chatrchyan:2012ufa}
{\scshape CMS} collaboration, \emph{{Observation of a New Boson at a Mass of
  125 GeV with the CMS Experiment at the LHC}},
  \href{https://doi.org/10.1016/j.physletb.2012.08.021}{\emph{Phys. Lett. B}
  {\bfseries 716} (2012) 30} [\href{https://arxiv.org/abs/1207.7235}{{\ttfamily
  1207.7235}}].

\bibitem{Sirunyan:2018koj}
{\scshape CMS} collaboration, \emph{{Combined measurements of Higgs boson
  couplings in proton\textendash{}proton collisions at $\sqrt{s}=13\,\text
  {Te}\text {V} $}},
  \href{https://doi.org/10.1140/epjc/s10052-019-6909-y}{\emph{Eur. Phys. J. C}
  {\bfseries 79} (2019) 421}
  [\href{https://arxiv.org/abs/1809.10733}{{\ttfamily 1809.10733}}].

\bibitem{Kaplan:1983fs}
D.~B. Kaplan and H.~Georgi, \emph{{SU(2) x U(1) Breaking by Vacuum
  Misalignment}},
  \href{https://doi.org/10.1016/0370-2693(84)91177-8}{\emph{Phys. Lett. B}
  {\bfseries 136} (1984) 183}.

\bibitem{Kaplan:1991dc}
D.~B. Kaplan, \emph{{Flavor at SSC energies: A New mechanism for dynamically
  generated fermion masses}},
  \href{https://doi.org/10.1016/S0550-3213(05)80021-5}{\emph{Nucl. Phys. B}
  {\bfseries 365} (1991) 259}.

\bibitem{Gripaios:2009pe}
B.~Gripaios, A.~Pomarol, F.~Riva and J.~Serra, \emph{{Beyond the Minimal
  Composite Higgs Model}},
  \href{https://doi.org/10.1088/1126-6708/2009/04/070}{\emph{JHEP} {\bfseries
  04} (2009) 070} [\href{https://arxiv.org/abs/0902.1483}{{\ttfamily
  0902.1483}}].

\bibitem{Panico:2016ull}
G.~Panico and A.~Pomarol, \emph{{Flavor hierarchies from dynamical scales}},
  \href{https://doi.org/10.1007/JHEP07(2016)097}{\emph{JHEP} {\bfseries 07}
  (2016) 097} [\href{https://arxiv.org/abs/1603.06609}{{\ttfamily
  1603.06609}}].

\bibitem{Antoniadis:1990ew}
I.~Antoniadis, \emph{{A Possible new dimension at a few TeV}},
  \href{https://doi.org/10.1016/0370-2693(90)90617-F}{\emph{Phys. Lett. B}
  {\bfseries 246} (1990) 377}.

\bibitem{Csaki:2003sh}
C.~Csaki, C.~Grojean, J.~Hubisz, Y.~Shirman and J.~Terning, \emph{{Fermions on
  an interval: Quark and lepton masses without a Higgs}},
  \href{https://doi.org/10.1103/PhysRevD.70.015012}{\emph{Phys. Rev. D}
  {\bfseries 70} (2004) 015012}
  [\href{https://arxiv.org/abs/hep-ph/0310355}{{\ttfamily hep-ph/0310355}}].

\bibitem{Cacciapaglia:2009pa}
G.~Cacciapaglia, A.~Deandrea and J.~Llodra-Perez, \emph{{A Dark Matter
  candidate from Lorentz Invariance in 6D}},
  \href{https://doi.org/10.1007/JHEP03(2010)083}{\emph{JHEP} {\bfseries 03}
  (2010) 083} [\href{https://arxiv.org/abs/0907.4993}{{\ttfamily 0907.4993}}].

\bibitem{ArkaniHamed:2002qx}
N.~Arkani-Hamed, A.~G. Cohen, E.~Katz, A.~E. Nelson, T.~Gregoire and J.~G.
  Wacker, \emph{{The Minimal moose for a little Higgs}},
  \href{https://doi.org/10.1088/1126-6708/2002/08/021}{\emph{JHEP} {\bfseries
  08} (2002) 021} [\href{https://arxiv.org/abs/hep-ph/0206020}{{\ttfamily
  hep-ph/0206020}}].

\bibitem{Abbas:2017vle}
G.~Abbas, \emph{{A model of spontaneous $CP$ breaking at low scale}},
  \href{https://doi.org/10.1016/j.physletb.2017.08.028}{\emph{Phys. Lett. B}
  {\bfseries 773} (2017) 252}
  [\href{https://arxiv.org/abs/1706.02564}{{\ttfamily 1706.02564}}].

\bibitem{Lavoura:1992np}
L.~Lavoura and J.~P. Silva, \emph{{The Oblique corrections from vector - like
  singlet and doublet quarks}},
  \href{https://doi.org/10.1103/PhysRevD.47.2046}{\emph{Phys. Rev. D}
  {\bfseries 47} (1993) 2046}.

\bibitem{delAguila:2000rc}
F.~del Aguila, M.~Perez-Victoria and J.~Santiago, \emph{{Observable
  contributions of new exotic quarks to quark mixing}},
  \href{https://doi.org/10.1088/1126-6708/2000/09/011}{\emph{JHEP} {\bfseries
  09} (2000) 011} [\href{https://arxiv.org/abs/hep-ph/0007316}{{\ttfamily
  hep-ph/0007316}}].

\bibitem{AguilarSaavedra:2005pv}
J.~A. Aguilar-Saavedra, \emph{{Pair production of heavy Q = 2/3 singlets at
  LHC}}, \href{https://doi.org/10.1016/j.physletb.2005.08.062}{\emph{Phys.
  Lett. B} {\bfseries 625} (2005) 234}
  [\href{https://arxiv.org/abs/hep-ph/0506187}{{\ttfamily hep-ph/0506187}}].

\bibitem{Cynolter:2008ea}
G.~Cynolter and E.~Lendvai, \emph{{Electroweak Precision Constraints on
  Vector-like Fermions}},
  \href{https://doi.org/10.1140/epjc/s10052-008-0771-7}{\emph{Eur. Phys. J. C}
  {\bfseries 58} (2008) 463} [\href{https://arxiv.org/abs/0804.4080}{{\ttfamily
  0804.4080}}].

\bibitem{AguilarSaavedra:2009es}
J.~A. Aguilar-Saavedra, \emph{{Identifying top partners at LHC}},
  \href{https://doi.org/10.1088/1126-6708/2009/11/030}{\emph{JHEP} {\bfseries
  11} (2009) 030} [\href{https://arxiv.org/abs/0907.3155}{{\ttfamily
  0907.3155}}].

\bibitem{Mrazek:2009yu}
J.~Mrazek and A.~Wulzer, \emph{{A Strong Sector at the LHC: Top Partners in
  Same-Sign Dileptons}},
  \href{https://doi.org/10.1103/PhysRevD.81.075006}{\emph{Phys. Rev. D}
  {\bfseries 81} (2010) 075006}
  [\href{https://arxiv.org/abs/0909.3977}{{\ttfamily 0909.3977}}].

\bibitem{Cacciapaglia:2012dd}
G.~Cacciapaglia, A.~Deandrea, L.~Panizzi, S.~Perries and V.~Sordini,
  \emph{{Heavy Vector-like quark with charge 5/3 at the LHC}},
  \href{https://doi.org/10.1007/JHEP03(2013)004}{\emph{JHEP} {\bfseries 03}
  (2013) 004} [\href{https://arxiv.org/abs/1211.4034}{{\ttfamily 1211.4034}}].

\bibitem{Berger:2012ec}
J.~Berger, J.~Hubisz and M.~Perelstein, \emph{{A Fermionic Top Partner:
  Naturalness and the LHC}},
  \href{https://doi.org/10.1007/JHEP07(2012)016}{\emph{JHEP} {\bfseries 07}
  (2012) 016} [\href{https://arxiv.org/abs/1205.0013}{{\ttfamily 1205.0013}}].

\bibitem{Okada:2012gy}
Y.~Okada and L.~Panizzi, \emph{{LHC signatures of vector-like quarks}},
  \href{https://doi.org/10.1155/2013/364936}{\emph{Adv. High Energy Phys.}
  {\bfseries 2013} (2013) 364936}
  [\href{https://arxiv.org/abs/1207.5607}{{\ttfamily 1207.5607}}].

\bibitem{DeSimone:2012fs}
A.~De~Simone, O.~Matsedonskyi, R.~Rattazzi and A.~Wulzer, \emph{{A First Top
  Partner Hunter's Guide}},
  \href{https://doi.org/10.1007/JHEP04(2013)004}{\emph{JHEP} {\bfseries 04}
  (2013) 004} [\href{https://arxiv.org/abs/1211.5663}{{\ttfamily 1211.5663}}].

\bibitem{Falkowski:2013jya}
A.~Falkowski, D.~M. Straub and A.~Vicente, \emph{{Vector-like leptons: Higgs
  decays and collider phenomenology}},
  \href{https://doi.org/10.1007/JHEP05(2014)092}{\emph{JHEP} {\bfseries 05}
  (2014) 092} [\href{https://arxiv.org/abs/1312.5329}{{\ttfamily 1312.5329}}].

\bibitem{Buchkremer:2013bha}
M.~Buchkremer, G.~Cacciapaglia, A.~Deandrea and L.~Panizzi, \emph{{Model
  Independent Framework for Searches of Top Partners}},
  \href{https://doi.org/10.1016/j.nuclphysb.2013.08.010}{\emph{Nucl. Phys. B}
  {\bfseries 876} (2013) 376}
  [\href{https://arxiv.org/abs/1305.4172}{{\ttfamily 1305.4172}}].

\bibitem{AguilarSaavedra:2013qpa}
J.~Aguilar-Saavedra, R.~Benbrik, S.~Heinemeyer and M.~P\'erez-Victoria,
  \emph{{Handbook of vectorlike quarks: Mixing and single production}},
  \href{https://doi.org/10.1103/PhysRevD.88.094010}{\emph{Phys. Rev. D}
  {\bfseries 88} (2013) 094010}
  [\href{https://arxiv.org/abs/1306.0572}{{\ttfamily 1306.0572}}].

\bibitem{Matsedonskyi:2014mna}
O.~Matsedonskyi, G.~Panico and A.~Wulzer, \emph{{On the Interpretation of Top
  Partners Searches}},
  \href{https://doi.org/10.1007/JHEP12(2014)097}{\emph{JHEP} {\bfseries 12}
  (2014) 097} [\href{https://arxiv.org/abs/1409.0100}{{\ttfamily 1409.0100}}].

\bibitem{Matsedonskyi:2015dns}
O.~Matsedonskyi, G.~Panico and A.~Wulzer, \emph{{Top Partners Searches and
  Composite Higgs Models}},
  \href{https://doi.org/10.1007/JHEP04(2016)003}{\emph{JHEP} {\bfseries 04}
  (2016) 003} [\href{https://arxiv.org/abs/1512.04356}{{\ttfamily
  1512.04356}}].

\bibitem{Panella:2017spx}
O.~Panella, R.~Leonardi, G.~Pancheri, Y.~Srivastava, M.~Narain and U.~Heintz,
  \emph{{Production of exotic composite quarks at the LHC}},
  \href{https://doi.org/10.1103/PhysRevD.96.075034}{\emph{Phys. Rev. D}
  {\bfseries 96} (2017) 075034}
  [\href{https://arxiv.org/abs/1703.06913}{{\ttfamily 1703.06913}}].

\bibitem{Barducci:2017xtw}
D.~Barducci and L.~Panizzi, \emph{{Vector-like quarks coupling discrimination
  at the LHC and future hadron colliders}},
  \href{https://doi.org/10.1007/JHEP12(2017)057}{\emph{JHEP} {\bfseries 12}
  (2017) 057} [\href{https://arxiv.org/abs/1710.02325}{{\ttfamily
  1710.02325}}].

\bibitem{Aaboud:2017zfn}
{\scshape ATLAS} collaboration, \emph{{Search for pair production of heavy
  vector-like quarks decaying to high-p$_{T}$ W bosons and b quarks in the
  lepton-plus-jets final state in pp collisions at $ \sqrt{s}=13 $ TeV with the
  ATLAS detector}}, \href{https://doi.org/10.1007/JHEP10(2017)141}{\emph{JHEP}
  {\bfseries 10} (2017) 141}
  [\href{https://arxiv.org/abs/1707.03347}{{\ttfamily 1707.03347}}].

\bibitem{Aaboud:2017qpr}
{\scshape ATLAS} collaboration, \emph{{Search for pair production of
  vector-like top quarks in events with one lepton, jets, and missing
  transverse momentum in $ \sqrt{s}=13 $ TeV $pp$ collisions with the ATLAS
  detector}}, \href{https://doi.org/10.1007/JHEP08(2017)052}{\emph{JHEP}
  {\bfseries 08} (2017) 052}
  [\href{https://arxiv.org/abs/1705.10751}{{\ttfamily 1705.10751}}].

\bibitem{Aaboud:2018xuw}
{\scshape ATLAS} collaboration, \emph{{Search for pair production of up-type
  vector-like quarks and for four-top-quark events in final states with
  multiple $b$-jets with the ATLAS detector}},
  \href{https://doi.org/10.1007/JHEP07(2018)089}{\emph{JHEP} {\bfseries 07}
  (2018) 089} [\href{https://arxiv.org/abs/1803.09678}{{\ttfamily
  1803.09678}}].

\bibitem{Aaboud:2018saj}
{\scshape ATLAS} collaboration, \emph{{Search for pair- and single-production
  of vector-like quarks in final states with at least one $Z$ boson decaying
  into a pair of electrons or muons in $pp$ collision data collected with the
  ATLAS detector at $\sqrt{s} = 13$ TeV}},
  \href{https://doi.org/10.1103/PhysRevD.98.112010}{\emph{Phys. Rev. D}
  {\bfseries 98} (2018) 112010}
  [\href{https://arxiv.org/abs/1806.10555}{{\ttfamily 1806.10555}}].

\bibitem{Aaboud:2018xpj}
{\scshape ATLAS} collaboration, \emph{{Search for new phenomena in events with
  same-charge leptons and $b$-jets in $pp$ collisions at $\sqrt{s}= 13$ TeV
  with the ATLAS detector}},
  \href{https://doi.org/10.1007/JHEP12(2018)039}{\emph{JHEP} {\bfseries 12}
  (2018) 039} [\href{https://arxiv.org/abs/1807.11883}{{\ttfamily
  1807.11883}}].

\bibitem{Aaboud:2018wxv}
{\scshape ATLAS} collaboration, \emph{{Search for pair production of heavy
  vector-like quarks decaying into hadronic final states in $pp$ collisions at
  $\sqrt{s} = 13$ TeV with the ATLAS detector}},
  \href{https://doi.org/10.1103/PhysRevD.98.092005}{\emph{Phys. Rev. D}
  {\bfseries 98} (2018) 092005}
  [\href{https://arxiv.org/abs/1808.01771}{{\ttfamily 1808.01771}}].

\bibitem{Aaboud:2018pii}
{\scshape ATLAS} collaboration, \emph{{Combination of the searches for
  pair-produced vector-like partners of the third-generation quarks at
  $\sqrt{s} =$ 13 TeV with the ATLAS detector}},
  \href{https://doi.org/10.1103/PhysRevLett.121.211801}{\emph{Phys. Rev. Lett.}
  {\bfseries 121} (2018) 211801}
  [\href{https://arxiv.org/abs/1808.02343}{{\ttfamily 1808.02343}}].

\bibitem{Sirunyan:2017pks}
{\scshape CMS} collaboration, \emph{{Search for pair production of vector-like
  quarks in the bW$\overline{\mathrm{b}}$W channel from proton-proton
  collisions at $\sqrt{s} =$ 13 TeV}},
  \href{https://doi.org/10.1016/j.physletb.2018.01.077}{\emph{Phys. Lett. B}
  {\bfseries 779} (2018) 82}
  [\href{https://arxiv.org/abs/1710.01539}{{\ttfamily 1710.01539}}].

\bibitem{Sirunyan:2018qau}
{\scshape CMS} collaboration, \emph{{Search for vector-like quarks in events
  with two oppositely charged leptons and jets in proton-proton collisions at
  $\sqrt{s} =$ 13 TeV}},
  \href{https://doi.org/10.1140/epjc/s10052-019-6855-8}{\emph{Eur. Phys. J. C}
  {\bfseries 79} (2019) 364}
  [\href{https://arxiv.org/abs/1812.09768}{{\ttfamily 1812.09768}}].

\bibitem{Sirunyan:2018omb}
{\scshape CMS} collaboration, \emph{{Search for vector-like T and B quark pairs
  in final states with leptons at $\sqrt{s} =$ 13 TeV}},
  \href{https://doi.org/10.1007/JHEP08(2018)177}{\emph{JHEP} {\bfseries 08}
  (2018) 177} [\href{https://arxiv.org/abs/1805.04758}{{\ttfamily
  1805.04758}}].

\bibitem{Sirunyan:2019sza}
{\scshape CMS} collaboration, \emph{{Search for pair production of vectorlike
  quarks in the fully hadronic final state}},
  \href{https://doi.org/10.1103/PhysRevD.100.072001}{\emph{Phys. Rev. D}
  {\bfseries 100} (2019) 072001}
  [\href{https://arxiv.org/abs/1906.11903}{{\ttfamily 1906.11903}}].

\bibitem{Sirunyan:2018yun}
{\scshape CMS} collaboration, \emph{{Search for top quark partners with charge
  5/3 in the same-sign dilepton and single-lepton final states in proton-proton
  collisions at {$\sqrt{s}=13$} TeV}},
  \href{https://doi.org/10.1007/JHEP03(2019)082}{\emph{JHEP} {\bfseries 03}
  (2019) 82} [\href{https://arxiv.org/abs/1810.03188}{{\ttfamily 1810.03188}}].

\bibitem{Sirunyan:2020qvb}
{\scshape CMS} collaboration, \emph{{A search for bottom-type, vector-like
  quark pair production in a fully hadronic final state in proton-proton
  collisions at $\sqrt{s} =$ 13 TeV}},
  \href{https://doi.org/10.1103/PhysRevD.102.112004}{\emph{Phys. Rev. D}
  {\bfseries 102} (2020) 112004}
  [\href{https://arxiv.org/abs/2008.09835}{{\ttfamily 2008.09835}}].

\bibitem{Aaboud:2018ifs}
{\scshape ATLAS} collaboration, \emph{{Search for single production of
  vector-like quarks decaying into $Wb$ in $pp$ collisions at $\sqrt{s} = 13$
  TeV with the ATLAS detector}},
  \href{https://doi.org/10.1007/JHEP05(2019)164}{\emph{JHEP} {\bfseries 05}
  (2019) 164} [\href{https://arxiv.org/abs/1812.07343}{{\ttfamily
  1812.07343}}].

\bibitem{Sirunyan:2017ynj}
{\scshape CMS} collaboration, \emph{{Search for single production of a
  vector-like T quark decaying to a Z boson and a top quark in proton-proton
  collisions at $\sqrt s$ = 13 TeV}},
  \href{https://doi.org/10.1016/j.physletb.2018.04.036}{\emph{Phys. Lett. B}
  {\bfseries 781} (2018) 574}
  [\href{https://arxiv.org/abs/1708.01062}{{\ttfamily 1708.01062}}].

\bibitem{Sirunyan:2018fjh}
{\scshape CMS} collaboration, \emph{{Search for single production of
  vector-like quarks decaying to a b quark and a Higgs boson}},
  \href{https://doi.org/10.1007/JHEP06(2018)031}{\emph{JHEP} {\bfseries 06}
  (2018) 031} [\href{https://arxiv.org/abs/1802.01486}{{\ttfamily
  1802.01486}}].

\bibitem{Sirunyan:2018ncp}
{\scshape CMS} collaboration, \emph{{Search for single production of
  vector-like quarks decaying to a top quark and a W boson in proton-proton
  collisions at $\sqrt{s} =$ 13 TeV}},
  \href{https://doi.org/10.1140/epjc/s10052-019-6556-3}{\emph{Eur. Phys. J. C}
  {\bfseries 79} (2019) 90} [\href{https://arxiv.org/abs/1809.08597}{{\ttfamily
  1809.08597}}].

\bibitem{Sirunyan:2019xeh}
{\scshape CMS} collaboration, \emph{{Search for electroweak production of a
  vector-like T quark using fully hadronic final states}},
  \href{https://doi.org/10.1007/JHEP01(2020)036}{\emph{JHEP} {\bfseries 01}
  (2020) 036} [\href{https://arxiv.org/abs/1909.04721}{{\ttfamily
  1909.04721}}].

\bibitem{Aaboud:2018uek}
{\scshape ATLAS} collaboration, \emph{{Search for pair production of heavy
  vector-like quarks decaying into high-$p_T$ $W$ bosons and top quarks in the
  lepton-plus-jets final state in $pp$ collisions at $\sqrt{s}=13$ TeV with the
  ATLAS detector}}, \href{https://doi.org/10.1007/JHEP08(2018)048}{\emph{JHEP}
  {\bfseries 08} (2018) 048}
  [\href{https://arxiv.org/abs/1806.01762}{{\ttfamily 1806.01762}}].

\bibitem{Serra:2015xfa}
J.~Serra, \emph{{Beyond the Minimal Top Partner Decay}},
  \href{https://doi.org/10.1007/JHEP09(2015)176}{\emph{JHEP} {\bfseries 09}
  (2015) 176} [\href{https://arxiv.org/abs/1506.05110}{{\ttfamily
  1506.05110}}].

\bibitem{Aguilar-Saavedra:2017giu}
J.~A. Aguilar-Saavedra, D.~E. L\'opez-Fogliani and C.~Mu\~noz, \emph{{Novel
  signatures for vector-like quarks}},
  \href{https://doi.org/10.1007/JHEP06(2017)095}{\emph{JHEP} {\bfseries 06}
  (2017) 095} [\href{https://arxiv.org/abs/1705.02526}{{\ttfamily
  1705.02526}}].

\bibitem{Chala:2017xgc}
M.~Chala, \emph{{Direct bounds on heavy toplike quarks with standard and exotic
  decays}}, \href{https://doi.org/10.1103/PhysRevD.96.015028}{\emph{Phys. Rev.
  D} {\bfseries 96} (2017) 015028}
  [\href{https://arxiv.org/abs/1705.03013}{{\ttfamily 1705.03013}}].

\bibitem{Bizot:2018tds}
N.~Bizot, G.~Cacciapaglia and T.~Flacke, \emph{{Common exotic decays of top
  partners}}, \href{https://doi.org/10.1007/JHEP06(2018)065}{\emph{JHEP}
  {\bfseries 06} (2018) 065}
  [\href{https://arxiv.org/abs/1803.00021}{{\ttfamily 1803.00021}}].

\bibitem{Han:2018hcu}
H.~Han, L.~Huang, T.~Ma, J.~Shu, T.~M.~P. Tait and Y.~Wu, \emph{{Six Top
  Messages of New Physics at the LHC}},
  \href{https://doi.org/10.1007/JHEP10(2019)008}{\emph{JHEP} {\bfseries 10}
  (2019) 008} [\href{https://arxiv.org/abs/1812.11286}{{\ttfamily
  1812.11286}}].

\bibitem{Xie:2019gya}
K.-P. Xie, G.~Cacciapaglia and T.~Flacke, \emph{{Exotic decays of top partners
  with charge 5/3: bounds and opportunities}},
  \href{https://doi.org/10.1007/JHEP10(2019)134}{\emph{JHEP} {\bfseries 10}
  (2019) 134} [\href{https://arxiv.org/abs/1907.05894}{{\ttfamily
  1907.05894}}].

\bibitem{Benbrik:2019zdp}
R.~Benbrik et~al., \emph{{Signatures of vector-like top partners decaying into
  new neutral scalar or pseudoscalar bosons}},
  \href{https://doi.org/10.1007/JHEP05(2020)028}{\emph{JHEP} {\bfseries 05}
  (2020) 028} [\href{https://arxiv.org/abs/1907.05929}{{\ttfamily
  1907.05929}}].

\bibitem{Cacciapaglia:2019zmj}
G.~Cacciapaglia, T.~Flacke, M.~Park and M.~Zhang, \emph{{Exotic decays of top
  partners: mind the search gap}},
  \href{https://doi.org/10.1016/j.physletb.2019.135015}{\emph{Phys. Lett. B}
  {\bfseries 798} (2019) 135015}
  [\href{https://arxiv.org/abs/1908.07524}{{\ttfamily 1908.07524}}].

\bibitem{Aguilar-Saavedra:2019ghg}
J.~A. Aguilar-Saavedra, J.~Alonso-Gonz\'alez, L.~Merlo and J.~M. No,
  \emph{{Exotic vectorlike quark phenomenology in the minimal linear
  \ensuremath{\sigma} model}},
  \href{https://doi.org/10.1103/PhysRevD.101.035015}{\emph{Phys. Rev. D}
  {\bfseries 101} (2020) 035015}
  [\href{https://arxiv.org/abs/1911.10202}{{\ttfamily 1911.10202}}].

\bibitem{Singer:1980sw}
M.~Singer, J.~W.~F. Valle and J.~Schechter, \emph{{Canonical Neutral Current
  Predictions From the Weak Electromagnetic Gauge Group SU(3) X $u$(1)}},
  \href{https://doi.org/10.1103/PhysRevD.22.738}{\emph{Phys. Rev. D} {\bfseries
  22} (1980) 738}.

\bibitem{Valle:1983dk}
J.~W.~F. Valle and M.~Singer, \emph{{Lepton Number Violation With Quasi Dirac
  Neutrinos}}, \href{https://doi.org/10.1103/PhysRevD.28.540}{\emph{Phys. Rev.
  D} {\bfseries 28} (1983) 540}.

\bibitem{Pisano:1991ee}
F.~Pisano and V.~Pleitez, \emph{{An SU(3) x U(1) model for electroweak
  interactions}}, \href{https://doi.org/10.1103/PhysRevD.46.410}{\emph{Phys.
  Rev. D} {\bfseries 46} (1992) 410}
  [\href{https://arxiv.org/abs/hep-ph/9206242}{{\ttfamily hep-ph/9206242}}].

\bibitem{Frampton:1992wt}
P.~H. Frampton, \emph{{Chiral dilepton model and the flavor question}},
  \href{https://doi.org/10.1103/PhysRevLett.69.2889}{\emph{Phys. Rev. Lett.}
  {\bfseries 69} (1992) 2889}.

\bibitem{Foot:1994ym}
R.~Foot, H.~N. Long and T.~A. Tran, \emph{{$SU(3)_L \otimes U(1)_N$ and
  $SU(4)_L \otimes U(1)_N$ gauge models with right-handed neutrinos}},
  \href{https://doi.org/10.1103/PhysRevD.50.R34}{\emph{Phys. Rev. D} {\bfseries
  50} (1994) R34} [\href{https://arxiv.org/abs/hep-ph/9402243}{{\ttfamily
  hep-ph/9402243}}].

\bibitem{Hoang:1995vq}
H.~N. Long, \emph{{The 331 model with right handed neutrinos}},
  \href{https://doi.org/10.1103/PhysRevD.53.437}{\emph{Phys. Rev. D} {\bfseries
  53} (1996) 437} [\href{https://arxiv.org/abs/hep-ph/9504274}{{\ttfamily
  hep-ph/9504274}}].

\bibitem{Corcella:2017dns}
G.~Corcella, C.~Corian\`o, A.~Costantini and P.~H. Frampton, \emph{{Bilepton
  Signatures at the LHC}},
  \href{https://doi.org/10.1016/j.physletb.2017.09.015}{\emph{Phys. Lett. B}
  {\bfseries 773} (2017) 544}
  [\href{https://arxiv.org/abs/1707.01381}{{\ttfamily 1707.01381}}].

\bibitem{Corcella:2018eib}
G.~Corcella, C.~Corian\`o, A.~Costantini and P.~H. Frampton, \emph{{Exploring
  Scalar and Vector Bileptons at the LHC in a 331 Model}},
  \href{https://doi.org/10.1016/j.physletb.2018.08.015}{\emph{Phys. Lett. B}
  {\bfseries 785} (2018) 73}
  [\href{https://arxiv.org/abs/1806.04536}{{\ttfamily 1806.04536}}].

\bibitem{Corcella:2021upj}
G.~Corcella, C.~Corian\`o, A.~Costantini and P.~H. Frampton,
  \emph{{Non-Leptonic Decays of Bileptons}},
  \href{https://arxiv.org/abs/2106.14748}{{\ttfamily 2106.14748}}.

\bibitem{FeynRules}
A.~Alloul, N.~D. Christensen, C.~Degrande, C.~Duhr and B.~Fuks,
  \emph{{FeynRules 2.0 - A complete toolbox for tree-level phenomenology}},
  \href{https://doi.org/10.1016/j.cpc.2014.04.012}{\emph{Comput. Phys. Commun.}
  {\bfseries 185} (2014) 2250}
  [\href{https://arxiv.org/abs/1310.1921}{{\ttfamily 1310.1921}}].

\bibitem{Fuks:2016ftf}
B.~Fuks and H.-S. Shao, \emph{{QCD next-to-leading-order predictions matched to
  parton showers for vector-like quark models}},
  \href{https://doi.org/10.1140/epjc/s10052-017-4686-z}{\emph{Eur. Phys. J.}
  {\bfseries C77} (2017) 135}
  [\href{https://arxiv.org/abs/1610.04622}{{\ttfamily 1610.04622}}].

\bibitem{UFO}
C.~Degrande, C.~Duhr, B.~Fuks, B.~Grellscheid, O.~Mattelaer and T.~Reiter,
  \emph{{UFO} – the universal feynrules output},
  \href{https://doi.org/10.1016/j.cpc.2012.01.022}{\emph{Computer Physics
  Communications} {\bfseries 183} (2012) 1201 }.

\bibitem{Alwall:2014hca}
J.~Alwall, R.~Frederix, S.~Frixione, V.~Hirschi, F.~Maltoni, O.~Mattelaer
  et~al., \emph{{The automated computation of tree-level and next-to-leading
  order differential cross sections, and their matching to parton shower
  simulations}}, \href{https://doi.org/10.1007/JHEP07(2014)079}{\emph{JHEP}
  {\bfseries 07} (2014) 079} [\href{https://arxiv.org/abs/1405.0301}{{\ttfamily
  1405.0301}}].

\bibitem{Frederix:2018nkq}
R.~Frederix, S.~Frixione, V.~Hirschi, D.~Pagani, H.~S. Shao and M.~Zaro,
  \emph{{The automation of next-to-leading order electroweak calculations}},
  \href{https://doi.org/10.1007/JHEP07(2018)185}{\emph{JHEP} {\bfseries 07}
  (2018) 185} [\href{https://arxiv.org/abs/1804.10017}{{\ttfamily
  1804.10017}}].

\bibitem{Berdine:2007uv}
D.~Berdine, N.~Kauer and D.~Rainwater, \emph{{Breakdown of the Narrow Width
  Approximation for New Physics}},
  \href{https://doi.org/10.1103/PhysRevLett.99.111601}{\emph{Phys. Rev. Lett.}
  {\bfseries 99} (2007) 111601}
  [\href{https://arxiv.org/abs/hep-ph/0703058}{{\ttfamily hep-ph/0703058}}].

\bibitem{Ball:2014uwa}
{\scshape NNPDF} collaboration, \emph{{Parton distributions for the LHC Run
  II}}, \href{https://doi.org/10.1007/JHEP04(2015)040}{\emph{JHEP} {\bfseries
  04} (2015) 040} [\href{https://arxiv.org/abs/1410.8849}{{\ttfamily
  1410.8849}}].

\bibitem{Pythia8}
T.~Sjöstrand, S.~Ask, J.~R. Christiansen, R.~Corke, N.~Desai, P.~Ilten et~al.,
  \emph{{An Introduction to PYTHIA 8.2}},
  \href{https://doi.org/10.1016/j.cpc.2015.01.024}{\emph{Comput. Phys. Commun.}
  {\bfseries 191} (2015) 159}
  [\href{https://arxiv.org/abs/1410.3012}{{\ttfamily 1410.3012}}].

\bibitem{Delphes}
{\scshape DELPHES 3} collaboration, \emph{{DELPHES 3, A modular framework for
  fast simulation of a generic collider experiment}},
  \href{https://doi.org/10.1007/JHEP02(2014)057}{\emph{JHEP} {\bfseries 02}
  (2014) 057} [\href{https://arxiv.org/abs/1307.6346}{{\ttfamily 1307.6346}}].

\bibitem{MadAnalysis5}
E.~Conte and B.~Fuks, \emph{{Confronting new physics theories to LHC data with
  MADANALYSIS 5}}, \href{https://doi.org/10.1142/S0217751X18300272}{\emph{Int.
  J. Mod. Phys. A} {\bfseries 33} (2018) 1830027}
  [\href{https://arxiv.org/abs/1808.00480}{{\ttfamily 1808.00480}}].

\bibitem{DVN/DQZWYL_2021}
J.~Salko and L.~Panizzi, \emph{{Implementation of a search for vector-like
  quarks with charge 5/3 in same-sign dilepton final states (35.9 fb-1; 13 TeV;
  CMS-B2G-17-014)}},  2021.
\newblock 10.14428/DVN/DQZWYL.

\bibitem{Bhattiprolu:2020mwi}
P.~N. Bhattiprolu, S.~P. Martin and J.~D. Wells, \emph{{Criteria for projected
  discovery and exclusion sensitivities of counting experiments}},
  \href{https://doi.org/10.1140/epjc/s10052-020-08819-6}{\emph{Eur. Phys. J. C}
  {\bfseries 81} (2021) 123}
  [\href{https://arxiv.org/abs/2009.07249}{{\ttfamily 2009.07249}}].

\bibitem{Moretti:2016gkr}
S.~Moretti, D.~O'Brien, L.~Panizzi and H.~Prager, \emph{{Production of extra
  quarks at the Large Hadron Collider beyond the Narrow Width Approximation}},
  \href{https://doi.org/10.1103/PhysRevD.96.075035}{\emph{Phys. Rev. D}
  {\bfseries 96} (2017) 075035}
  [\href{https://arxiv.org/abs/1603.09237}{{\ttfamily 1603.09237}}].

\bibitem{Moretti:2017qby}
S.~Moretti, D.~O'Brien, L.~Panizzi and H.~Prager, \emph{{Production of extra
  quarks decaying to Dark Matter beyond the Narrow Width Approximation at the
  LHC}}, \href{https://doi.org/10.1103/PhysRevD.96.035033}{\emph{Phys. Rev. D}
  {\bfseries 96} (2017) 035033}
  [\href{https://arxiv.org/abs/1705.07675}{{\ttfamily 1705.07675}}].

\bibitem{Carvalho:2018jkq}
A.~Carvalho, S.~Moretti, D.~O'Brien, L.~Panizzi and H.~Prager, \emph{{Single
  production of vectorlike quarks with large width at the Large Hadron
  Collider}}, \href{https://doi.org/10.1103/PhysRevD.98.015029}{\emph{Phys.
  Rev. D} {\bfseries 98} (2018) 015029}
  [\href{https://arxiv.org/abs/1805.06402}{{\ttfamily 1805.06402}}].

\bibitem{Crivellin:2018ahj}
A.~Crivellin, M.~Ghezzi, L.~Panizzi, G.~M. Pruna and A.~Signer, \emph{{Low- and
  high-energy phenomenology of a doubly charged scalar}},
  \href{https://doi.org/10.1103/PhysRevD.99.035004}{\emph{Phys. Rev. D}
  {\bfseries 99} (2019) 035004}
  [\href{https://arxiv.org/abs/1807.10224}{{\ttfamily 1807.10224}}].

\bibitem{Cacciapaglia:2018qep}
G.~Cacciapaglia, A.~Carvalho, A.~Deandrea, T.~Flacke, B.~Fuks, D.~Majumder
  et~al., \emph{{Next-to-leading-order predictions for single vector-like quark
  production at the LHC}},
  \href{https://doi.org/10.1016/j.physletb.2019.04.056}{\emph{Phys. Lett. B}
  {\bfseries 793} (2019) 206}
  [\href{https://arxiv.org/abs/1811.05055}{{\ttfamily 1811.05055}}].

\bibitem{Deandrea:2021vje}
A.~Deandrea, T.~Flacke, B.~Fuks, L.~Panizzi and H.-S. Shao, \emph{{Single
  production of vector-like quarks: the effects of large width, interference
  and NLO corrections}},  \href{https://arxiv.org/abs/2105.08745}{{\ttfamily
  2105.08745}}.

\bibitem{Buras:2013dea}
A.~J. Buras, F.~De~Fazio and J.~Girrbach, \emph{{331 models facing new $b \to
  s\mu^+ \mu^-$ data}},
  \href{https://doi.org/10.1007/JHEP02(2014)112}{\emph{JHEP} {\bfseries 02}
  (2014) 112} [\href{https://arxiv.org/abs/1311.6729}{{\ttfamily 1311.6729}}].

\bibitem{Costantini:2020xrn}
A.~Costantini, M.~Ghezzi and G.~M. Pruna, \emph{{Theoretical constraints on the
  Higgs potential of the general $331$ model}},
  \href{https://doi.org/10.1016/j.physletb.2020.135638}{\emph{Phys. Lett. B}
  {\bfseries 808} (2020) 135638}
  [\href{https://arxiv.org/abs/2001.08550}{{\ttfamily 2001.08550}}].

\bibitem{Read:2000ru}
A.~L. Read, \emph{{Modified frequentist analysis of search results (The CL(s)
  method)}},  in \emph{{Workshop on Confidence Limits}}, 8, 2000.

\bibitem{Read:2002hq}
A.~L. Read, \emph{{Presentation of search results: The CL(s) technique}},
  \href{https://doi.org/10.1088/0954-3899/28/10/313}{\emph{J. Phys. G}
  {\bfseries 28} (2002) 2693}.

\bibitem{MadAnalysis5:Expert}
E.~Conte, B.~Dumont, B.~Fuks and C.~Wymant, \emph{{Designing and recasting LHC
  analyses with MadAnalysis 5}},
  \href{https://doi.org/10.1140/epjc/s10052-014-3103-0}{\emph{Eur. Phys. J.}
  {\bfseries C74} (2014) 3103}
  [\href{https://arxiv.org/abs/1405.3982}{{\ttfamily 1405.3982}}].

\bibitem{Cacciari:2008gp}
M.~Cacciari, G.~P. Salam and G.~Soyez, \emph{{The anti-$k_t$ jet clustering
  algorithm}}, \href{https://doi.org/10.1088/1126-6708/2008/04/063}{\emph{JHEP}
  {\bfseries 04} (2008) 063} [\href{https://arxiv.org/abs/0802.1189}{{\ttfamily
  0802.1189}}].

\bibitem{Cacciari:2011ma}
M.~Cacciari, G.~P. Salam and G.~Soyez, \emph{{FastJet User Manual}},
  \href{https://doi.org/10.1140/epjc/s10052-012-1896-2}{\emph{Eur. Phys. J. C}
  {\bfseries 72} (2012) 1896}
  [\href{https://arxiv.org/abs/1111.6097}{{\ttfamily 1111.6097}}].

\bibitem{Artoisenet:2012st}
P.~Artoisenet, R.~Frederix, O.~Mattelaer and R.~Rietkerk, \emph{{Automatic
  spin-entangled decays of heavy resonances in Monte Carlo simulations}},
  \href{https://doi.org/10.1007/JHEP03(2013)015}{\emph{JHEP} {\bfseries 03}
  (2013) 015} [\href{https://arxiv.org/abs/1212.3460}{{\ttfamily 1212.3460}}].

\bibitem{Ball:2012cx}
R.~D. Ball et~al., \emph{{Parton distributions with LHC data}},
  \href{https://doi.org/10.1016/j.nuclphysb.2012.10.003}{\emph{Nucl. Phys. B}
  {\bfseries 867} (2013) 244}
  [\href{https://arxiv.org/abs/1207.1303}{{\ttfamily 1207.1303}}].

\bibitem{Czakon:2011xx}
M.~Czakon and A.~Mitov, \emph{{Top++: A Program for the Calculation of the
  Top-Pair Cross-Section at Hadron Colliders}},
  \href{https://doi.org/10.1016/j.cpc.2014.06.021}{\emph{Comput. Phys. Commun.}
  {\bfseries 185} (2014) 2930}
  [\href{https://arxiv.org/abs/1112.5675}{{\ttfamily 1112.5675}}].

\bibitem{Sjostrand:2014zea}
T.~Sj\"ostrand, S.~Ask, J.~R. Christiansen, R.~Corke, N.~Desai, P.~Ilten
  et~al., \emph{{An introduction to PYTHIA 8.2}},
  \href{https://doi.org/10.1016/j.cpc.2015.01.024}{\emph{Comput. Phys. Commun.}
  {\bfseries 191} (2015) 159}
  [\href{https://arxiv.org/abs/1410.3012}{{\ttfamily 1410.3012}}].

\bibitem{1697570}
{\scshape CMS} collaboration, \emph{{Search for top quark partners with charge
  5/3 in the same-sign dilepton and single-lepton final states in proton-proton
  collisions at $ \sqrt{s}=13 $ TeV}},  2019.
\newblock 10.17182/hepdata.85767.

\end{thebibliography}\endgroup

\end{document}